\documentclass[a4paper,11pt]{scrbook}

\usepackage[utf8]{inputenc}
\usepackage[T1]{fontenc}

\usepackage{lmodern}
\usepackage{microtype}
\usepackage{amsmath,amsfonts,amssymb,amsthm}
\usepackage{mathrsfs}
\usepackage{graphicx}
\usepackage{tensor}
\usepackage{braket}
\usepackage{mleftright} \mleftright
\usepackage{comment}
\usepackage{array}
\usepackage{booktabs}
\usepackage{mathtools} 
\usepackage{mdframed}
\usepackage{cite}
\usepackage{eucal}
\usepackage{slashed}
\usepackage[ngerman,english]{babel}

\usepackage[hidelinks,hypertexnames=false,linktoc=all]{hyperref}
\usepackage{cleveref}
\usepackage{autonum}

\crefname{equation}{}{}

\newtheorem{theorem}{Theorem}[chapter]
\newtheorem{example}[theorem]{Example}
\newtheorem{definition}[theorem]{Definition}
\newtheorem{assumption}[theorem]{Assumption}

\newcommand*{\dd}{\mathop{}\!d}
\newcommand*{\pd}{\mathop{}\!\partial}
\newcommand*{\cd}{\mathop{}\!\nabla}
\newcommand*{\hodge}{\mathop{}\!\ast}
\newcommand*{\bdry}{\mathop{}\!\partial}
\newcommand*{\scri}{\ensuremath{\mathscr{I}}}
\newcommand*{\lied}{\mathop{}\!\mathcal{L}}
\newcommand*{\defn}{\mathrel{\stackrel{\mathclap{\mathrm{def}}}=}}
\newcommand*{\dv}{\mathop{}\!\delta}
\newcommand*{\bdryeq}{\mathrel{\hat=}}
\newcommand*{\pb}[2]{\{#1, #2\}}
\newcommand*{\BMS}{\ensuremath{\textrm{BMS}}}
\newcommand*{\R}{{\mathbb{R}}}
\newcommand*{\contract}{\mathbin{\cdot}}
\newcommand*{\confSpace}{\mathcal{F}}
\newcommand*{\solutions}{\bar{\mathcal{F}}}
\newcommand*{\symplecticStructure}{\Omega}
\newcommand*{\conformalFactor}{{\tilde\Omega}}
\newcommand*{\otherConformalFactor}{{\bar\Omega}}
\newcommand*{\sOfLambda}{{s}}

\newcommand*{\sia}{\mu}
\newcommand*{\sib}{\nu}
\newcommand*{\sic}{\kappa}
\newcommand*{\sid}{\sigma}
\newcommand*{\sie}{\rho}
\newcommand*{\sif}{\lambda}
\newcommand*{\sig}{\chi}
\newcommand*{\sih}{\eta}

\newcommand*{\fia}{i}

\begin{document}

\begin{titlepage}
\vspace*{30mm}
\begin{flushleft}
\sffamily
\Huge
Symmetries and \\
Asymptotically Flat Space \\
\vspace*{0.5em}
\LARGE
\hrule height 1pt \relax
\vspace*{1em}
Friedrich Schöller \\
\end{flushleft}
\end{titlepage}

\begin{titlepage}
\vspace*{\fill}
\noindent Thesis submitted in partial fulfillment of the requirements for the degree of Doctor of Natural Sciences (Dr.\ rer.\ nat.) to the faculty of physics at Technische Universität Wien.
\vspace*{\fill}
\end{titlepage}

\chapter*{Kurzfassung}

\begin{otherlanguage}{ngerman}
Ein ausstehendes Problem in der theoretischen Physik ist die Konstruktion einer Theorie der Quantengravitation.
Für die Lösung dieses Problems ist es nützlich, Naturgesetze zu verstehen, von denen erwartet wird, dass sie in Regimen gelten, die dem Experiment zur Zeit noch unzugänglich sind.
Solche fundamentalen Gesetze können mitunter durch Betrachtung des klassischen Pendants einer Quantentheorie gefunden werden.
Beispielsweise stammen Erhaltungsgrößen in Quantentheorien oft von Erhaltungsgrößen der entsprechenden klassischen Theorie.
Mit dem Ziel derartige Gesetze zu konstruieren, behandelt diese Dissertation den Zusammenhang zwischen Symmetrien und Erhaltungsgrößen von klassischen Feldtheorien und betrachtet Anwendungen auf asymptotisch flache Raumzeiten.

Zu Beginn dieser Arbeit steht die Einführung von Symmetrien in Feldtheorien unter besonderer Berücksichtigung von Variationssymmetrien und deren dazugehörigen Erhaltungsgrößen.
Randbedingungen der allgemeinen Relativitätstheorie auf dreidimensionalen, asymptotisch flachen Raumzeiten in lichtartiger Unendlichkeit werden mithilfe von konformer Vervollständigung der Raumzeit formuliert.
Erhaltungsgrößen, die zu asymptotischen Symmetrien gehören, werden in manifest koordinatenunabhängiger Form konstruiert und untersucht.
In einem separaten Schritt wird ein Koordinatensystem eingeführt, welches den Vergleich mit bestehender Literatur ermöglicht.
Als Nächstes werden all jene asymptotisch flachen Raumzeiten betrachtet, die sowohl eine zukünftige, als auch eine vergangene, lichtartige Unendlichkeit beinhalten.
Die an diesen beiden unzusammenhängenden Gebieten auftretenden asymptotischen Symmetrien werden im dreidimensionalen Fall miteinander verbunden und die entsprechenden Erhaltungsgrößen abgeglichen.
Zuletzt wird gezeigt, wie asymptotische Symmetrien zum Auftreten von verschiedenartigen Minkowski-Räumen führen, welche durch ihre Erhaltungsgrößen differenziert werden können.
\end{otherlanguage}

\chapter*{Abstract}

The construction of a theory of quantum gravity is an outstanding problem that can benefit from better understanding the laws of nature that are expected to hold in regimes currently inaccessible to experiment.
Such fundamental laws can be found by considering the classical counterparts of a quantum theory.
For example, conservation laws in a quantum theory often stem from conservation laws of the corresponding classical theory.
In order to construct such laws, this thesis is concerned with the interplay between symmetries and conservation laws of classical field theories and their application to asymptotically flat spacetimes.

This work begins with an explanation of symmetries in field theories, with a focus on variational symmetries and their associated conservation laws.
Boundary conditions for general relativity are then formulated on three-dimensional asymptotically flat spacetimes at null infinity using the method of conformal completion.
Conserved quantities related to asymptotic symmetry transformations are derived and their properties are studied.
This is done in a manifestly coordinate independent manner.
In a separate step, a coordinate system is introduced such that the results can be compared to existing literature.
Next, asymptotically flat spacetimes which contain both future as well as past null infinity are considered.
Asymptotic symmetries occurring at these disjoint regions of three-dimensional asymptotically flat spacetimes are linked and the corresponding conserved quantities are matched.
Finally, it is shown how asymptotic symmetries lead to the notion of distinct Minkowski spaces that can be differentiated by conserved quantities.

\chapter*{Acknowledgments}

First, I would like to thank my supervisor Daniel Grumiller for his continuous support.
I am very grateful for the opportunity to work in his research group and for the possibility to pursue my own research interests and questions.

I also had great pleasure working with Harald Skarke, who gave me insight into a different and exciting area of research.
Thank you for the many enlightening discussions about our work and physics in general.

I would like to thank Glenn Barnich and Robert McNees for agreeing to referee this thesis.

I am grateful to the office crew consisting of Maria Irakleidou, Stefan Prohazka, and Jakob Salzer for making the office a place that felt like home.
You made my studies a unique journey.

A big thank you to all my friends outside academia for their support and for an unforgettable time.

Last but not least I would like to thank my family, especially my mother Anna, my sister Sabine and my cousin Monika, as well as my girlfriend Raphaela for their love, for believing in me and for encouraging me in all my endeavors.

\chapter*{Note to the Reader}

This thesis is based on the central part of the author's doctoral studies and includes (partially verbatim) the contents of the following publications:
\begin{itemize}
\item[\cite{Prohazka:2017equ}]
S.~Prohazka, J.~Salzer, and F.~Schöller, ``{Linking Past and Future Null Infinity in Three Dimensions},''
\href{http://dx.doi.org/10.1103/PhysRevD.95.086011}{{\em Phys. Rev.} {\bfseries D95} no.~8, (2017) 086011},
\href{http://arxiv.org/abs/1701.06573}{{\ttfamily arXiv:1701.06573 [hep-th]}}.
\item[\cite{Scholler:2017uni}]
F.~Schöller, ``{Distinct Minkowski spaces from Bondi-Metzner-Sachs supertranslations},''
\href{http://dx.doi.org/10.1103/PhysRevD.97.046009}{{\em Phys. Rev.} {\bfseries D97} no.~4, (2018) 046009},
\href{http://arxiv.org/abs/1711.02670}{{\ttfamily arXiv:1711.02670 [gr-qc]}}.
\end{itemize}
In addition to the topics covered here, the author worked on thermodynamics of dilaton gravity in two dimensions
and on the classification of Calabi--Yau fourfolds, which led to the following publications:
\begin{itemize}
\item[\cite{Bagchi:2014ava}]
\noindent A.~Bagchi, D.~Grumiller, J.~Salzer, S.~Sarkar, and F.~Schöller, ``{Flat space cosmologies in two dimensions --- Phase transitions and asymptotic mass-domination},''
\href{http://dx.doi.org/10.1103/PhysRevD.90.084041}{{\em Phys. Rev.} {\bfseries D90} no.~8, (2014) 084041},
\href{http://arxiv.org/abs/1408.5337}{{\ttfamily arXiv:1408.5337 [hep-th]}}.
\item[\cite{Scholler:2018apc}]
F.~Schöller and H.~Skarke, ``{All Weight Systems for Calabi--Yau Fourfolds from Reflexive Polyhedra},''
\href{http://dx.doi.org/10.1007/s00220-019-03331-9}{{\em Commun. Math. Phys.} (2019)},
\href{http://arxiv.org/abs/1808.02422}{{\ttfamily arXiv:1808.02422 [hep-th]}}.
\end{itemize}

\cleardoublepage
\pdfbookmark{\contentsname}{toc}
\setcounter{tocdepth}{1}
\tableofcontents
\setcounter{tocdepth}{2} 

\chapter{Introduction}
\label{cha:introduction}

Experimental guidance towards a solution to the problem of combining general relativity with quantum mechanics is currently out of reach.
Not knowing which features the new theory exhibits in regimes currently inaccessible,
one is confined to looking for a mathematically consistent theory that obeys the laws of quantum mechanics and reduces to general relativity in the classical limit.
This process of quantization --- of constructing a quantum theory from a given classical theory --- is neither unique nor always possible.
The standard way to quantize~\cite{Dirac:1925jy} is by constructing a representation of a subalgebra of the Poisson algebra of the classical system as self-adjoint operators in a Hilbert space subject to conditions forced by the uncertainty relation (for a modern formulation see for example~\cite{woodhouse_geometric_1997}).
While this method is not well understood in general, there are properties of a theory that are often preserved by quantization.
It is those properties that give us an understanding of how a quantized theory might look without actually formulating one.
Among them are conservation laws, which are a central theme of this work.

The road that led to the definition of conserved quantities in general relativity was riddled with obstacles.
First, there is no known local expression for the energy of the gravitational field --- the energy momentum tensor $T_{\sia\sib}$ describes the sum of energy densities of non-gravitational fields only.
This can be anticipated by the fact that, while $T_{\sia\sib}$ is divergence-less ($\cd^\sia T_{\sia\sib} = 0$), a vector is needed,
instead of a two-tensor to get a conserved quantity.
The vector $j^\sia = T^\sia_\sib \xi^\sib$ is conserved when $\xi^\sia$ is a Killing vector, but a general spacetime does not admit any Killing vectors.
The solution to this problem is the study of isolated systems, i.e.\ systems that are far away from any other gravitational sources.
This can be formalized by specifying falloff conditions on the metric tensor.
In general relativity with vanishing cosmological constant, we can demand falloff conditions on the metric $g_{\sia\sib}$ such that its Cartesian components with respect to some background Minkowski metric $\eta_{\sia\sib}$ fall off as
\begin{align}
  g_{\sia\sib} = \eta_{\sia\sib} + O(1/r)
  \,,
\end{align}
where $r$ is the distance with respect to $\eta_{\sia\sib}$ to some arbitrary center point.
It turns out that $\eta_{\sia\sib}$ is not at all unique --- if we make an angle dependent time translation, the components of $\eta_{\sia\sib}$ will change by terms of order $1/r$.
The asymptotic symmetry group is not only the Poincaré group, but the Poincaré groups for different choices of $\eta_{\sia\sib}$ combined --- the Bondi--Metzner--Sachs (BMS) group.
This group is a semidirect product between the Lorentz group and angle dependent translations called \textit{supertranslations}.
Noether's theorem suggests that conserved quantities can be found for each generator of the BMS group.
Straightforward application of Noether's theorem to the Einstein--Hilbert action associates with each vector field $\xi$ that generates a symmetry the \textit{Komar integral}~\cite{Komar:1958wp}
\begin{align}
  Q
  &= \frac{1}{\kappa} \int_\Sigma \cd_\sib { \cd^{[\sib} \xi^{\sia]} } \dd S_\sia
    \,.
\end{align}
One would hope that the Komar integral corresponds to physically relevant conserved quantities, but this is generally not the case.
When studying properties of radiation, one lets $\Sigma$ approach null infinity, i.e.\ the asymptotic region approached by null geodesics.
In this case the Komar integral does not reproduce the correct energy of well known four-dimensional spacetimes.
It also depends on subleading terms of $\xi^\sia$ which correspond to trivial BMS transformations and as such the integral depends on an arbitrary gauge choice~\cite{Geroch:1981ut}.

Bondi, van der Burg and Metzner~\cite{Bondi:1960jsa,Bondi:1962px} found a satisfactory expression for the energy of asymptotically flat spacetimes.
By studying gravitational waves they deduced a conservation law for spacetimes with axially symmetric metric of the form
\begin{multline}
  \dd s^2
  = \left( r^{-1} V e^{2\beta} - r^2 e^{2\gamma} U^2 \right) \dd u^2
    + 2 e^{2 \beta} \dd u \dd r
  \\
    + 2 r^2 U e^{2\gamma} \dd u \dd \theta
    - r^2 \left( e^{2\gamma} \dd \theta^2 + e^{-2\gamma} \sin^2 \theta \dd \varphi^2 \right)
  \,,
\end{multline}
where $U$, $V$, $\beta$, and $\gamma$ are functions of $u$, $r$, and $\theta$.
They defined the \textit{Bondi mass}
\begin{align}
  m(u)
  &= \frac{1}{4} \lim_{r \to \infty} \int_0^{\pi} (r - V) \sin \theta \dd \theta
  \,,
\end{align}
and showed that it is conserved when no gravitational radiation is present.
The analysis was generalized by Sachs~\cite{Sachs:1962wk} to asymptotically flat spacetimes that are not necessarily axially symmetric.
Penrose~\cite{Penrose:1962ij} extended the definition of the Bondi mass to a four-momentum --- the \textit{Bondi energy-momentum}.
The search for quantities associated with Lorentz transformations was completed by Tamburino \& Winicour~\cite{Tamburino:1966zz}, who modified the Komar integral to make it independent of the subleading terms of $\xi^\sia$.
The modification was later shown to be equivalent~\cite{Geroch:1981ut} to the Komar integral together with the gauge condition $\cd_\sia \xi^\sia = 0$.
In the meantime, Geroch~\cite{Geroch:1977jn} discovered the last piece of the puzzle --- quantities associated with arbitrary supertranslations given basically by the integral over the Coulomb part of the Weyl tensor.
The expressions found by Geroch together with the modified Komar integral for Lorentz transformations have the properties that they are diffeomorphism invariant,
lead to a reasonable flux at infinity when radiation is present, match the Bondi energy-momentum for translations, and are zero for Minkowski space.
Penrose~\cite{Penrose:1982wp} introduced a method relying on twistors to rederive the conserved quantities corresponding to Poincaré transformations.
The twistor method was used by Dray \& Streubel~\cite{Dray:1984rfa} to give the first unified derivation of conserved quantities associated with all infinitesimal BMS transformations.

From this history one can anticipate that the definition of conserved quantities with physically reasonable properties is not trivial and that we can benefit from a unified method of deriving them.
A way that achieves this is the use of Hamiltonian mechanics where Hamiltonian functions generate symmetries.
These Hamiltonian functions represent conserved quantities as long as there is no flux of matter or radiation present at infinity.
They are closely related to the conserved quantities obtained by application of Noether's theorem, differing only by terms containing integrals over expressions involving the fields at infinity.
These terms are, however, essential in obtaining correct results.
Wald \& Zoupas~\cite{Wald:1999wa} rederived the expressions of Dray \& Streubel using covariant Hamiltonian mechanics.
This is the method used to obtain conserved quantities in this work.

Any construction of conserved quantities that are defined as integrals over infinitely large regions of spacetime relies on the introduction of boundary conditions.
In general relativity boundary conditions are not only necessary to define how fast fields fall off towards infinity, but since the metric itself is a dynamical quantity they are required to know where infinity even is.
Boundary conditions serve many additional purposes.
They are often imposed so that a quantized version of a theory can be defined.
For this, a well-defined variational principle is typically needed, which means that solutions to the equations of motion correspond to extrema of the action.
A related requirement in the formulation of a quantum theory is that one can turn the classical phase space into a Banach manifold and define a Poisson structure or a symplectic structure on it.
Boundary conditions are important in making these structures well-defined.
Boundary conditions are also necessary for many classical considerations.
They are used to formulate isolated systems where the surroundings are considered to be fixed.
Asymptotically flat spacetimes formalize systems with small cosmological constant that are far away from other gravitational sources.
Boundary conditions are also required to make an initial value problem well posed if the spacetime manifold is not globally hyperbolic.
Regardless of the reason why they are introduced, they have profound impact on symmetries and conservation laws.
Time evolution of gauge symmetries, which are symmetries that can be parameterized by arbitrary functions of time, might get fixed by boundary conditions.
The symmetries cease to be proper gauge symmetries, which makes it possible to define associated conserved quantities.
For this reason, diffeomorphisms can lead to meaningful conservation laws in general relativity as soon as boundary conditions are imposed.

In this work the interplay between symmetries, boundary conditions and conserved quantities of classical field theories is studied.
In \cref{cha:symmetries,cha:conservation-laws} well known results are reviewed to present a coherent image of the concepts involved.
While \cref{cha:symmetries} gives an introduction to different kinds of symmetries in field theories, \cref{cha:conservation-laws} focuses on variational symmetries and their associated conservation laws.
Two kinds of conserved quantities are discussed and related to each other.
The first ones arise in the Lagrangian formulation of a field theory due to Noether's theorem.
The second ones are Hamiltonian functions that generate symmetries on phase space.
After the concepts are developed they are used in \cref{cha:three-dimens-asympt} to derive conservation laws of general relativity on three-dimensional asymptotically flat spacetimes.
Boundary conditions are formulated in a coordinate free manner by conformal completion of spacetime.
Hamiltonian functions generating BMS transformations are derived and their properties are studied.
While the central derivations are performed without reference to any coordinates,
it is shown that a coordinate system can easily be adapted to the boundary conditions.
In \cref{cha:linking-past-future} the BMS transformations occurring at disjoint regions of three-dimensional asymptotically flat spacetimes are linked and the corresponding conserved quantities are matched.
In \cref{cha:dist-mink-spac} it is shown how BMS transformations can be extended into the bulk of Minkowski space so that they are well-defined everywhere.
This leads to the notion of distinct Minkowski spaces differentiated by the values of their conserved quantities.

\chapter{Symmetries}
\label{cha:symmetries}

This chapter serves as an introduction to symmetries of field theories.
Systems of point particles and rigid bodies can be considered as field theories formulated on a one-dimensional manifold that corresponds to time.
In \cref{sec:form-mech} the language used to describe field theories in this work is introduced and Lagrangian mechanics is formulated.
Symmetries of the equations of motion and gauge symmetries are defined in \cref{sec:symm-field-theor} and \cref{sec:gauge-symmetries}, respectively.
Symmetries of the Lagrangian and their relation to symmetries of the equations of motion are reviewed in \cref{sec:vari-symm}.
Finally, the special case of asymptotic symmetries in general relativity is presented in \cref{sec:asympt-symm-gener}.

\section{Field Theory}
\label{sec:form-mech}

The starting point for constructing a classical field theory is a spacetime manifold $M$.
The space of field configurations of a theory is the space of smooth maps from $M$ to some target manifold $N$ (or more generally, the smooth sections of a fiber bundle over $M$), that obey certain boundary conditions.
We assume that the boundary conditions are strong enough to give the space $\confSpace$ of field configurations the structure of an infinite-dimensional Banach manifold (i.e.\ a space that is locally isomorphic to a Banach space in the same sense as a finite dimensional manifold is locally isomorphic to $\R^n$).
This makes it possible to define an exterior derivative on $\confSpace$.
For many of the following arguments this requirement can be relaxed by rephrasing expressions involving the field locally in terms of jet bundles, but it is often more cumbersome to do so.
Given a system of differential equations involving the fields (the equations of motion), we define the space of solutions $\solutions$ as the subspace of $\confSpace$ that satisfies them.

We will mostly be concerned with Lagrangian mechanics,
where the equations of motion are the Euler--Lagrange equations that can be derived using a variational problem.
Lagrangian mechanics can be phrased in terms of differential forms on the product $M \times \confSpace$ of the spacetime manifold $M$ with the manifold $\confSpace$ of field configurations~\cite{MR0650113,Zuckerman:1989cx}.
We define the space $\Omega^{r,s}(M, \confSpace) = \Omega^r(M) \otimes_\R \Omega^s(\confSpace)$ of $(r,s)$-forms on $M \times \confSpace$ (see \cref{cha:diff-forms-prod} for remarks on product manifolds).
Denote the exterior derivatives on $M$ and $\confSpace$ by $\dd$ and $\dv$, respectively.
Of particular interest are local $(r,s)$-forms:
A form is local if its value at some point in $M$ depends on the fields and finitely many of their derivatives at that particular point only.
If the fields in $\confSpace$ are locally described by a basis of real functions $\phi^\fia$, a vector field on $\confSpace$ is uniquely defined by its action on $\phi^\fia$ --- its \textit{characteristic}.
The characteristic\footnote{
The characteristic is sometimes written as $\dv_X \phi^\fia$ in the literature.
We refrain from doing so to avoid ascribing different meanings to the symbol $\dv$ in this text.
} of a vector field $X$ on $\confSpace$ is the tuple $X^\fia$ of $(0,0)$-forms defined as
\begin{align}
  X^\fia
  \defn
  X \contract \dv \phi^\fia
  \equiv
  \lied_X \phi^\fia
  \,,
\end{align}
where the dot donates contraction.
This is analogous to the definition of the components of a vector on spacetime $\xi^\sia = \xi \contract \dd x^\sia$.
A vector field on $\confSpace$ is called \textit{local} if its contraction with local $(r,s)$-forms yields local $(r,s-1)$-forms.
Such vector fields will also just be called \textit{local vector fields} where it is implicit that they are vector fields on $\confSpace$.

The Lagrangian is a local $(n, 0)$-form, with $n$ being the dimension of $M$.
One of the central results~\cite{MR600611} required in this work is the fact that the expression $\dv L$ can always be decomposed into the sum
\begin{align}
  \dv L
  &= E + \dd \theta
  \,,
  \label{eq:dvL}
\end{align}
where $\theta$ is a local $(n-1,1)$-form and $E$ is a local $(n,1)$-form that can be expressed as
\begin{align}
  E
  &= E_\fia \wedge \dv \phi^\fia
  \label{eq:E}
    \,.
\end{align}
The Euler--Lagrange equations are then $E_\fia = 0$.
This decomposition is known to the physicist from the derivation of the Euler--Lagrange equations, where all derivatives of the variation of the fields are, by integration by parts, moved into a boundary term corresponding to $\theta$.
For a review on the fact that it is always possible to achieve decomposition~\cref{eq:dvL} globally see for example Anderson~\cite{MR1188434}.
While $\theta$ is defined up to addition of a local $\dd$-closed form, $E$ is uniquely defined by~\cref{eq:dvL,eq:E} together with the locality requirement.
It immediately follows that a $\dd$-exact Lagrangian leads to no equations of motion.

It is essential to require $\theta$ to be local. The Lagrangian itself, having maximal spacetime form degree, is by the Poincaré lemma always $\dd$-exact (assuming that the topology of spacetime is trivial).
Consequently, if we dropped the locality requirement of $\theta$, we could choose $E$ to vanish.
Consider for example the one-dimensional case where $L = \mathcal{L} \dd t$. We can always write $\mathcal{L} = \frac{\dd}{\dd t} \mathcal{F}$ with non-local $\mathcal{F} = \int_0^t \mathcal{L}(t') \dd t'$.
Then we can choose $\theta = \dv \mathcal{F}$ and $E = 0$.

\begin{example}
  For a massless scalar field the Lagrangian is $L = - \frac12 \pd_\sia \phi \pd^\sia \phi \dd^4 x = - \frac12 \dd \phi \wedge \hodge {\dd \phi}$.
  $\dv L = - \dd {\dv \phi} \wedge \hodge {\dd \phi}$, so that $E = - \dv \phi \wedge \dd { \hodge { \dd \phi } } $ and $\theta = - \dv \phi \wedge \hodge {\dd \phi}$.
\end{example}

\begin{example}
  In electrodynamics the field is the potential one-form $A$ and the Lagrangian is $L = - \frac12 \dd A \wedge \hodge { \dd A }$.
  $\dv L = - \dd {\dv A} \wedge \hodge {\dd A}$, so that $E = - \dv A \wedge \dd { \hodge { \dd A } } $ and $\theta = - \dv A \wedge \hodge {\dd A}$.
\end{example}

\section{Symmetries in Field Theories}
\label{sec:symm-field-theor}

There exists a variety of notions of symmetries in a field theory.
We start with a very general one:
A symmetry group of the equations of motion is a group that acts on the space of field configurations $\confSpace$ in a way that leaves the subspace $\solutions$ of solutions to the equations of motion $E_\fia = 0$ invariant.
Accordingly, a symmetry algebra is a Lie algebra of vector fields $X$ on $\confSpace$ that are tangent to $\solutions$.
We demand from now on that the equations of motion are chosen such that they satisfy the following regularity condition:
\begin{assumption}
A vector field $X$ is tangential to $\solutions$ if and only if it is annihilated by all $\dv E_\fia$ (i.e.\ $X \contract \dv E_\fia \equiv \lied_X E_\fia = 0$) at any point of $\solutions$.
\end{assumption}
If $X$ is annihilated by all $\dv E_\fia$ at any point of $\solutions$ it is said that $X$ satisfies the linearized equations of motion.
We do not require the equations of motion to be the Euler--Lagrange equations of a variational problem,
but we assume that they are local, i.e.\ exclude equations of motion relating fields at different points on $M$.
In order for $\lied_X E_\fia$ to be local as well, we require $X$ to be local.
Local infinitesimal symmetries are also called generalized symmetries (see~\cite{olver2000applications} for an excellent introduction).

\begin{example}
Consider a scalar field $\phi$ on Minkowski space with the single equation of motion $E_1 = \pd^\sia { \pd_\sia \phi }$.
A translation along constant $\xi^\sia$ is generated by the local vector field $X$ with characteristic $X^1 = - \xi^\sia \pd_\sia \phi$.
$X$ is an infinitesimal symmetry: $\lied_X E_1 = - \xi^\sib \pd_\sib { \pd^\sia { \pd_\sia \phi } }$ vanishes for any $\phi$ for which $E_1 = 0$.
More generally, a vector field $X$ with characteristic $X^1 = - \xi^{\sia_1 \cdots \sia_k} \pd_{\sia_1} {\cdots \pd_{\sia_k} \phi}$ and constant $\xi^{\sia_1 \cdots \sia_k}$ is a local infinitesimal symmetry.
\end{example}

\begin{example}
All local infinitesimal symmetries of the vacuum Einstein equations in four spacetime dimensions are given by $X$, such that its characteristic is given by
\begin{align}
  \lied_X g_{\sia\sib}
  &= c g_{\sia\sib} + \cd_\sia \xi_\sib + \cd_\sib \xi_\sia
    \,,
\end{align}
with some constant $c$ and some vector field $\xi^\sia$ that depends locally on the metric and finitely many of its derivatives~\cite{Anderson:1994eg}.
\end{example}

\begin{example}
\label{ex:diffeos}
Consider diffeomorphisms generated by a vector field $\xi$ on the spacetime manifold $M$ and the corresponding local vector field $X(\xi)$ with characteristic
\begin{align}
  \lied_{X(\xi)} \phi^\fia
  &= \lied_\xi \phi^\fia
  \,,
\end{align}
where the Lie derivative on the right hand side is the Lie derivative on $M$.
The action of the commutator of two local vector fields $X(\xi)$ and $X(\zeta)$ is given by
\begin{align}
  \lied_{[X(\xi), X(\zeta)]} \phi^\fia
  &= \lied_{X(\xi)} \lied_{X(\zeta)} \phi^\fia - \lied_{X(\zeta)} \lied_{X(\xi)} \phi^\fia
  \\
  &= \lied_{X(\xi)} \lied_\zeta \phi^\fia - \lied_{X(\zeta)} \lied_\xi \phi^\fia
  \\
  &= \lied_\zeta \lied_{X(\xi)} \phi^\fia - \lied_\xi \lied_{X(\zeta)} \phi^\fia
  \\
  &= \lied_\zeta \lied_\xi \phi^\fia - \lied_\xi \lied_\zeta \phi^\fia
  \\
  &= - \lied_{[\xi, \zeta]} \phi^\fia
  \\
  &= - \lied_{X([\xi, \zeta])} \phi^\fia
  \,,
\end{align}
so that
\begin{align}
  [X(\xi), X(\zeta)]
  &= - X([\xi, \zeta])
  \,.
\end{align}
If $\xi$ is not only a vector field on spacetime but also a function of the fields $\phi^\fia$ then
\begin{align}
  \lied_{[X(\xi), X(\zeta)]} \phi^\fia
  &= \lied_{X(\xi)} \lied_\zeta \phi^\fia - \lied_{X(\zeta)} \lied_\xi \phi^\fia
  \\
  &= \lied_\zeta \lied_\xi \phi^\fia - \lied_\xi \lied_\zeta \phi^\fia
  + \lied_{\lied_{X(\xi)} \zeta} \phi^\fia - \lied_{\lied_{X(\zeta)} \xi} \phi^\fia
  \\
  &= - \lied_{[\xi, \zeta] - \lied_{X(\xi)} \zeta + \lied_{X(\zeta)} \xi} \phi^\fia
  \,,
\end{align}
so that
\begin{align}
  [X(\xi), X(\zeta)]
  &= - X([\xi, \zeta] + \lied_{X(\xi)} \zeta - \lied_{X(\zeta)} \xi)
  \,,
\end{align}
where the Lie derivative acts componentwise on $\xi$ and $\zeta$.
\end{example}

\section{Gauge Symmetries}
\label{sec:gauge-symmetries}

Gauge theories are field theories in which time evolution is not unique because there are fewer independent equations of motion than there are field components.
We say that equations of motion $E_\fia = 0$ are dependent if there exist local differential operators $\mathcal{D}^\fia$ which are not all zero, such that
\begin{align}
  \mathcal{D}^\fia E_\fia
  = 0
  \,.
\end{align}
A local differential operator is an operator $\mathcal{D}$ of the form
\begin{align}
  \mathcal{D}
  &= t + t^{\sia} \pd_\sia + t^{\sia\sib} \pd_\sia{\pd_\sib} + \dots
    \,,
\end{align}
containing finitely many summands, where the components of $t$, $t^\sia$, $t^{\sia\sib}$, and so on, are local functions.
\begin{example}
  Maxwell's equations where $E^\sia = \pd_\sib F^{\sib\sia} - j^\sia$ with some conserved current $j^\sia$ are dependent since $\pd_\sia E^\sia = 0$.
  There is one more field component than independent equations of motion.
\end{example}
\begin{example}
  Einstein's vacuum equations where $E^{\sia\sib} = G^{\sia\sib}$ are dependent since $\cd_\sia G^{\sia\sib} = 0$ by Bianchi's second identity.
  On a spacetime with dimension $n$ there are $n$ more field components than independent equations of motion.
\end{example}

A symmetry of the equations of motion is a map that sends solutions to solutions.
Since time evolution of a gauge theory is not unique, there are symmetries that map one solution to a different solution with the same initial conditions:
the gauge symmetries.
These symmetries have the property that they act independently on different points in spacetime.
We make this notion of the infinitesimal version of a gauge symmetry precise following the lines of~\cite{Lee:1990nz}:
\begin{definition}
\label{def:gauge-symmetry}
An infinitesimal gauge symmetry is a local infinitesimal symmetry $X$, such that
for any two disjoint, closed subsets $C_1, C_2 \subset M$ of spacetime, there exists another local infinitesimal symmetry $Y$ whose characteristic $Y^\fia$ satisfies
\begin{alignat}{2}
  Y^\fia
  &= X^\fia
  & &\quad \text{on} \quad C_1
  \\
  Y^\fia
  &= 0
  & &\quad\text{on} \quad C_2
    \,.
\end{alignat}
\end{definition}
This definition expresses the fact that gauge symmetries can be freely deformed on spacetime.
The action of a gauge symmetry at one region does not specify its action at a disjoint region.

\section{Variational Symmetries}
\label{sec:vari-symm}

Infinitesimal symmetries of Lagrangians are of particular importance because they lead to conservation laws by Noether's theorem as discussed in \cref{sec:noether-thm1}.
We now define the symmetries of a Lagrangian, called variational symmetries, and observe that they are a subset of the infinitesimal symmetries of the equations of motion.

Addition of an exact form to any Lagrangian form does not change the corresponding Euler--Lagrange equations.
This leads to the definition of a variational symmetry as a local vector field $X$ on $\confSpace$ satisfying
\begin{align}
  \lied_X L
  &= \dd \alpha
    \,,
    \label{eq:symmL}
\end{align}
with some local $(n-1, 0)$-form $\alpha$.

Any variational symmetry is an infinitesimal symmetry of the equations of motion.
To show this, take a Lagrangian $L$ with the aforementioned decomposition of its derivative
\begin{align}
  \dv L
  &= E + \dd \theta
    \,,
    \label{eq:dvL2}
\end{align}
such that $E = E_\fia \wedge \dv \phi^\fia$.
Following~\cite{olver2000applications} we define for any local vector field $X$ a new Lagrangian $\tilde L$ as
\begin{align}
  \tilde L
  &= X \contract E
    \,,
    \label{eq:tildeL}
\end{align}
which is decomposed in the same way,
\begin{align}
  \dv \tilde L
  &= \tilde E + \dd \tilde \theta
    \,,
    \label{eq:dvTildeL}
\end{align}
such that $\tilde E = \tilde E_\fia \wedge \dv \phi^\fia$.
For any local vector field $Y$ we then have
\begin{align}
  Y \contract \tilde E
  &=
    Y \contract \dv ( X \contract E )
    - \dd ( Y \contract \tilde \theta )
  \\
  &=
    Y \contract \lied_X E
    - Y \contract X \contract \dv E
    - \dd ( Y \contract \tilde \theta )
  \\
  &=
    Y \contract \lied_X E
    + \dd ( Y \contract X \contract \dv \theta - Y \contract \tilde \theta )
  \\
  &=
    Y^\fia \lied_X E_\fia
    + Y \contract \lied_X \dv \phi^\fia E_\fia
    + \dd ( Y \contract X \contract \dv \theta - Y \contract \tilde \theta )
    \,
\end{align}
where we used~\cref{eq:tildeL,eq:dvTildeL}, Cartan's formula in field space $\lied_X E = X \contract \dv E + \dv ( X \contract E)$, and $\dv E + \dv { \dd \theta } = \dv^2 L = 0$.
Restricting to the solution subspace $\solutions$ this can be written more succinctly as
\begin{align}
  Y^\fia \tilde E_\fia
  &\approx
    Y^\fia \lied_X E_\fia
    + \dd ( Y \contract X \contract \dv \theta - Y \contract \tilde \theta )
  \,,
\end{align}
where $\approx$ denotes equality when both sides are pulled back to $\solutions$ (see conventions in \cref{cha:conventions}).
Since this holds for any $Y$ it follows that
\begin{align}
  \tilde E_\fia
  &\approx
    \lied_X E_\fia
    \,.
\end{align}
Consider now the case that $X$ is a variational symmetry of $L$.
By using~\cref{eq:dvL2,eq:symmL} we find that $\tilde L$ is given by
\begin{align}
  \tilde L
  &= X \contract ( \dv L - \dd \theta )
  = \dd ( \alpha - X \contract \theta )
  \,.
\end{align}
From the fact that $\tilde L$ is exact it follows that $\tilde E$ vanishes and we find that $X$ is an infinitesimal symmetry of the equations of motion
\begin{align}
  \lied_X E_\fia
  &\approx 0
    \,,
\end{align}
or equivalently,
\begin{align}
  \lied_X E
  &= 0
    \quad \text{at}
    \quad \solutions
    \,.
\end{align}

To summarize, any variational symmetry of a Lagrangian is a symmetry of the corresponding Euler--Lagrange equations as well.
The converse is not true. A symmetry of the equations of motion is not necessarily a symmetry of the Lagrangian.
\begin{example}
  Consider a scalar field with Lagrangian $L = - \frac12 \pd_\sia \phi \pd^\sia \phi \dd^n x$.
  While scaling (with characteristic $\lied_X \phi = \phi$) is not a symmetry of the Lagrangian, it is a symmetry of the Euler--Lagrange equations:
  $\lied_X { \pd_\sia { \pd^\sia \phi } } = \pd_\sia { \pd^\sia \phi } \approx 0$.
\end{example}

\section{Asymptotic Symmetries in General Relativity}
\label{sec:asympt-symm-gener}

We now turn to the notion of asymptotic symmetries, which roughly means symmetries that act non-trivially at infinity.
Since we are dealing here with general relativity, a complication arises:
Unlike in theories with a fixed background metric, without boundary conditions there is no notion of infinity common to all solutions to Einstein's equations.
If any solution to the Einstein equations is permitted, a curve of finite length with respect to one metric can have infinite length with respect to another.
In general relativity, boundary conditions have to be imposed before one can even talk about the meaning of an asymptotic symmetry for this reason.

\subsection{Conformal Completion}
\label{sec:conf-comp}

In general relativity, a way to specify boundary conditions is by conformal completion of spacetime~\cite{Penrose:1962ij}.
Here one attaches a boundary to the spacetime at infinity and demands that an unphysical metric exists which can be extended to the boundary and which is related to the physical metric by a conformal transformation.
Depending on the cosmological constant, as well as on the falloff conditions of the energy momentum tensor, the boundary inherits a particular structure.
The boundary condition is the condition that this structure exists and matches a given one.

Conformal completion of a physical spacetime manifold $M$ with metric $g_{\sia\sib}$ is performed as follows.
The manifold $M$ is embedded into a bigger manifold with boundary $\tilde M$, such that the spacetime $M$ is the interior of $\tilde M$.
This manifold is called the unphysical spacetime.
The boundary of $\tilde M$ is referred to as $\scri$.
The unphysical spacetime $\tilde M$ is required to admit a smooth metric $\tilde g_{\sia\sib}$ that is in the interior related to the physical metric $g_{\sia\sib}$ by a conformal transformation
\begin{align}
  \tilde g_{\sia\sib}
  &= \conformalFactor^2 g_{\sia\sib} \,,
\end{align}
with $\conformalFactor$ vanishing at $\scri$.
It follows that, while $\tilde g_{\sia\sib}$ is regular on all of $\tilde M$, $g_{\sia\sib}$ blows up as one approaches $\scri$, formalizing the fact that $\scri$ is infinitely far away.
It is further demanded that the normal vector
\begin{align}
  \tilde n^\sia
  &= \tilde g^{\sia\sib} \cd_\sib \conformalFactor
    \,,
\end{align}
vanishes nowhere on $\scri$, which fixes the smooth structure of $\tilde M$.
It is possible and sometimes necessary (see for example~\cite{Chrusciel:1993hx}) to relax smoothness of the unphysical metric at $\scri$, but this direction will not be pursued in this work.

Depending on the conditions imposed on the energy-momentum tensor we obtain additional structure at infinity, which we call asymptotic structure.
Below, we denote by ``$\bdryeq$'' equality in the limit as $\scri$ is approached (see conventions in \cref{cha:conventions}).
Indices of tensors with and without tilde are raised and lowered with $\tilde g_{\sia\sib}$ and $g_{\sia\sib}$, respectively.
For spacetimes satisfying Einstein's equation with spacetime dimension $n>2$,
when the trace of the energy momentum tensor vanishes at $\scri$ ($T \bdryeq 0$) it follows (see \cref{cha:conformal-boundary}) that the norm of the normal vector is proportional to the cosmological constant:
\begin{align}
  \tilde n^\sia \tilde n_\sia
  &\bdryeq
    - \frac{2}{(n-2)(n-1)} \Lambda
    \label{eq:norm-n-from-lambda}
\end{align}

We now consider symmetries obtained by the action of diffeomorphisms from the unphysical spacetime to itself.
Diffeomorphisms that keep $\scri$ fixed are considered to be trivial.
The group of asymptotic symmetries is defined as the quotient of the group of diffeomorphisms that leave the asymptotic structure invariant by the subgroup of trivial diffeomorphisms.
An infinitesimal asymptotic symmetry is a vector $X$ that generates asymptotic symmetries.
It acts on tensor fields as the Lie derivative
\begin{align}
  \lied_X \phi^\fia
  &= \lied_\xi \phi^\fia
  \,,
\end{align}
where $\xi^\sia$ is a vector fields such that the asymptotic structure is invariant under the action.

\subsection{Asymptotically Anti-de~Sitter}

For a negative cosmological constant $\Lambda < 0$ it follows from~\cref{eq:norm-n-from-lambda} that $\scri$ is a timelike boundary.
The unphysical metric $\tilde g_{\sia\sib}$ induces a metric $\underline{\tilde g}{}_{\sia\sib}$ with Lorentzian signature on $\scri$, where the underline denotes the pullback to $\scri$ (see conventions in \cref{cha:conventions}).
Under a change of conformal factor $\conformalFactor \mapsto \lambda \conformalFactor$, the boundary metric changes accordingly as $\underline{\tilde g}{}_{\sia\sib} \mapsto \lambda^2 \underline{\tilde g}{}_{\sia\sib}$.
The equivalence class of metrics at $\scri$ modulo conformal transformations is independent of the choice of $\conformalFactor$, so the asymptotic structure is given by the manifold $\scri$ together with its conformal structure.

Since the boundary is timelike, there cannot be any Cauchy hypersurfaces and the spacetime is not globally hyperbolic.
It follows that boundary conditions have to be imposed in order to make time evolution well-defined.
Typical boundary conditions are that the boundary metric $\underline{\tilde g}{}_{\sia\sib}$ lies in the conformal equivalence class of the Einstein static Universe~\cite{Hollands:2005wt}
\begin{align}
  \underline{\tilde g}{}_{\sia\sib} \dd x^\sia \dd x^\sib
  \propto - \dd t^2 + \dd \sigma^2
  \,,
\end{align}
where $\dd \sigma^2$ is the line element of the unit sphere $S^{n-2}$.

An infinitesimal asymptotic symmetry $X$ is parameterized by some vector field $\xi^\sia$ tangent to $\scri$ (i.e.\ $\xi^\sia \tilde n_\sia \bdryeq 0$) that acts via the Lie derivative along $\xi^\sia$ on the metric
\begin{align}
  \lied_X g_{\sia\sib}
  &= \lied_\xi g_{\sia\sib}
    \,,
    \label{eq:ads-action-on-g}
\end{align}
such that the asymptotic structure is left invariant, i.e.
\begin{align}
  \lied_X \underline{\tilde g}_{\sia\sib}
  &\bdryeq \kappa^2 \underline{\tilde g}_{\sia\sib}
  \,,
\end{align}
with some smooth function $\kappa$.
By~\cref{eq:ads-action-on-g} this is equivalent to
\begin{align}
  \lied_\xi \underline{\tilde g}_{\sia\sib}
  - 2 \conformalFactor^{-1} \xi^\sic \tilde n_\sic \underline{\tilde g}_{\sia\sib}
  &\bdryeq \kappa^2 \underline{\tilde g}_{\sia\sib}
  \,.
\end{align}

\subsection{Asymptotically Flat}
\label{sec:asymptotically-flat}

For vanishing cosmological constant, $\scri$ is a null boundary by~\cref{eq:norm-n-from-lambda}.
Such spacetimes are called asymptotically flat at null infinity.
Since $\scri$ is null $\tilde n^\sia$ is tangent to $\scri$.
The asymptotic structure is an equivalence class of pairs $(\underline{\tilde g}{}_{\sia\sib}, \tilde n^\sia)$ evaluated at $\scri$, where the underline again denotes the pullback to $\scri$.
Two such pairs are equivalent if they are related by a conformal transformation $(\underline{\tilde g}{}_{\sia\sib}, \tilde n^\sia) \sim (\lambda^2 \underline{\tilde g}{}_{\sia\sib}, \lambda^{-1} \tilde n^\sia)$, with $\lambda$ being some smooth, nonvanishing function.

We define BMS transformations as asymptotic symmetries following Geroch~\cite{Geroch:1977jn}.
A BMS transformation is defined as a diffeomorphism around $\scri$ that preserves the asymptotic structure.
A trivial BMS transformation is a BMS transformation that keeps $\scri$ fixed.
Any BMS transformation can be combined with a trivial one, such that the pair $(\underline{\tilde g}{}_{\sia\sib}, \tilde n^\sia)$ is invariant, not only its conformal equivalence class.

A vector field $\xi^\sia$ that is tangent to $\scri$ ($n_\sia \xi^\sia \bdryeq 0$) is the generator of a BMS transformation if it acts as the Lie derivative
\begin{align}
  \lied_X g_{\sia\sib}
  &= \lied_\xi g_{\sia\sib}
    \,,
\end{align}
such that the asymptotic structure is left invariant, i.e.
\begin{align}
  \lied_X \underline{\tilde g}{}_{\sia\sib}
  &\bdryeq - 2 \kappa \underline{\tilde g}{}_{\sia\sib}
  \\
  \lied_X \tilde n^\sia
  &\bdryeq \kappa \tilde n{}^\sia
  \,,
\end{align}
with some smooth function $\kappa$.
Equivalently,
\begin{align}
  \lied_\xi \underline{\tilde g}{}_{\sia\sib}
  - 2 \conformalFactor^{-1} \xi^\sic \tilde n_\sic \underline{\tilde g}{}_{\sia\sib}
  &\bdryeq - 2 \kappa \underline{\tilde g}{}_{\sia\sib}
  \\
  \lied_\xi \tilde n^\sia
  + \conformalFactor^{-1} \xi^\sic \tilde n_\sic \tilde n^\sia
  &\bdryeq \kappa \tilde n{}^\sia
  \,.
\end{align}
By using the fact that $\tilde n^\sib \lied_X \tilde g_{\sia\sib} = - \tilde g_{\sia\sib} \lied_X \tilde n^\sib$
this is also equivalent to
\begin{align}
\label{eq:bmsDef1b}
\begin{aligned}
  \lied_\xi \tilde g_{\sia\sib}
  - 2 \conformalFactor^{-1} \xi^\sic \tilde n_\sic \tilde g_{\sia\sib}
  &\bdryeq - 2 \kappa \tilde g_{\sia\sib} + 2 \tilde n_{(\sia} \tilde t_{\sib)}
  \\
  \tilde t_\sia \tilde n^\sia
  &\bdryeq \kappa
  \,,
\end{aligned}
\end{align}
with the same smooth function $\kappa$ and some smooth one-form $\tilde t_\sia$.

As noted before we can add a trivial BMS transformation such that the boundary metric is fixed.
By replacing $\xi^\sia$ with $\xi^\sia + \conformalFactor v^\sia$, where $v^\sia$ is some smooth vector field such that  $\tilde n_\sia v^\sia = \kappa$,
the condition for $\xi^\sia$ to generate a BMS transformation becomes
\begin{align}
\label{eq:bmsDef2}
\begin{aligned}
  \lied_\xi \underline{\tilde g}{}_{\sia\sib}
  - 2 \conformalFactor^{-1} \xi^\sic \tilde n_\sic \underline{\tilde g}{}_{\sia\sib}
  &\bdryeq 0
  \\
  \lied_\xi \tilde n^\sia
  + \conformalFactor^{-1} \xi^\sic \tilde n_\sic \tilde n^\sia
  &\bdryeq 0
  \,.
\end{aligned}
\end{align}
One particular choice is to set $v^\sia = \tilde t^\sia$ from~\cref{eq:bmsDef1b}.
It follows that by fixing one subleading order of $\xi^\sia$ in the expansion around $\scri$ we can demand that a BMS symmetry satisfies
\begin{align}
  \lied_\xi \tilde g_{\sia\sib}
  - 2 \conformalFactor^{-1} \xi^\sic \tilde n_\sic \tilde g_{\sia\sib}
  &\bdryeq 0
  \,,
  \label{eq:bmsDef3}
\end{align}
or equivalently
\begin{align}
  \conformalFactor^2 \lied_\xi g_{\sia\sib}
  &\bdryeq 0
  \,.
\end{align}

The supertranslations are a normal subgroup of BMS transformations defined as follows.
A \textit{supertranslation} is a BMS transformation that is generated by a smooth vector field $\xi^\sia$ such that
\begin{align}
  \xi^\sia
  &\bdryeq h \tilde n^\sia
    \,,
\end{align}
with some smooth function $h$.
The supertranslations form an abelian normal subgroup of the BMS transformations~\cite{Sachs:1962zza}.
Assume that $\scri$ has topology $B \times \R$, such that by picking a $B$ slice, $\scri$ consists of the integral lines along $\tilde n^\sia$ that go through $B$.
Then the quotient of the BMS transformations by the supertranslations is isomorphic to the group of conformal transformations of $B$~\cite{Geroch:1977jn}.
We consider now the typical case, where $B$ is the sphere $\mathbb{S}^{n-2}$, with $n$ being the spacetime dimension.

In four spacetime dimensions the quotient is the Lorentz group~\cite{Sachs:1962zza}.
If we require the BMS transformations to be defined only locally on $\mathbb{S}^2$, the quotient is much bigger and called superrotations~\cite{Barnich:2009se}.
Consider the group of Poincaré transformations, which is a semidirect product between the Lorentz group and the translations.
Similarly, the group of BMS transformations is a semidirect product between the Lorentz group and the supertranslations.
In the Poincaré case there is not a single Lorentz subgroup, but there are many, one for each choice of base point around which to rotate or boost.
The different Lorentz subgroups are all related by translations.
The BMS case is similar: there is no unique Lorentz subgroup, there are many, each one related to another by a supertranslation.
In four spacetime dimensions there is exactly one four-dimensional normal subgroup of the BMS group: the translation group~\cite{Sachs:1962zza}.

In three spacetime dimensions the quotient of the BMS transformations by the supertranslations is the infinite-dimensional group of diffeomorphisms of $\mathbb{S}^1$.
In contrast to the four-dimensional case there is no way to single out a translation subgroup without introducing additional structure~\cite{Ashtekar:1996cd}.

\chapter{Conservation Laws}
\label{cha:conservation-laws}

Noether showed in her famous work~\cite{Noether:1918zz} that in a Lagrangian system there is a correspondence between variational symmetries and conserved currents.
Integrating these conserved currents over a hypersurface in spacetime gives quantities obeying certain conservation laws.
Adding appropriate boundary terms to the integrals gives Hamiltonian functions that generate the symmetries.
In \cref{sec:noether-thm1,sec:noether-thm2} Noether's first and second theorem are reviewed.
There it will be shown that dependent equations of motion in a Lagrangian system always lead to gauge symmetries that are also variational symmetries.
The notion of phase space is discussed in \cref{sec:phase-space,sec:covar-phase-space}.
The connection between Noether currents and Hamiltonian functions is explained in \cref{sec:local-hamilt-funct} and the ambiguities in their definition is discussed in \cref{sec:ambiguities}.

\section{Noether's First Theorem}
\label{sec:noether-thm1}

Consider the Euler--Lagrange equations $E_\fia = 0$, where $E_\fia$ is defined in~\cref{eq:E}.
Noether's first theorem~\cite{Noether:1918zz} states that for every variational symmetry $X$ the expression $E_\fia X^\fia$ is a local, $\dd$-exact form, i.e.
\begin{align}
  E_\fia X^\fia
  &= - \dd j
    \label{eq:eomExact}
    \,,
\end{align}
with some local $(n-1,0)$-form $j$.
Conversely, if~\cref{eq:eomExact} holds for some local vector field $X$ then $X$ is a variational symmetry.
Let $X$ be a variational symmetry, i.e.\ a local vector field satisfying~\cref{eq:symmL}.
From contracting~\cref{eq:dvL} with $X$ it follows that~\cref{eq:eomExact} holds with the local \textit{Noether current} $(n-1, 0)$-form
\begin{align}
  j
  &= X \contract \theta - \alpha
    \,.
    \label{eq:noether-current}
\end{align}
The converse follows by reversing the argument.

It is apparent that the Noether current is closed on-shell, i.e.\ $\dd j \approx 0$.
It also follows from its definition that the Noether current is not unique.
The form $\alpha$, and therefore $j$, is defined up to addition of a local $\dd$-closed term only.
Any $\dd$-exact term is $\dd$-closed, so there is the ambiguity
\begin{align}
  \label{eq:noetherAmbig}
  j
  &\mapsto j + \dd k
    \,,
\end{align}
where $k$ is a local $(n-2,0)$-form.
By the algebraic Poincaré lemma (see \cref{cha:cohom-local-forms}) there is, at least locally, no additional ambiguity apart from~\cref{eq:noetherAmbig}.

\section{Noether's Second Theorem}
\label{sec:noether-thm2}

Since there are as many equations of motion as fields in a Lagrangian system, the time evolution is not unique as soon as the equations of motion are dependent (see \cref{sec:gauge-symmetries}), i.e.
\begin{align}
  \mathcal{D}^\fia E_\fia
  = 0
  \,.
  \label{eq:eomdep}
\end{align}
This also follows from Noether's second theorem~\cite{Noether:1918zz}, which states that for each set of local operators $\mathcal{D}^\fia$ satisfying~\cref{eq:eomdep}
there are variational symmetries $X_\Lambda$ parameterized by a smooth function $\Lambda$ whose Noether current vanishes on-shell (up to addition of a local, $\dd$-closed form).
The function $\Lambda$ is restricted by the boundary conditions only.
Furthermore, $X_\Lambda^\fia = \tilde{\mathcal{D}}^\fia \Lambda$, where $\tilde{\mathcal{D}}^\fia$ are the formal adjoints of $\mathcal{D}^\fia$, i.e.\ the operators satisfying
\begin{align}
  \psi \mathcal{D}^\fia P_\fia
  = P_\fia \tilde{\mathcal{D}}^\fia \psi
  + \dd s
  \label{eq:adjoint}
\end{align}
for all local $(n, 0)$-forms $P_\fia$ and local functions $\psi$, where $s$ is some local $(n-1, 0)$-form that depends bilinearly on $P_\fia$ and $\psi$.
Locally, the operators $\tilde{\mathcal{D}}^\fia$ and the form $s$ can be constructed by using Leibniz's rule to move derivatives acting on $P_\fia$ to the other side such that they act on $\psi$ instead, while collecting total divergences in $s$.
The same can be achieved globally by first rewriting the operators $\mathcal{D}^\fia$ in terms of an arbitrarily chosen symmetric connection on the spacetime manifold.
The operators $\tilde{\mathcal{D}}^\fia$ are uniquely defined and independent of the choice of connection.
Noether's second theorem is proved by setting $P_\fia = E_\fia$ in~\cref{eq:adjoint} which gives that
\begin{align}
  E_\fia \tilde{\mathcal{D}}^\fia \Lambda
  = - \dd s_\Lambda
  \,,
\end{align}
where $s_\Lambda$ is a local $(n-1, 0$)-form.
This equation has the form of~\cref{eq:eomExact}, so we can immediately conclude that $X_\Lambda^\fia = \tilde{\mathcal{D}}^\fia \Lambda$ is a variational symmetry with corresponding Noether current $s_\Lambda$.
Since $\Lambda$ is arbitrary in the bulk, $X_\Lambda$ is a gauge symmetry, at least if $\Lambda$ vanishes in a neighborhood around the boundary.
From the way the Noether current was constructed it follows that it is linear in $\Lambda$ and vanishes on-shell,
\begin{align}
  s_\Lambda
  &\approx 0
    \,.
\end{align}
The ambiguity in the definition of the Noether current is such that any other Noether current of the symmetry $X_\Lambda$ is obtained by adding a closed form to $s_\Lambda$.
We see that the Noether current of a gauge symmetry is closed on-shell and, by the algebraic Poincaré lemma, locally exact on-shell.
If $j_\Lambda$ is constructed such that it depends linearly on $\Lambda$, it follows that it is globally exact on-shell and we write
\begin{align}
  j_\Lambda
  &\approx \dd q_\Lambda
  \,,
\end{align}
where $q_\Lambda$ is the local \textit{Noether charge} $(n-2, 0)$-form.
Global exactness can be proven by promoting $\Lambda$ to be a field in the space of field configurations of the theory without adding any new equations of motion.
Since $j_\Lambda$ is linear in $\Lambda$ one can then replace $\Lambda$ by $\dv \Lambda$ and use \cref{thm:horizontal-complex-exact-onshell} to show that $j_{\dv \Lambda}$ and therefore $j_\Lambda$ is exact on-shell.

\section{Symplectic Geometry}
\label{sec:phase-space}

A classical mechanical system is described by its space of possible states --- its \textit{phase space}.
The phase space is a symplectic manifold, i.e.\ a manifold with a closed, nondegenerate~2-form $\symplecticStructure$ --- its \textit{symplectic structure}.
Diffeomorphisms leaving $\symplecticStructure$ invariant are called symplectomorphisms or, traditionally, ``canonical transformations''.
Vector fields $X$ that generate symplectomorphisms have to satisfy $\lied_X \symplecticStructure = 0$ in order to preserve $\symplecticStructure$.
Since $\symplecticStructure$ is closed, an equivalent condition can be obtained using Cartan's formula and reads
\begin{align}
  \dv ( X \contract \symplecticStructure )
  &= 0
  \,,
\end{align}
where the exterior derivative on phase space is denoted by $\dv$ to be in line with the conventions of later sections.
By the Poincaré lemma, it follows that, locally
\begin{align}
  X \contract \symplecticStructure
  &= - \dv H
    \,,
\end{align}
for some real function $H$.
If the relation holds globally, $H$ is called a Hamiltonian function for $X$.
If $X$ generates time translations, $H$ is the Hamiltonian in the traditional sense.
Conversely, since $\symplecticStructure$ is nondegenerate, there is a unique vector field $X_H$ for any Hamiltonian function $H$ such that
\begin{align}
  X_H \contract \symplecticStructure
  &= - \dv H
    \label{eq:hamiltonian}
    \,.
\end{align}
The Poisson bracket $\pb{H}{K}$ of two Hamiltonian functions $H$ and $K$ is defined by
\begin{align}
  \pb{H}{K}
  = X_K \contract X_H \contract \symplecticStructure
  \equiv \symplecticStructure(X_H, X_K)
  \,.
\end{align}
By using~\cref{eq:hamiltonian} we find equivalent definitions
\begin{align}
  \pb{H}{K}
  = X_H \contract \dv K
  = - X_K \contract \dv H
  \label{eq:poisson-bracket-alt}
  \,.
\end{align}
It follows that the change of $K$ along integral curves of $X_H$ can be written using the Poisson bracket
\begin{align}
  \dot K
  &= \lied_{X_H} K
  = \pb{H}{K}
  \,.
\end{align}
The Poisson bracket is antisymmetric
\begin{align}
  \pb{H}{K} = - \pb{K}{H}
  \,,
\end{align}
satisfies the Jacobi identity
\begin{align}
  \pb{H}{\pb{K}{L}}
  + \pb{L}{\pb{H}{K}}
  + \pb{K}{\pb{L}{H}}
  = 0
  \,,
\end{align}
and makes the map that sends the Hamiltonian function $H$ to the vector field $X_H$ into a Lie algebra homomorphism
\begin{align}
  [X_H, X_K]
  = X_{\pb{H}{K}}
  \label{eq:hamiltonian-to-vector-hom}
  \,.
\end{align}
Antisymmetry follows from antisymmetry of $\Omega$.
To show that~\cref{eq:hamiltonian-to-vector-hom} holds we use the fact that the action of $\dv$ on the Poisson bracket can be expressed as follows
\begin{align}
  \label{eq:dv-pb}
  \dv \pb{H}{K}
  &= \dv ( X_K \contract X_H \contract \symplecticStructure )
  \\
  &= \lied_{X_K} ( X_H \contract \symplecticStructure )
  \\
  &= \lied_{X_K} X_H \contract \symplecticStructure
  \\
  &= - [X_H, X_K] \contract \symplecticStructure
  \,.
\end{align}
Replacing $H$ with $\pb{H}{K}$ in~\cref{eq:hamiltonian} gives then
\begin{align}
  X_{\pb{H}{K}} \contract \Omega
  &= [X_H, X_K] \contract \symplecticStructure
  \,,
\end{align}
which together with the fact that $\Omega$ is nondegenerate completes the proof.
That the Jacobi identity holds follows from the equation
\begin{align}
  \pb{H}{\pb{K}{L}}
  + \pb{L}{\pb{H}{K}}
  + \pb{K}{\pb{L}{H}}
  &= - \lied_{X_{\pb{K}{L}}} H
  - \lied_{X_L} { \lied_{X_K} H }
  + \lied_{X_K} { \lied_{X_L} H }
  \nonumber \\
  &=
  - \lied_{X_{\pb{K}{L}}} H
  + \lied_{[X_K, X_L]} H
  \,,
\end{align}
whose right hand side vanishes by~\cref{eq:hamiltonian-to-vector-hom}.
\begin{example}
Consider a particle moving in $n$-dimensional, Euclidean space.
Its position is given by the coordinates $q_i$, its momentum by $p_i$.
Here $i$ runs from 1 to $n$.
The phase space is $\R^{2n}$ with coordinates $q_i$ and $p_i$.
The symplectic structure is
\begin{align}
  \symplecticStructure
  &= \sum_i \dv p_i \dv q_i
    \,.
\end{align}
It follows that the vector field $X_H$ is given by
\begin{align}
  X_H
  &= \frac{\pd H}{\pd p_i} \frac{\pd}{\pd q_i}
    - \frac{\pd H}{\pd q_i} \frac{\pd}{\pd p_i}
    \,,
\end{align}
and an integral curve $(q_i(t), p_i(t))$ of the vector field $X_H$ has to satisfy
\begin{align}
  \dot q_i
  &= \frac{\pd H}{\pd p_i}
  \\
  \dot p_i
  &= - \frac{\pd H}{\pd q_i}
    \,,
\end{align}
which are Hamilton's equations.
The Poisson bracket is given by
\begin{align}
  \pb{H}{K}
  &= \frac{\pd H}{\pd p_i} \frac{\pd K}{\pd q_i}
    - \frac{\pd H}{\pd q_i} \frac{\pd K}{\pd p_i}
    \,.
\end{align}
\end{example}

Suppose there is a map $X \mapsto H_X$ from some subalgebra of vector fields to the real functions with the property that each $H_X$ is a Hamiltonian function for $X$, i.e.
\begin{align}
  X \contract \symplecticStructure
  &= - \dv H_X
  \,.
\end{align}
Consider two vector fields $X$ and $Y$.
The Poisson bracket of the Hamiltonian functions $H_X$ and $H_Y$ is related to the Hamiltonian function $H_{[X, Y]}$ via
\begin{align}
  \pb{H_X}{H_Y}
  &= H_{[X, Y]} + C(X, Y)
  \,,
  \label{eq:poisson-central-extension}
\end{align}
where $C(X, Y)$ is some constant, i.e.\ $\dv C(X, Y) = 0$.
This can be seen by acting with $\dv$ on~\cref{eq:poisson-central-extension} to obtain the relation
\begin{align}
  \dv \pb{H_X}{H_Y}
  &= \dv H_{[X, Y]} + \dv C(X, Y)
  \\
  &= - [X, Y] \contract \symplecticStructure + \dv C(X, Y)
  \,,
\end{align}
from which it follows using~\cref{eq:dv-pb} that $\dv C(X, Y) = 0$.

When the phase space is infinite-dimensional an additional distinction arises:
The form $\symplecticStructure$ is called weakly nondegenerate if the map $X \mapsto X \contract \symplecticStructure$ is injective, i.e.\ $X \contract \symplecticStructure = 0$ implies that $X = 0$.
It is called (strongly) nondegenerate if the map $X \mapsto X \contract \symplecticStructure$ is an isomorphism.
If $\symplecticStructure$ is only weakly nondegenerate the vector field $X_H$ does not necessary exist.
For details in the case of infinite-dimensional phase spaces see for example Chernoff \& Marsden~\cite{MR0650113}.

\begin{example}
  Consider the space of smooth functions $\phi(x)$ on the interval $[0, 1]$.
  The symplectic structure $\symplecticStructure = \int_0^1 \pd_x { \dv \phi(x) } \dv \phi(x) \dd x$ is weakly nondegenerate.
  There is no vector field with Hamiltonian function $H = \int_0^1 \phi(x) \dd x$.
  Set $f = X_H \contract \dv \phi$, then~\cref{eq:hamiltonian} reads
  $2 \int_0^1 \dv \phi(x) \pd_x f(x) \dd x - \dv \phi(x) f(x) |_0^1 = - \int_0^1 \dv \phi(x) \dd x$,
  which is equivalent to the incompatible conditions $\pd_x f = - \frac12$ and $f(0) = f(1) = 0$.
\end{example}

It is often useful to work with an enlarged description of phase space where different points can correspond to the same physical states.
To do this we relax a condition on the symplectic structure, allow it to be degenerate and call it \textit{presymplectic structure}.
What changes is that the vector fields $X_H$ are not uniquely defined anymore.
Moreover the Poisson bracket is only defined for Hamiltonian functions that are constant in directions annihilated by the symplectic structure.
In the following we will work with such a presymplectic structure.
Its null directions correspond to gauge symmetries.
The fact that $X_H$ is no longer unique corresponds to the nonuniqueness of time evolution in gauge systems.

\section{Covariant Phase Space}
\label{sec:covar-phase-space}

There is a nice way of constructing the phase space of a theory without introducing the split of space and time that usually occurs when going to a Hamiltonian formulation of a system.
The Lagrangian is used as a starting point to construct the \textit{covariant phase space}~\cite{Crnkovic:1986ex,Zuckerman:1989cx,Ashtekar:1990gc,Lee:1990nz}.
Recall the splitting in~\cref{eq:dvL}, i.e.
\begin{align}
  \dv L
  &= E + \dd \theta
    \,.
    \label{eq:dvLagain}
\end{align}
We define the \textit{symplectic current} $(n-1,2)$-form $\omega$ as
\begin{align}
  \omega
  &= \dv \theta
    \label{eq:omega}
    \,.
\end{align}
The pullback of the symplectic current to the space of solutions $\solutions$ is $\dd$-closed,
which can be seen by acting with $\dv$ on~\cref{eq:dvLagain} yielding
\begin{align}
  \dd \omega
  &= - \dv E \approx 0
    \,.
    \label{eq:conservedOmega}
\end{align}
Define $\symplecticStructure_\Sigma$ by integrating the symplectic current $\omega$ over a hypersurface $\Sigma$.
\begin{align}
  \symplecticStructure_\Sigma
  &= \int_\Sigma \omega
    \label{eq:symplecticStructure}
    \,.
\end{align}
From~\cref{eq:omega} it follows that
\begin{align}
  \dv \symplecticStructure_\Sigma = 0
  \,,
\end{align}
so $\symplecticStructure_\Sigma$ is a closed 2-form on $\confSpace$, which makes it a presymplectic structure.
The subspace $\solutions$ of field configurations that satisfy the Euler--Lagrange equations is the covariant phase space.
For a fixed boundary $\bdry \Sigma$ it follows from~\cref{eq:conservedOmega} that the pullback of $\symplecticStructure_\Sigma$ to $\solutions$ depends only on the homology class of $\Sigma$.
When there is no gauge symmetry, there is a one-to-one correspondence between initial data at a given time and solutions to the equations of motion, and $\solutions$ is just the ordinary phase space spanned by the positions and momenta of a theory.
In general it can be shown~\cite{Barnich:1991tc} that the formulation in terms of the covariant phase space with the presymplectic structure $\symplecticStructure_\Sigma$ is equivalent to the Hamiltonian formulation by Dirac~\cite{Dirac:1958sq}.

\section{Conserved Symplectic Structure}
\label{sec:cons-sympl-struct}

\newcommand*{\futureBdry}{{\Sigma_2}}
\newcommand*{\pastBdry}{{\Sigma_1}}
\newcommand*{\spatialBdry}{{\mathcal{T}}}
\newcommand*{\spacetime}{B}

\begin{figure}
  \centering
  \includegraphics{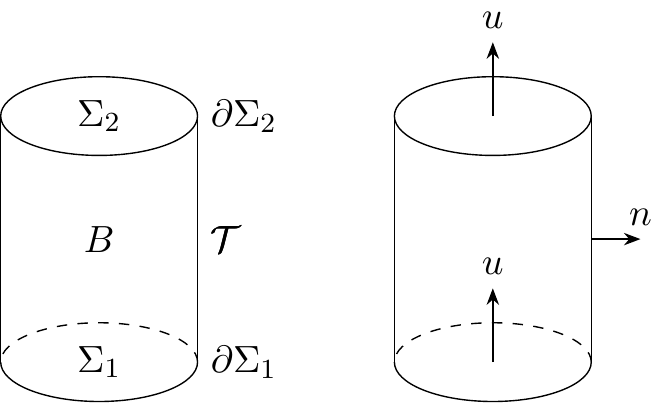}
  \caption{Spacetime with boundaries.}
  \label{fig:boundaries}
\end{figure}

In this section an implication~\cite{Iyer:1995kg} of a well-defined variational principle on the presymplectic structure is reviewed.
Consider the region $\spacetime$ in spacetime lying between two time-slices $\Sigma_1$ and $\Sigma_2$.
Assume that $\spacetime$ is orientable and has additionally to the boundaries $\Sigma_1$ and $\Sigma_2$ the boundary $\spatialBdry$ (typically at infinity) as illustrated in \cref{fig:boundaries}.
From the definition~\cref{eq:symplecticStructure} of $\symplecticStructure_\Sigma$ it follows using stokes theorem that
\begin{align}
  \symplecticStructure_\futureBdry - \symplecticStructure_\pastBdry
  &= \int_\futureBdry \omega
  - \int_\pastBdry \omega
  = \int_\spacetime \dd \omega
  - \int_\spatialBdry \omega
  \approx - \int_\spatialBdry \omega
  \,.
\end{align}
If $\spatialBdry$ is infinitely far away, the integral there can sometimes be neglected so that the presymplectic structure is conserved, i.e.\ it does not depend on $\Sigma$.
In general however, there is a flux of symplectic current through $\spatialBdry$ which changes the presymplectic structure with time.
The action is defined as
\begin{align}
  S
  &= \int_\spacetime L
    \,.
\end{align}
Using~\cref{eq:dvL} its variation along a vector field $X$ is given by
\begin{align}
  X \contract \dv S
  &= \int_\spacetime X \contract E
    + \int_\futureBdry X \contract \theta
    - \int_\pastBdry X \contract \theta
    + \int_\spatialBdry X \contract \theta
    \,.
\end{align}
We restrict $X$ such that the fields at $\Sigma_1$ and $\Sigma_2$ are held fixed.
It follows that $X \contract \theta$ vanishes there, so the variation of the action vanishes for all solutions to the equations of motion and all allowed $X$ if the pullback to $\spatialBdry$ of $\theta$, denoted by $\underline \theta$, is $\dd$-exact at the solution subspace
\begin{align}
  \underline \theta
  &= \dd \kappa
    \quad \text{at}
    \quad \solutions
    \,.
\end{align}
Use the freedom of shifting $\theta$ by a $\dd$-exact term to obtain
\begin{align}
  \underline \theta
  &= 0
  \quad \text{at}
  \quad \solutions
  \,.
\end{align}
It follows that $\underline \omega \approx 0$ and
\begin{align}
  \symplecticStructure_\futureBdry \approx \symplecticStructure_\pastBdry
  \,.
\end{align}
Because the Lagrangian gives rise to a well-defined variational principle, we were able to define a symplectic current such that the presymplectic structure is conserved.

\section{Local Hamiltonian Functions}
\label{sec:local-hamilt-funct}

Since variational symmetries lead to Noether currents and symplectomorphisms to Hamiltonian functions it is natural to study their relation.
It turns out that Noether currents can be used to construct Hamiltonian functions.

A Hamiltonian function $H$ is a real valued, smooth function on phase space, $H \in C^\infty(\solutions)$.
The relation between the vector field $X$ and its Hamiltonian function in accordance with~\cref{eq:hamiltonian} is given by
\begin{align}
  \dv H
  &\approx - X \contract \symplecticStructure_\Sigma
  \label{eq:covariantHamiltonian}
  \,.
\end{align}
Here we just demand equality when pulled back to $\solutions$ because that is the phase space under consideration.

Let $X$ be a variational symmetry, i.e.\ a local vector field satisfying
\begin{align}
  \lied_X L
  &= \dd \alpha
    \,,
\end{align}
with some local $(n-1, 0)$-form $\alpha$, then from~\cref{eq:dvL}
\begin{align}
  \lied_X E
  &= \lied_X ( \dv L - \dd \theta )
  \\
  &= \dv { \lied_X L } - \lied_X { \dd \theta }
  \\
  &= \dd ( \dv \alpha - \lied_X \theta )
  \,.
\end{align}
Since $X$ is also a symmetry of the equations of motion (see \cref{sec:vari-symm})
\begin{align}
  \dd ( \lied_X \theta - \dv \alpha )
  = 0
  \quad \text{at}
  \quad \solutions
  \,.
\end{align}
By a variant of the Poincaré lemma (\cref{thm:horizontal-complex-exact-onshell}),
it follows that there exists a local $(n-2, 1)$-form $\beta$ such that
\begin{align}
  \lied_X \theta - \dv \alpha
  &= \dd \beta
  \quad \text{at}
  \quad \solutions
  \,.
\end{align}
Acting with $\dv$ on the Noether current~\cref{eq:noether-current} gives
\begin{align}
  \dv j
  &= \dv ( X \contract \theta - \alpha)
  \\
  &= \lied_X \theta
    - \dv \alpha
    - X \contract \omega
    \,,
\end{align}
and we find that
\begin{align}
  X \contract \omega
  + \dv j
  - \dd \beta = 0
  \quad \text{at}
  \quad \solutions
  \,.
\end{align}
Inserting this into the  defining property of a Hamiltonian function~\cref{eq:covariantHamiltonian} gives
\begin{align}
  \dv H
  &\approx \int_\Sigma \dv j
  - \int_{\bdry \Sigma} \beta
    \,.
\end{align}
We see that the Hamiltonian function is always given by the integral over the Noether current plus a field independent function and a boundary term.
The condition for integrability of the Hamiltonian is
\begin{align}
  \int_{\bdry \Sigma} \dv \beta
  \approx 0
  \,.
\end{align}

\begin{example}
\label{ex:diffeoLcharge}
Consider diffeomorphisms generated by a vector field $\xi$ on the spacetime manifold $M$ and the corresponding local vector field $X$ as in \cref{ex:diffeos} on page \pageref{ex:diffeos}.
Assume that the Lagrangian and $\theta$ are both covariant, i.e.\ they satisfy
\begin{align}
  \lied_{X} L
  &= \lied_\xi L \equiv \dd (\xi \contract L)
  \\
  \lied_{X} \theta
  &= \lied_\xi \theta
  \,,
\end{align}
so that $\alpha = \xi \contract L$.
It follows from
\begin{align}
  \lied_{X} \theta - \dv \alpha
  &= \lied_\xi \theta - \xi \contract \dv L
  \\
  &= \lied_\xi \theta - \xi \contract ( E + \dd \theta )
  \\
  &= \dd ( \xi \contract \theta ) - \xi \contract E
  \,,
\end{align}
that we can set $\beta = \xi \contract \theta$ and find the relation
\begin{align}
  X \contract \omega
  + \dv j
  - \dd ( \xi \contract \theta )
  + \xi \contract E
  &= 0
  \,.
\end{align}
Hamiltonian functions obey
\begin{align}
  \dv H
  \approx \int_\Sigma \dv j
  - \int_{\bdry \Sigma} \xi \contract \theta
    \,.
\end{align}
This can be integrated to obtain a Hamiltonian function if the corresponding integrability condition is satisfied, i.e.
\begin{align}
  \int_{\bdry \Sigma} \xi \contract \omega
  &\approx 0
  \,.
\end{align}
If $\xi$ is tangent to $\bdry \Sigma$ and $\omega$ is finite, this is automatically satisfied.
If $\xi$ is finite and we have a well-defined variational principle, i.e.\ $\underline \theta \approx 0$ (see \cref{sec:cons-sympl-struct}), the Hamiltonian function can be integrated to satisfy
\begin{gather}
  H \approx \int_\Sigma j
  \,.
\end{gather}
\end{example}

From~\cref{eq:covariantHamiltonian} it follows that if the presymplectic structure $\symplecticStructure_\Sigma$ pulled back to $\solutions$ does not depend on $\Sigma$, neither does the pullback of the $\dv$-derivative of a Hamiltonian function $\dv H$.
The pullback of the Hamiltonian function $H$ itself on the other hand can still depend on the hypersurface $\Sigma$.

If the symmetry under consideration is a gauge symmetry according to \cref{def:gauge-symmetry} and $\dv H_\Sigma$ pulled back to $\solutions$ does not depend on $\Sigma$, then
\begin{align}
  \dv H_\Sigma \approx 0
  \,.
  \label{eq:vanishing-charge-variation}
\end{align}
This can be seen by considering $\dv H_{\Sigma_1}$ and $\dv H_{\Sigma_2}$ and deforming the symmetry such that it stays the same around $\Sigma_1$ and vanishes around $\Sigma_2$.
It follows that $\dv H_{\Sigma_2}$ vanishes and since $\dv H_\Sigma$ is independent of $\Sigma$ also~\cref{eq:vanishing-charge-variation} holds.

\section{Ambiguities}
\label{sec:ambiguities}

In the definition of the presymplectic structure and the Hamiltonian functions certain ambiguities arise.
Addition of a $\dd$-exact form to the Lagrangian $L \mapsto L + \dd A$ changes the potential $\theta \mapsto \theta + \dv A$ but does not affect the symplectic current.
Addition of a $\dd$-exact form to the potential has the effect
\begin{align}
  \theta
  &\mapsto \theta + \dd B
  \\
  \omega
  &\mapsto \omega + \dd { \dv B }
  \\
  \symplecticStructure_\Sigma
  &\mapsto \symplecticStructure_\Sigma + \dv { \int_{\bdry \Sigma} B }
  \\
  \dv H
  &\mapsto \dv H - X_H \contract \dv \int_{\bdry \Sigma} B
  \,.
\end{align}
Here, $B$ is a local $(n-2,1)$-form.
Demanding that the presymplectic structure is conserved ($\underline\omega \approx 0$, see \cref{sec:cons-sympl-struct}) reduces the ambiguity by restricting $B$ to satisfy
\begin{align}
  \dd \underline { \dv B }
  &= 0
  \,.
\end{align}
Demanding the stronger condition that $\underline\theta = 0$ at $\solutions$ (as before in the case of a well-defined variational principle) restricts $B$ as follows
\begin{align}
  \underline{\dd B}
  &= 0
  \quad \text{at}
  \quad \solutions
  \,,
\end{align}
so that by \cref{thm:horizontal-complex-exact-onshell}
\begin{align}
  \underline{B}
  &= \underline{\dd C}
  \quad \text{at}
  \quad \solutions
  \,,
\end{align}
where $C$ is some local $(n-2,0)$-form.
Under the assumption that there are no corner contributions to the integral at $\bdry^2 \Sigma$, the presymplectic structure is fixed uniquely
\begin{align}
  \symplecticStructure_\Sigma
  &\mapsto \symplecticStructure_\Sigma + \dv { \int_{\bdry \Sigma} \dd C }
  = \symplecticStructure_\Sigma
  \,.
\end{align}


\chapter{Three-dimensional Asymptotically Flat Space at Null Infinity}
\label{cha:three-dimens-asympt}

We now turn to general relativity in three dimensions with asymptotically flat boundary conditions according to \cref{sec:asymptotically-flat}.
This case is particularly interesting as the $\dv$-derivatives of the Hamiltonian functions are conserved in vacuum, but there is no way to define conserved Hamiltonian functions themselves without introducing additional structure on the spacetime manifold.
We introduce boundary conditions in~\cref{sec:3d-boundary-conditions} to make all expressions well defined and to make the Hamiltonian functions integrable.
Hamiltonian functions for the theory without matter are derived in \cref{sec:3d-gravitational-part}.
They are studied for the theory with an additional scalar in \cref{sec:3d-scalar-field}.
The quantities are expressed in a specially constructed coordinate system in \cref{sec:3d-expr-coord-syst}.
Invariance and the Poisson algebra of the Hamiltonian functions are studied in \cref{sec:3d-invariance}.
Additional background structure is introduced in \cref{sec:3d-addit-backgr-struct} to change their conservation law.
The equations of motion for general relativity coupled to a scalar field are solved order by order in \cref{sec:3d-scalar-sol}.

\section{Boundary Conditions}
\label{sec:3d-boundary-conditions}

We assume that the spacetime can be conformally completed as in \cref{sec:conf-comp}.
This means in particular that $\tilde g_{\sia\sib} = \conformalFactor^2 g_{\sia\sib}$ has a smooth limit to \scri{}.
We further keep the metric $\tilde g_{\sia\sib}$ fixed at \scri{}, so that also $\tilde\tau_{\sia\sib}$, defined as
\begin{align}
  \tilde\tau_{\sia\sib}
  &\defn \conformalFactor \dv g_{\sia\sib}
  \,,
\end{align}
has a smooth limit to \scri{}.
For later convenience we define $\tilde\tau \defn \tilde g^{\sia\sib} \tilde\tau_{\sia\sib}$ and $\tilde\tau_\sia \defn \tilde\tau_{\sia\sib} \tilde n^\sib$.
Since we are interested in asymptotically flat space, we set the cosmological constant to zero.
In the first part of the derivation we assume $\conformalFactor^{-1} R \bdryeq 0$, or equivalently $\conformalFactor^{-1} G_{\sia\sib} g^{\sia\sib} \bdryeq 0$, where $G_{\sia\sib}$ is the Einstein tensor.
It then follows from~\cref{eq:norm-n-from-R} that
$\tilde n^\sia \tilde n_\sia \bdryeq 0$.
Defining
\begin{align}
  f \defn \conformalFactor^{-1} \tilde n_\sia \tilde n^\sia
  \label{eq:deff}
\end{align}
we get from~\cref{eq:nnDivN} that
\begin{align}
  f
  &\bdryeq \frac{2}{3} \tilde\cd_\sia \tilde n^\sia
  \label{eq:f3d}
  \,,
\end{align}
where $\tilde\cd$ is the Levi-Civita connection with respect the unphysical metric $\tilde g_{\sia\sib}$.
By acting with $\dv$ on this equation it follows that
\begin{align}
  \tilde n^\sia \tilde\tau_\sia
  &\bdryeq 0 \label{eq:n-n-tau}
    \,.
\end{align}
Using $\conformalFactor^{-1} R \bdryeq 0$ and~\cref{eq:f3d} in~\cref{eq:bla} we further find that
\begin{align}
  \conformalFactor R_{\sia\sib}
  &\bdryeq
    \tilde\cd_\sia \tilde n_\sib
    - \frac{1}{2} f \tilde g_{\sia\sib}
    \,.
    \label{eq:Rfdivn}
\end{align}
Later in the derivation we will require the stronger condition that $\conformalFactor V^\sia G_{\sia\sib} \bdryeq 0$ for all smooth $V^\sia$ tangential to $\scri$.
\footnote{
  This is slightly stronger than the condition imposed in~\cite{Ashtekar:1996cd} that $G_{\sia\sib} \tilde n^\sia V^\sib \bdryeq 0$ for all smooth $V^\sia$ tangential to $\scri$.
  The condition is erroneously stated as $\conformalFactor^{-1} G_{\sia\sib} \tilde n^\sia V^\sib \bdryeq 0$ in~\cite{Ashtekar:1996cd}.
}
Because $\conformalFactor R_{\sia\sib} \bdryeq \conformalFactor G_{\sia\sib}$ and by symmetry of $R_{\sia\sib}$ this is equivalent to demanding that
\begin{align}
  \conformalFactor R_{\sia\sib}
  &\bdryeq A \tilde n_\sia \tilde n_\sib
  \label{eq:defA}
  \,,
\end{align}
with some smooth function $A$.
Plugging this into~\cref{eq:Rfdivn} yields the relation
\begin{align}
  \tilde \cd_\sia \tilde n_\sib
  &\bdryeq
  A \tilde n_\sia \tilde n_\sib
    + \frac12 f \tilde g_{\sia\sib}
    \label{eq:strongerBCconseqence}
    \,.
\end{align}
Later we will make use of the fact that, by rescaling $\conformalFactor$, it is possible to make $f$ equal to any chosen smooth function at \scri{}.

\section{Gravitational Part}
\label{sec:3d-gravitational-part}

In this section we consider only the gravitational part of the Lagrangian.
We do not use the equations of motion to set the Einstein tensor to zero, instead we just require that it satisfies the boundary conditions from the previous section.
This makes the analysis quite general and allows addition of matter at a later stage, as will be seen in \cref{sec:3d-scalar-field}.
The Lagrangian is taken to be
\begin{align}
  L
  &= \frac{1}{2\kappa} R \epsilon
    \,,
\end{align}
where $\epsilon$ is the natural volume form.
Decomposition of $\dv L$ as in $\cref{eq:dvL}$ yields
\begin{align}
  \theta_{\sia\sib}
  &= \frac{1}{2\kappa} \left( g^{\sie\sic} g^{\sif\sid} - g^{\sie\sid} g^{\sif\sic} \right) \cd_\sid { \dv g_{\sif\sic} } \, \epsilon_{\sie\sia\sib}
    \,,
\end{align}
and we find the symplectic current to be
\begin{align}
  \omega_{\sia\sib}
  &= \dv \theta_{\sia\sib}
    = - \frac{1}{2\kappa} P^{\sic\sih\sig\sid\sie\sif} \dv g_{\sih\sig} \wedge \cd_\sid { \dv g_{\sie\sif} } \, \epsilon_{\sic\sia\sib}
    \,,
\end{align}
where
\begin{align}
  P^{\sic\sih\sig\sid\sie\sif}
  &=
    g^{\sic\sie} g^{\sih\sif} g^{\sig\sid}
    - \frac{1}{2} g^{\sic\sid} g^{\sih\sie} g^{\sig\sif}
    - \frac{1}{2} g^{\sic\sih} g^{\sig\sid} g^{\sie\sif}
    \nonumber \\
  &\quad
    - \frac{1}{2} g^{\sih\sig} g^{\sic\sie} g^{\sid\sif}
    + \frac{1}{2} g^{\sih\sig} g^{\sic\sid} g^{\sie\sif}
    \,.
\end{align}
By using the conformal relation between the metrics $g_{\sia\sib}$ and $\tilde g_{\sia\sib}$,
the action of the covariant derivative $\cd_\sia$ (compatible with $g_{\sia\sib}$) on a tensor $\gamma_{\sia\sib}$ can be rewritten in terms of the covariant derivative $\tilde \cd_\sia$ (compatible with $\tilde g_{\sia\sib}$).
\begin{align}
  \cd_\sia \gamma_{\sib\sic}
  &= \tilde \cd_\sia \gamma_{\sib\sic}
    + 2 C\indices{^\sid_\sia_{(\sib}} \gamma_{\sic)\sid}
  \\
  C\indices{^\sid_\sia_\sib}
  &= 2 \conformalFactor^{-1} \delta^\sid_{(\sia} \tilde n^{\vphantom{\sid}}_{\sib)}
    - \conformalFactor^{-1} \tilde n^\sid \tilde g_{\sia\sib}
\end{align}
This can be used to rewrite $\theta$ and $\omega$ in terms of quantities that are well-defined at \scri{},
\begin{align}
  \theta_{\sia\sib}
  &=
    \frac{1}{2\kappa} \left(
    \tilde\cd_\sid \tilde\tau^{\sic\sid}
    - \tilde\cd{}^\sic \tilde\tau
    - 2 \conformalFactor^{-1} \tilde\tau^\sic
    \right) \tilde\epsilon_{\sic\sia\sib}
  \\
  \omega_{\sia\sib}
  &= \frac{1}{4\kappa} \big(
    {-} 2 \conformalFactor \tilde P^{\sic\sih\sig\sid\sie\sif}
    \tilde \tau_{\sih\sig} \wedge \tilde \cd_\sid \tilde \tau_{\sie\sif}
    + \tilde\tau^\sic \wedge \tilde\tau
    + 2 \tilde\tau^\sic_\sid \wedge \tilde\tau^\sid
    \big) \tilde\epsilon_{\sic\sia\sib}
    \,.
\end{align}
We observe that while $\omega$ has a smooth limit to \scri{}, $\theta$ cannot be extended to $\scri$ because the last term diverges.
Since the pullback to $\scri$ of the contraction of any smooth vector field $\tilde t^\sia$ with the volume form ($\underline{\tilde t \contract \tilde\epsilon}$) is proportional to $\tilde t^\sia \tilde n_\sia$,
and by using~\cref{eq:n-n-tau} we see that the pullback to \scri{} of the symplectic current vanishes,
\begin{align}
  \underline\omega
  &= 0
  \label{eq:pullback-omega-vanishes}
  \,.
\end{align}
It follows that the presymplectic structure is conserved.

\subsection{BMS Symmetries}

Consider a vector field $\xi$ that generates BMS transformations and satisfies condition~\cref{eq:bmsDef3}.
We choose $\xi^\sia$ to be independent of the fields, which is consistent with~\cref{eq:bmsDef3} as can be seen by acting with $\dv$.
The infinitesimal BMS symmetry $X$ is defined by its action as follows:
\begin{align}
  \lied_X g_{\sia\sib}
  &= \lied_\xi g_{\sia\sib}
\end{align}
To calculate the Noether current~\cref{eq:noether-current} we use
\begin{align}
  X \contract \theta_{\sia\sib}
  &= \frac{1}{\kappa} \left(
    \cd_\sid { \cd^{(\sid} \xi^{\sic)} }
    - \cd^\sic { \cd_\sid \xi^\sid }
    \right) \epsilon_{\sic\sia\sib}
  \\
  &= \frac{1}{\kappa} \left(
    \cd_\sid { \cd^{[\sid} \xi^{\sic]} }
    + R^{\sic\sid} \xi_\sid
    \right) \epsilon_{\sic\sia\sib}
\end{align}
and
\begin{align}
  \lied_X L
  &= \lied_\xi L
  = \dd ( \xi \contract L )
  \,.
\end{align}
In accordance with \cref{sec:noether-thm2} we find that the Noether current is exact
\begin{align}
  (j_\xi)_{\sia\sib}
  &= (X \contract \theta - \xi \contract L)_{\sia\sib}
  \\
  &= \frac{1}{\kappa} \left(
    \cd_\sid { \cd^{[\sid} \xi^{\sic]} }
    + G^{\sic\sid} \xi_\sid
    \right) \epsilon_{\sic\sia\sib}
  \\
  &\approx
    (\dd q_\xi)_{\sia\sib}
    \,,
\end{align}
with the Noether charge $(1,0)$-form $q_\xi$ given by
\begin{align}
\label{eq:noether-charge-form}
\begin{aligned}
  (q_\xi)_\sia
  &=
    - \frac{1}{2\kappa} \cd^\sic \xi^\sib \epsilon_{\sic\sib\sia}
  \\
  &=
    - \frac{1}{2\kappa} \conformalFactor \tilde\cd{}^\sic ( \conformalFactor^{-2} \xi^\sib ) \, \tilde\epsilon_{\sic\sib\sia}
    \,.
\end{aligned}
\end{align}
Since $L$ and $\theta$ are both covariant as in \cref{ex:diffeoLcharge} we see that
\begin{align}
  X \contract \omega
  + \dd ( \dv q - \xi \contract \theta )
  &\approx 0
  \,,
  \label{eq:3d-charge-law}
\end{align}
such that the Hamiltonian functions obey
\begin{align}
  \dv H_\xi
  &\approx \int_\Sigma \dv j
  - \int_{\bdry \Sigma} \xi \contract \theta
  = \int_{\bdry\Sigma} ( \dv q_\xi - \xi \contract \theta )
    \,.
\end{align}
This is a condition on the Hamiltonian at $\solutions$ only.
By using the freedom to extend $H_\xi$ away from $\solutions$ in an arbitrary way we require that the condition holds everywhere:
\begin{align}
  \dv H_\xi
  &= \int_{\bdry\Sigma} ( \dv q_\xi - \xi \contract \theta )
  \label{eq:var-charge-def}
\end{align}
Because $\omega$ has a smooth limit to $\scri$ this is also well-defined and independent of how the $\conformalFactor \to 0$ limit is taken.
What remains is to integrate this equation to obtain $H_\xi$.
The integration constants could in principle be fixed by demanding that $H_\xi$ vanishes (for all $\Sigma$ and $\xi$) on a reference spacetime such as Minkowski space~\cite{Wald:1999wa}.
In this case the Hamiltonian functions $H_\xi$ are by construction independent of any background structure like, for example, $\conformalFactor$.
If all $H_\xi$ vanished on a particular spacetime also $\lied_Y H_\xi$ would vanish on that spacetime where $Y$ is another arbitrary infinitesimal BMS symmetry.
One can check using the coordinate expressions of \cref{sec:3d-expr-coord-syst} that no vacuum spacetime exists such that $\lied_Y H_\xi$ vanishes for all $Y$ and $\xi$.
It follows that we should expect $H_\xi$ to depend on some background structure.

In the following we integrate~\cref{eq:var-charge-def} using only $\conformalFactor$ as background structure.
We compare this in \cref{sec:3d-addit-backgr-struct} to earlier work~\cite{Barnich:2010eb} where more background structure was introduced.

\subsection{Supertranslations}

We first consider the case where the vector field $\xi^\sia$ generates supertranslations.
Such a vector field has the form
\begin{align}
  \xi^\sia
  &= h \tilde n^\sia + \conformalFactor \tilde v^\sia
    \,,
    \label{eq:supertransl}
\end{align}
with arbitrary smooth $h$ and some smooth $\tilde v^\sia$.
To rewrite the expressions appearing in~\cref{eq:var-charge-def} we introduce smooth covectors $\tilde m_\sia$ and $\tilde l_\sia$ which are required to satisfy the conditions
\begin{align}
  \tilde m^\sia \tilde m_\sia
  &= 1
  \\
  \tilde n^\sia \tilde m_\sia
  &= 0
  \\
  \tilde n^\sia \tilde l_\sia
  &= 1
    \,.
\end{align}
At $\scri$ this defines $\tilde m_\sia$ up to sign and up to addition of some covector field along $\tilde n_\sia$ ($\tilde m_\sia \mapsto \pm \tilde m_\sia + \alpha \tilde n_\sia + O(\conformalFactor)$).
Since the $n$-form $\tilde\sigma = \tilde n \wedge \tilde l \wedge \tilde m$ satisfies
\begin{align}
  \tilde\sigma^{\sia\sib\sic} \tilde\sigma_{\sia\sib\sic}
  &=
  \big( 6 \, \tilde n^{[\sia} \tilde l^{\sib\vphantom]} \tilde m^{\sic]} \big)
  \big( 6 \, \tilde n_{[\sia} \tilde l_{\sib\vphantom]} \tilde m_{\sic]} \big)
  \\
  &= -6
    + 6 \, \tilde n^\sia \tilde n_\sia \left( \tilde l^\sib \tilde l_\sib - {(\tilde l^\sib \tilde m_\sib)}^2 \right)
  \\
  &= -6 + O(\conformalFactor)
  \,,
\end{align}
it follows that by choosing the sign of $m_\sia$ appropriately, in some neighborhood around \scri{}, we have
\begin{align}
  \tilde\epsilon
  &= \left(
    1
    + O(\conformalFactor)
    \right) \, \tilde n \wedge \tilde l \wedge \tilde m
    \,.
\end{align}
The pullback of $\tilde m_\sia$ to $\scri$ is now unambiguously defined. The unphysical metric at \scri{} can be decomposed as
\begin{align}
  \tilde g_{\sia\sib}
  &\bdryeq
    \tilde m_\sia \tilde m_\sib
    - 2 \tilde l^\sic \tilde m_\sic \tilde m_{(\sia} \tilde n_{\sib)}
    + 2 \tilde l_{(\sia} \tilde n_{\sib)}
    + \left( {( \tilde l^\sic \tilde m_\sic )}^2 - \tilde l^\sic \tilde l_\sic \right) \tilde n_\sia \tilde n_\sib
    \label{eq:metric-decomposition}
    \,.
\end{align}
Inserting the supertranslation~\cref{eq:supertransl} into~\cref{eq:bmsDef3} leads to the relations
\begin{align}
  \conformalFactor^2 {\lied_\xi g_{\sia\sib}}
  &\bdryeq
    2 \tilde n_{(\sia} \tilde \cd_{\sib)} h
    + 2 \tilde \cd_\sia \tilde n_\sib h
    + 2 \tilde n_{(\sia} \tilde v_{\sib)}
    \nonumber \\
    &\quad- 2 \tilde n_\sic \tilde v^\sic \tilde g_{\sia\sib}
    - 2 h f \tilde g_{\sia\sib}
    \bdryeq 0
  \\
  \tilde n^\sia \conformalFactor^2 {\lied_\xi g_{\sia\sib}}
  &\bdryeq
    \tilde n_\sib \big(
    \tilde n^\sia \tilde \cd_\sia h
    - \tilde n_\sia \tilde v^\sia
    - h f
    \big)
    \bdryeq 0
  \\
  \tilde g^{\sia\sib} \conformalFactor^2 {\lied_\xi g_{\sia\sib}}
  &\bdryeq
    2 \tilde n^\sia \tilde \cd_\sia h
    - 4 \tilde n_\sia \tilde v^\sia
    - 3 h f
    \bdryeq 0
  \\
  \tilde l^\sia \conformalFactor^2 {\lied_\xi g_{\sia\sib}}
  &\bdryeq
    \tilde v_\sib
    + \tilde \cd_\sib h
    + \tilde n_\sib l^\sia \tilde \cd_\sia h
    + 2 \tilde l^\sia \tilde \cd_\sia \tilde n_\sib h
    \nonumber \\
    &\quad+ \tilde n_\sib \tilde l^\sia \tilde v_\sia
    - 2 \tilde n_\sia \tilde v^\sia \tilde l_\sib
    - 2 h f \tilde l_\sib
    \bdryeq 0
  \\
  \tilde l^\sia \tilde l^\sib \conformalFactor^2 {\lied_\xi g_{\sia\sib}}
  &\bdryeq
    2 \tilde l^\sia \tilde \cd_\sia h
    + 2 \tilde l^\sia \tilde l^\sib \tilde \cd_\sia \tilde n_\sib h
    + 2 \tilde l^\sia \tilde v_\sia
    \nonumber \\
    &\quad- 2 \tilde l^\sia \tilde l_\sia \tilde n_\sib \tilde v^\sib
    - 2 h f \tilde l^\sia \tilde l_\sia
    \bdryeq 0
    \,,
\end{align}
from which we find that
\begin{align}
  \tilde n^\sia \tilde\cd_\sia h
  &\bdryeq \frac12 h f
  \\
  \tilde v^\sia
  &\bdryeq
    - \tilde \cd{}^\sia h
    - 2 h \tilde l^\sib \tilde \cd_\sib \tilde n^\sia
    + h \tilde n^\sia \tilde l^\sib \tilde l^\sic \tilde \cd_\sib \tilde n_\sic
    - \frac12 h f \tilde l^\sib \tilde l_\sib \tilde n^\sia
    + h f \tilde l^\sia
    \label{eq:n-v-tau}
  \\
  \tilde n_\sia \tilde v^\sia
  &\bdryeq - \frac12 h f
    \,.
\end{align}
Using the relation
\begin{align}
  \tilde\epsilon_{\sia\sib\sic}
  &= 2 \big( 1 + O(\conformalFactor) \big) \left(
    \tilde n_{[\sia} \tilde l_{\sib]} \tilde m_\sic
    + \tilde m_{[\sia} \tilde n_{\sib]} \tilde l_\sic
    + \tilde l_{[\sia} \tilde m_{\sib]} \tilde n_\sic
    \right)
    \label{eq:volExpanded}
    \,,
\end{align}
we find
\begin{align}
  \xi \contract \theta
  &=
    \frac{1}{2\kappa} \left(
    \tilde n_\sia \tilde\cd_\sib \tilde\tau^{\sia\sib}
    - \tilde n^\sia \tilde\cd_\sia \tilde\tau
    - 2 \conformalFactor^{-1} \tilde n^\sia \tilde\tau_\sia
    + f \tilde l^\sia \tilde\tau_\sia
    \right) h \tilde m
    - \frac{1}{2\kappa} h f \tilde m^\sia \tilde\tau_\sia \tilde l
  \nonumber \\
  &\quad
    + (\dots) \tilde n
    + O(\conformalFactor)
    \,,
\end{align}
and observe that the pullback of $\xi \contract \theta$ to constant $\conformalFactor$ hypersurfaces has a smooth limit to \scri{} because of~\cref{eq:n-n-tau}.
The $\dv$-derivative of the Noether charge form~\cref{eq:noether-charge-form} satisfies
\begin{align}
  (\dv q_\xi)_\sia
  &= - \frac{1}{2\kappa} \bigg(
    \tilde n^\sid \left(
    - \frac{1}{2} \tilde \tau \tilde v^\sib
    + \tilde \tau^{\sib\sic} \tilde v_\sic
    + \tilde \tau^{\sib\sic} \tilde \cd_\sic h
    - \conformalFactor^{-1} \tilde \tau^\sib h
    - \frac{1}{2} \tilde \tau \tilde \cd{}^\sib h
    \right)
  \nonumber \\
  &\hspace{4em}
    + \tilde \tau^\sid \tilde v^\sib
    + h \tilde \cd{}^\sid \tilde \tau^\sib
    \bigg)
    \tilde \epsilon_{\sid\sib\sia}
    + O(\conformalFactor)
\end{align}
where the expression for the volume form~\cref{eq:volExpanded} was used again.
By the relations for $\tilde v^\sia$~\cref{eq:n-v-tau} this can be written as
\begin{align}
  \dv q_\xi
  &= - \frac{1}{2\kappa} \left(
    h \tilde n^\sia l^\sib \tilde \cd_\sia \tilde \tau_\sib
    + h \tilde l^\sia \tilde \tau^\sib \tilde \cd_\sia \tilde n_\sib
    - \tilde v^\sia \tilde \tau_\sia
    - \tilde \tau^\sia \tilde \cd_\sia h
    - \frac12 h f \tilde l^\sia \tilde \tau_\sia
    \right)
    \tilde m
  \\
  &\quad
    + \frac{1}{2\kappa} \tilde n^\sia \tilde m^\sib \tilde \cd_\sia \tilde \tau_\sib h \tilde l
    + (\dots) \tilde n
    + O(\conformalFactor)
  \\
  &= - \frac{1}{2\kappa} \left(
    \tilde n^\sia l^\sib \tilde \cd_\sia \tilde \tau_\sib
    + 3 \tilde l^\sia \tilde \tau^\sib \tilde \cd_\sia \tilde n_\sib
    - \frac32 f \tilde l^\sia \tilde \tau_\sia
    \right)
    h \tilde m
  \\
  &\quad
    + \frac{1}{2\kappa} \tilde n^\sia \tilde m^\sib \tilde \cd_\sia \tilde \tau_\sib h \tilde l
    + (\dots) \tilde n
    + O(\conformalFactor)
    \,.
\end{align}
Define now the Schouten tensors
\begin{align}
\label{eq:s}
\begin{aligned}
  S_{\sia\sib}
  &\defn R_{\sia\sib} - \frac14 R g_{\sia\sib}
  \\
  \tilde S_{\sia\sib}
  &\defn \tilde R_{\sia\sib} - \frac14 \tilde R \tilde g_{\sia\sib}
  \,.
\end{aligned}
\end{align}
Using~\cref{eq:ricci-conf} we find that their difference is given by
\begin{align}
  S_{\sia\sib} - \tilde S_{\sia\sib}
  &= \conformalFactor^{-1} \tilde \cd_\sia \tilde n_\sib
  - \frac12 \, \conformalFactor^{-2} \tilde n_\sic \tilde n^\sic \tilde g_{\sia\sib}
  \,.
\end{align}
Acting with $\dv$ on the terms in this difference and contracting with $\tilde m^\sia$ gives
\begin{align}
  \tilde m^\sia \tilde m^\sib \dv ( \tilde g_{\sia\sib} \tilde n^\sic \tilde n_\sic )
  &= \conformalFactor \tilde m^\sia \tilde m^\sib \tilde \tau_{\sia\sib} \tilde n^\sic \tilde n_\sic
    - \conformalFactor \tilde n^\sic \tilde \tau_\sic
  \\
  \tilde m^\sia \tilde m^\sib \dv ( \tilde \cd_\sia \tilde n_\sib )
  &= - \frac12 \conformalFactor \tilde m^\sia \tilde m^\sib \tilde n^\sic \left(
    2 \tilde \cd_\sia \tilde \tau_{\sib\sic}
    - \tilde \cd_\sic \tilde \tau_{\sia\sib}
    \right)
    + \frac12 \tilde n^\sic \tilde n_\sic \tilde m^\sia \tilde m^\sib \tilde \tau_{\sia\sib}
    \,,
\end{align}
from which it follows that
\begin{align}
  2 \tilde m^\sia \tilde m^\sib \dv ( S_{\sia\sib} - \tilde S_{\sia\sib} ) =
  \tilde m^\sia \tilde m^\sib \left(
    \tilde n^\sic \tilde \cd_\sic \tilde \tau_{\sia\sib}
    - 2 \tilde \cd_\sia \tilde \tau_\sib
    + 2 \tilde \tau_{\sib\sic} \tilde \cd_\sia \tilde n^\sic
    \right)
    + \conformalFactor^{-1} \tilde n^\sic \tilde \tau_\sic
    \,.
\end{align}
From now on we assume the additional boundary condition that led to~\cref{eq:strongerBCconseqence}.
It follows using $\dv f \bdryeq - \tilde n^\sia \tilde\tau_\sia$ and~\cref{eq:n-n-tau} that
\begin{align}
  \dv (\tilde \cd_\sia \tilde n_\sib)
  &\bdryeq
  \dv A \tilde n_\sia \tilde n_\sib
  \,,
\end{align}
and because in general $\dv ( \tilde\cd_\sia \tilde n_\sib ) \bdryeq - \tilde n_{(\sia} \tilde \tau_{\sib)}$ also that
\begin{align}
  \label{eq:tau-from-a}
  \tilde \tau_\sia
  &\bdryeq - \dv A \tilde n_\sia
  \,.
\end{align}
Using
\begin{align}
  \tilde\cd_\sia \tilde\tau^\sia
  &\bdryeq
    - \tilde n^\sia \tilde\cd_\sia { \dv A }
    + \conformalFactor^{-1} \tilde\tau^\sia \tilde n_\sia
    - f \dv A
    - \frac12 f \tilde l^\sia \tilde\tau_\sia
  \\
  \tilde \tau
  &\bdryeq
    \tilde m^\sia \tilde m^\sib \tilde \tau_{\sia\sib}
    - 2 \dv A
    \,,
\end{align}
allows us to find the relations
\begin{align}
  \dv \underline q_\xi
  &\bdryeq \frac{1}{2\kappa} \left(
    \tilde n^\sia \tilde \cd_\sia \dv A
    + \frac12 f \dv A
    \right) h \underline{\tilde m}
  \\
  \underline{\xi \contract \theta}
  &\bdryeq
    \frac{1}{2\kappa} \Big(
    \tilde n^\sic \tilde\cd_\sic \left(
    \dv A
    - \tilde m^\sia \tilde m^\sib \tilde \tau_{\sia\sib}
    \right)
    - \conformalFactor^{-1} \tilde n^\sia \tilde\tau_\sia
    - \frac12 f \dv A
    - \frac12 f \tilde m^\sia \tilde m^\sib \tilde \tau_{\sia\sib}
    \Big) h \underline{\tilde m}
    \,,
\end{align}
and
\begin{align}
  \frac{1}{\kappa} \tilde m^\sia \tilde m^\sib \dv ( S_{\sia\sib} - \tilde S_{\sia\sib} )
  &\bdryeq \frac{1}{2\kappa} \left(
    \tilde m^\sia \tilde m^\sib \tilde n^\sic \tilde \cd_\sic \tilde \tau_{\sia\sib}
    + \conformalFactor^{-1} \tilde n^\sic \tilde \tau_\sic
    + f \dv A
    + f \tilde m^\sia \tilde m^\sib \tilde \tau_{\sia\sib}
    \right)
    \,.
\end{align}
Combining these equations and using $\tilde \tau_{\sia\sib} \tilde n^\sic \tilde \cd_\sic ( \tilde m^\sia \tilde m^\sib ) \bdryeq 0$ leads to
\begin{align}
  \dv \underline q_\xi - \underline{\xi \contract \theta}
  &\bdryeq
  \frac{1}{\kappa} \tilde m^\sia \tilde m^\sib \left(
  \dv ( S_{\sia\sib} - \tilde S_{\sia\sib} )
  - \frac{1}{4} f \tilde \tau_{\sia\sib}
  \right) h \underline{\tilde m}
  \,.
\end{align}
By construction, the left hand side is independent of $\conformalFactor$ and therefore independent of $f$.
To integrate this equation, we assume that $\conformalFactor$ is chosen such that $f \bdryeq 0$.
We find the Hamiltonian function to be
\begin{align}
  \label{eq:hamiltonian-supertrans}
  H^T_\xi
  &= \frac{1}{\kappa} \int_{\bdry \Sigma}
    \tilde m^\sia \tilde m^\sib ( S_{\sia\sib} - \tilde S_{\sia\sib} )
    h \tilde m
    \,.
\end{align}
This is independent of the choice of $\tilde m_\sia$ because
\begin{align}
  \tilde n^\sib ( S_{\sia\sib} - \tilde S_{\sia\sib} )
  &= \frac12 \tilde \cd_\sia f
  \bdryeq \frac12 \conformalFactor^{-1} f \tilde n_\sia
  \,.
\end{align}
The expression is --- up to metric independent terms --- the same as the one given before by Ashtekar~\cite{Ashtekar:1996cd}.
There, an additional metric-independent term to remove dependence on $\conformalFactor$ was introduced, which came with the necessity to add an additional boundary condition.

\subsection{Superrotations}

We turn now to the case where the vector field $\xi^\sia$ generates a superrotation.
There is no invariant notion of superrotation without supertranslation.
We can however choose a class of superrotations by demanding that the vector field $\xi^\sia$ is tangent to $\bdry\Sigma$.
All other superrotations can be obtained by combination with a supertranslation.
For such a choice of $\xi^\sia$, the pullback of $\xi \contract \theta$ to $\bdry \Sigma$, denoted by a double underline, is just given by
\begin{align}
  \underline{\underline{ \xi^\sib \theta_{\sib\sia} }}
  &\bdryeq
    - \frac{1}{\kappa} \conformalFactor^{-1} \underline{\underline{
    \tilde\tau^\sic
    \xi^\sib \tilde\epsilon_{\sic\sib\sia} }}
  \,,
\end{align}
which can be expressed using expansion~\cref{eq:volExpanded} for $\epsilon$,~\cref{eq:tau-from-a,eq:defA} as
\begin{align}
  \underline{\underline{ \xi \contract \theta }}
    &\bdryeq - \frac{1}{\kappa} \conformalFactor^{-1} \dv A \, \xi^\sia \tilde n_\sia \underline{\underline{\tilde m}}
    \\
    &\bdryeq - \frac{1}{\kappa} \dv \big(
    \tilde l^\sia \tilde l^\sib R_{\sia\sib} \xi^\sic \tilde n_\sic \underline{\underline{\tilde m}}
    \big)
  \,.
\end{align}
This can be integrated such that the Hamiltonian function is given by
\begin{align}
  \label{eq:hamiltonian-superrot}
  H^R_\xi
  &= \int_{\bdry\Sigma} \left(
  \frac{1}{\kappa} \tilde l^\sia \tilde l^\sib R_{\sia\sib} \xi^\sic \tilde n_\sic \tilde m
  + q_\xi
  \right)
  \,.
\end{align}
The first term in the integral is finite due to the boundary conditions.
Finiteness of the second part of the integral is shown by checking finiteness of the expression on Minkowski space and using the fact that $\dv H^R_\xi$ is finite.
It can be checked that $\dd q_\xi$ has a smooth limit to $\scri$ on Minkowski space.
From this follows that the integral over $q_\xi$ is well-defined and independent of how the $\conformalFactor \to 0$ limit is taken on Minkowski space.
We conclude that $H^R_\xi$ is well-defined and independent of how the $\conformalFactor \to 0$ limit is taken.

A trivial BMS transformation has the form $\xi^\sia = \conformalFactor^2 v^\sia$ and leads to a vanishing $H^R_\xi$.
So in contrast to the four-dimensional case~\cite{Geroch:1981ut}, the integral does not depend on the chosen BMS representative.

\subsection{Combined Hamiltonian}

We found the Hamiltonian functions $H^T_\xi$~\cref{eq:hamiltonian-supertrans} and $H^R_\xi$~\cref{eq:hamiltonian-superrot} corresponding to supertranslations and superrotations, respectively.
Since any diffeomorphism generated by a vector field $\xi$ that is a BMS transformation satisfying~\cref{eq:bmsDef3} is composed of a supertranslation plus a superrotation we proceed to construct the Hamiltonian function for arbitrary BMS transformations.
Split $\xi^\sia = \xi_R^\sia + \xi_T^\sia$ into a vector field $\xi_R^\sia$ tangent to $\Sigma$ and a vector field of the form $\xi_T^\sia \bdryeq h \tilde n^\sia$.
A Hamiltonian satisfying~\cref{eq:covariantHamiltonian} is then given by
\begin{align}
  H_\xi
  &= H^T_{\xi_T} + H^R_{\xi_R}
  \\
  &= \int_{\bdry\Sigma} \left(
    \frac{1}{\kappa} \tilde m^\sia \tilde m^\sib ( S_{\sia\sib} - \tilde S_{\sia\sib} )
    h \tilde m
    + \frac{1}{\kappa} \tilde l^\sia \tilde l^\sib R_{\sia\sib} \xi^\sid \tilde n_\sid \tilde m
    + q_{\xi_R}
    \right)
    \label{eq:flat3dhamiltonian}
    \,,
\end{align}
where $S_{\sia\sib}$ and $\tilde S_{\sia\sib}$ are defined in~\cref{eq:s}, $\tilde l^\sia$ is some vector field satisfying $\tilde l^\sia \tilde n_\sia \bdryeq 1$, and
\begin{align}
  (q_{\xi_R})_\sia
  &=
    - \frac{1}{2\kappa} \cd^\sic \xi_R^\sib \epsilon_{\sic\sib\sia}
  \,.
\end{align}
While $\dv H_\xi$ is independent of the conformal factor, $H_\xi$ itself depends on $\conformalFactor$.
When $\conformalFactor$ is rescaled by a constant factor, the quantities $h$ and $\tilde m_\sia$ change inversely to $\tilde m^\sia$ while $S_{\sia\sib} - \tilde S_{\sia\sib}$ remains unchanged.
It follows that the Hamiltonian depends only on the equivalence class $\conformalFactor \sim \alpha \conformalFactor$ with any non-zero constant $\alpha$.

Since the Hamiltonian functions are given by integrals over the boundary of $\Sigma$ only, they obey a simple conservation law:
\begin{align}
  H_{\xi,\Sigma_2} - H_{\xi,\Sigma_1}
  &= - \int_{\spatialBdry} F_\xi
  \label{eq:3d-flux-law}
\end{align}
Here, $\spatialBdry$ is the region of $\scri$ bounded by $\Sigma_1$ and $\Sigma_2$ and the flux $F_\xi$ is given by
\begin{align}
  F_\xi
  &= \frac{1}{\kappa} \Big({
    - \conformalFactor^{-1} \xi^\sia \tilde n^\sib G_{\sia\sib}
    + {\left( \tilde m^\sia \tilde \cd_\sia \right)}^3 \left( \tilde m_\sib \xi^\sib \right)
    }\Big) \tilde l \wedge \tilde m
    \label{eq:3d-flux}
  \,,
\end{align}
which can be checked by using the expressions of \cref{sec:3d-expr-coord-syst}.

The Hamiltonian functions were derived for the Lagrangian of general relativity without matter and are therefore valid for vacuum solutions.
We were however careful not to set $G_{\sia\sib}$ to zero in the derivation, so the Hamiltonian functions are still interesting quantities when matter is introduced.
In this case the Hamiltonian functions do not generate BMS symmetries.
To reflect this we simply call them charges.
The flux formula~\cref{eq:3d-flux} is still valid after introduction of matter.
In the next section we study the meaning of the charges when a scalar field is added to the theory.

\section{Scalar Field}
\label{sec:3d-scalar-field}

The functions~\cref{eq:flat3dhamiltonian} cease to generate BMS symmetries, but correspond to the ``conserved quantities'' given by Wald and Zoupas~\cite{Wald:1999wa}, as we now show.
Supplement the Lagrangian from before with a massless scalar field
\begin{align}
  L
  &= \frac{1}{2\kappa} R \, \epsilon - \frac{1}{2} \cd^\sia \Phi \cd_\sia \Phi \, \epsilon
    \,.
\end{align}
No additional terms appear in the Noether charge 1-form $q_\xi$.
We assume $\tilde\Phi$ to be smooth where
\begin{align}
  \tilde\Phi &= \conformalFactor^{-1/2} \Phi
  \label{eq:scalar-field-falloff}
  \,,
\end{align}
and find that on-shell (see \cref{sec:scalar-field-conf})
\begin{align}
  \conformalFactor^{-1} R
  &\bdryeq 0
  \\
  \conformalFactor R_{\sia\sib}
  &\bdryeq \frac{\kappa}{4} \tilde\Phi^2 \tilde n_\sia \tilde n_\sib
    \,,
\end{align}
so that the boundary conditions are satisfied.
The potential for the symplectic current gets an additional contribution
\begin{align}
  \theta^\Phi_{\sia\sib}
  &=
    - \cd^\sic \Phi \dv \Phi \, \epsilon_{\sic\sia\sib}
  =
    - \left(
    \tilde\cd^\sic \tilde\Phi
    + \frac{1}{2} \conformalFactor^{-1} \tilde n^\sic \tilde\Phi
    \right) \dv \tilde\Phi \, \tilde\epsilon_{\sic\sia\sib}
  \,,
\end{align}
where $\conformalFactor$ was again assumed to be chosen such that $f \bdryeq 0$.
The potential cannot be extended to \scri{}, however, its pullback to constant $\conformalFactor$ hypersurfaces admits the smooth limit
\begin{align}
  \underline\theta^\Phi
  &\bdryeq - 
    \tilde n^\sia \tilde\cd_\sia \tilde\Phi
    \dv \tilde\Phi \, \underline{\tilde l} \wedge \underline{\tilde m}
    \label{eq:scalar-potential-pullback}
    \,.
\end{align}
Because $\underline{\dv \theta^\Phi}$ does not vanish at $\scri{}$ the relation~\cref{eq:var-charge-def} with the $\theta^\Phi$ contribution cannot be integrated for $\xi$ that is not tangent to $\bdry\Sigma$, i.e.\ supertranslations.
Following~\cite{Wald:1999wa} we can still define a quantity $H^\Phi_\xi$ satisfying the condition that $\dv H^\Phi_\xi$ is conserved for stationary spacetimes.
Introduce a smooth $(n-1, 1)$-form $\Theta$ on \scri{} that vanishes for stationary spacetimes, such that $\underline{\dv \theta} \bdryeq \dv \Theta$.
Modify~\cref{eq:var-charge-def} by adding the term $\xi \contract \Theta$ on the right hand side, i.e.
\begin{align}
  \dv H^\Phi_\xi
  &= \int_{\bdry\Sigma} ( \dv q_\xi - \xi \contract \theta + \xi \contract \Theta )
  \,.
\end{align}
If there exists a supertranslation $\eta^\sia$, nonvanishing on $\scri$, such that $\lied_\eta \Phi = 0$ it follows that
\begin{align}
  \eta^\sia \tilde \cd_\sia \tilde \Phi
  &=
  \conformalFactor^{-1/2} \lied_\eta \Phi
  - \frac12 \conformalFactor^{-1} \eta^\sia \tilde n_\sia \tilde\Phi
  \bdryeq 0
  \,.
\end{align}
Since $\eta^\sia$ is non-vanishing and proportional to $\tilde n^\sia$ at \scri{} it follows that $\tilde n^\sia \tilde \cd_\sia \tilde \Phi \bdryeq 0$.
This means that $\underline{\theta}^\Phi$ itself vanishes for stationary spacetime, and we can set $\Theta = \underline{\theta}^\Phi$.
It follows that $\dv H^\Phi_\xi = \dv H_\xi$ which we can integrate to $H^\Phi_\xi = H_\xi$.
Also the equation from before for the flux~\cref{eq:3d-flux} holds and is given explicitly by
\begin{align}
  F_\xi
  &\approx \Big({
    - \xi^\sia \tilde\cd_\sia \tilde\Phi \, \tilde n^\sib \tilde\cd_\sib \tilde\Phi
    + \frac{1}{\kappa} {\left( \tilde m^\sia \tilde \cd_\sia \right)}^3 \left( \tilde m_\sib \xi^\sib \right)
    }\Big) \tilde l \wedge \tilde m
    \,.
\end{align}

\section{Expressions in a Coordinate System}
\label{sec:3d-expr-coord-syst}

Now we construct a coordinate system that is adapted to the boundary conditions.
This is done similarly as in~\cite{Tamburino:1966zz} with the modification that only the unphysical metric $\tilde g_{\sia\sib}$ at $\scri$ and the conformal factor $\conformalFactor$ is used in the construction here.
These two are kept fixed by our boundary conditions so the coordinate system does not depend on the particular metric under consideration.
For the construction an additional metric $\bar g_{\sia\sib}$ is defined as
\begin{align}
  \bar g_{\sia\sib}
  &= \frac{y^2}{\conformalFactor^2} \tilde g_{\sia\sib}
  \,,
\end{align}
where $y$ is one of the coordinates constructed below.
From the condition $R \bdryeq 0$ we know that $\scri$ is a null hypersurface.
We proceed by following the steps:
\begin{enumerate}
\item Pick a spacelike slice of $\scri$ and assign a coordinate $\varphi$ to the slice.
  \label{item:coords2}
\item There is a unique null curve through each point of the slice. Assign to each point on a curve the same value of $\varphi$. Take a parameter of the curves as the second coordinate $u$.
  \label{item:coords3}
\item Pick a family of curves that span a hypersurface intersecting $\scri$ in a particular slice of constant $u$ such that each curve is null at $\scri$.
  Apply this procedure for all slices and assign the same values of $u$ and $\varphi$ to all points of any one curve.
  \label{item:coords4}
\item Define $y = \conformalFactor \sqrt{ \tilde g^{\sia\sib} \cd_\sia \varphi \cd_\sib \varphi \big|_\scri }$.
  Here, ``$|_\scri$'' means that the expression is evaluated at the point at $\scri$ with the same value of $u$ and $\varphi$.
  \label{item:coords5}
\item At the beginning there was a freedom of changing the parameter $u$. Use this freedom to set $\bar g_{uy} \bdryeq 1$.
  \label{item:coords6}
\end{enumerate}
This defines our coordinate system $(u, y, \varphi)$.
Since the tangent vectors to the curves in \cref{item:coords3} are null and therefore orthogonal to all vectors in \scri{} it follows that $\bar g_{uu} \bdryeq \bar g_{u\varphi} \bdryeq 0$.
The tangent vectors to the curves in \cref{item:coords4} are null at \scri{} and therefore orthogonal to the constant $u$ hypersurfaces at \scri{}, so $\bar g_{yy} \bdryeq g_{y\varphi} \bdryeq 0$.
\Cref{item:coords5,item:coords6} fix the metric components $\bar g_{\varphi\varphi} = \bar g_{uy} \bdryeq 1$.
The weaker boundary condition $\conformalFactor^{-1} R \bdryeq 0$ is now equivalent to $\pd_y \bar g_{uu} \bdryeq 0$.
The stronger boundary condition that $\conformalFactor V^\sia R_{\sia\sib} \bdryeq 0$ for all $V^\sia$ tangential to $\scri$ is equivalent to
$\pd_y \bar g_{uu} \bdryeq \pd_y \bar g_{u\varphi} \bdryeq 0$.
The curves of constant $\varphi$ at $\scri$ are geodesics with respect to $\tilde g_{\sia\sib}$ and $u$ is an affine parameter.

To summarize, given any metric satisfying the boundary conditions, a coordinate system can be constructed such that the components of the metric obey
\begin{align}
  g_{\sia\sib}
  =
    \begin{pmatrix}
      g_{uu} & g_{uy} & g_{u\varphi} \\
      g_{uy} & g_{yy} & g_{y\varphi} \\
      g_{u\varphi} & g_{y\varphi} & g_{\varphi\varphi}
    \end{pmatrix}
  &= \frac{1}{y^2}
    \begin{pmatrix}
      O(y^2) & 1 + O(y) & O(y^2) \\
      1 + O(y) & O(y) & O(y) \\
      O(y^2) & O(y) & 1 + O(y)
    \end{pmatrix}
    \,.
\end{align}
The construction depends on the unphysical metric at \scri{} and the conformal factor $\conformalFactor$ only.
For convenience, we introduce the function $A$, $B$, $C$, $M$, $N$, and $F$, which depend on $u$ and $\varphi$, and write the metric as
\begin{align}
  g_{\sia\sib}
  &=
    \begin{pmatrix}
      M - \pd_u F + O(y) & \cdots & \cdots \\
      y^{-2} + y^{-1} B + O(1) & y^{-1} A + O(1) & \cdots \\
      \tfrac12 ( N - \pd_\varphi F + \pd_u C ) + O(y) & y^{-1} C + O(1) & y^{-2} + y^{-1} F + O(1)
    \end{pmatrix}
    \label{eq:flat3dMetric}
    \,.
\end{align}
We further take the range of $\varphi$ to be from $0$ to $2 \pi$.
While it might be more intuitive to use the radial coordinate $r$ instead of $y$,
we refrain from doing so because the coordinate system $(r, u, \varphi)$ does not cover $\scri$.
It is difficult to see if the vector $r \pd_r$ vanishes at $\scri$ where $r$ becomes infinite.
Expressing it as $- y \pd_y$ makes it clear that it does indeed vanish.

Since the Hamiltonian functions are given by integrals of differential forms we have to introduce an orientation of spacetime.
This orientation should be in accordance with Stokes' theorem for any boundaries.
We start from the standard orientation such that an integral of $\dd u \wedge \dd r \wedge \dd \varphi$ with $r = 1/y$ is positive.
It follows that the orientations have to be chosen such that the forms in \cref{tab:orientation} are positive.
\begin{table}
\centering
\begin{tabular}{@{}cl@{}}
  \toprule
  Manifold & Positive form
  \\
  \midrule
  $M$
  & $- \dd u \dd y \dd \varphi$
  \\
  $\Sigma$
  & $- \dd y \dd \varphi$
  \\
  $\bdry \Sigma$
  & $\dd \varphi$
  \\
  $\spatialBdry$
  & $- \dd u \dd \varphi$
  \\
  $\bdry \spatialBdry$
  & $- \dd \varphi$
  \\
  \bottomrule
\end{tabular}
\caption{Forms giving rise to an orientation that is in accordance with Stokes theorem.}
\label{tab:orientation}
\end{table}

We now write the conformal factor as
\begin{align}
  \conformalFactor
  &= e^{\lambda} y
  \,,
\end{align}
where $\lambda$ only depends on $\varphi$ since $f \bdryeq 0$.
It follows that
\begin{align}
  \tilde n
  &= e^{\lambda} \left(
    \dd y
    + y \pd_\varphi \lambda \dd \varphi
    \right)
    \,,
\end{align}
and $\tilde l$ and $\tilde m$ can be chosen such that
\begin{align}
  \tilde l
  &\bdryeq e^\lambda \dd u
  \\
  \tilde m
  &\bdryeq e^\lambda \dd \varphi
  \,.
\end{align}
The generators of BMS symmetries have to satisfy~\cref{eq:bmsDef3}.
We choose the subleading terms to vanish such that the generators read
\begin{align}
  \xi
  &=
  ( T + u \pd_\varphi Y ) \pd_u
  + y \pd_\varphi Y \pd_y
  + \big( Y - y \pd_\varphi ( T + u \pd_\varphi Y ) \big) \pd_\varphi
  \,,
\end{align}
where $T$ and $Y$ are functions of $\varphi$.
We want to evaluate the Hamiltonian functions at $u = u_0(\varphi)$.
For that we split $\xi^\sia = \xi_T^\sia + \xi_R^\sia$ into a supertranslation $\xi_T^\sia \bdryeq h \tilde n^\sia$ and a superrotation $\xi_R^\sia$ that is tangent to the $u = u_0(\varphi)$ hypersurface.
The condition for $\xi_R^\sia$ to be tangent to $u = u_0(\varphi)$ is equivalent to $\xi_R^\sia \pd_\sia ( u - u_0 ) \bdryeq 0$.
This fixes the leading terms in the decomposition and we set
\begin{align}
  h
  &= ( T + u_0 \pd_\varphi Y - Y \pd_\varphi u_0 ) e^\lambda
  \\
  \xi_R
  &= \big( (u - u_0) \pd_\varphi Y + Y \pd_\varphi u_0 \big) \pd_u
  + y \pd_\varphi Y \pd_y
  \nonumber \\
  &\quad+ \big( Y - y \pd_\varphi ( (u - u_0) \pd_\varphi Y + Y \pd_\varphi u_0 ) \big) \pd_\varphi
  \,.
\end{align}
Evaluating contributions to the Hamiltonian functions~\cref{eq:hamiltonian-supertrans,eq:hamiltonian-superrot} at $u = u_0(\varphi)$ gives
\begin{align}
  H^R_{\xi_R}
  &= \int_{\bdry\Sigma} \left(
    \frac{1}{\kappa} \tilde l^\sia \tilde l^\sib R_{\sia\sib} \xi^\sic \tilde n_\sic \tilde m
    + q_{\xi_R}
    \right)
  \\
  &\;
  \begin{aligned}
    = \frac{1}{2\kappa} \int_{\bdry\Sigma} \Big[
    \!&\big(
    N Y
    - \pd_\varphi u_0 \pd_u B Y
    - 2 B \pd_\varphi \lambda Y
    - B \pd_\varphi Y
    - \pd_\varphi (F Y)
    - 2 y^{-1} \pd_\varphi Y
    \big) \dd \varphi
    \\
    &+ \big( 2 M Y - \pd_u F Y - 2 \pd_\varphi^2 Y \big) \dd u
    \Big]
  \end{aligned}
  \\
  &\;
  \begin{aligned}
    = \frac{1}{2\kappa} \int_{\bdry\Sigma} \Big[
    \!&\big(
    N
    + \pd_\varphi B
    - 2 B \pd_\varphi \lambda
    \big) Y \dd \varphi
    + 2 \big(
    M Y
    - \pd_\varphi^2 Y
    \big) \dd u
    \Big]
  \end{aligned}
  \\
  H^T_{\xi_T}
  &= \frac{1}{\kappa} \int_{\bdry \Sigma}
    \tilde m^\sia \tilde m^\sib ( S_{\sia\sib} - \tilde S_{\sia\sib} )
    h \tilde m
  \\
  &= \frac{1}{2\kappa} \int_{\bdry\Sigma}
    \big(
    M
    + 2 \sOfLambda
    \big) \big(
    T
    + u_0 \pd_\varphi Y
    - Y \pd_\varphi u_0
    \big) \dd \varphi
  \\
  &= \frac{1}{2\kappa} \int_{\bdry\Sigma} \Big[
    \big( M + 2 \sOfLambda \big) T \dd \varphi
    - \big( u \pd_\varphi M + 2 \, u \pd_\varphi \sOfLambda \big) Y \dd \varphi
    - \big( 2 M + u \pd_u M + 4 \sOfLambda \big) Y \dd u
  \Big]
  \,,
\end{align}
with
\begin{align}
  \sOfLambda
  &= \lambda'' - \frac12 \lambda'^2
  \,.
\end{align}
Combined, these two expressions give the Hamiltonian
\begin{alignat}{3}
  H_\xi
  &= \frac{1}{2\kappa} \int_{\bdry\Sigma} &\Big[
    \!&\big( M + 2 \sOfLambda \big) T \dd \varphi
    + \big(
    N
    - u \pd_\varphi M
    + \pd_\varphi B
    - 2 B \pd_\varphi \lambda
    - 2 u \pd_\varphi \sOfLambda
    \big) Y \dd \varphi
  \nonumber \\
  &&&
  - \big(
  u \pd_u M Y
  + 2 \pd_\varphi^2 Y
  + 4 \sOfLambda Y
  \big) \dd u
  \Big]
  \\
  &= \frac{1}{2\kappa} \int_{\bdry\Sigma} &\Big[
    \!&\big( M + 2 \sOfLambda \big) T
    + \big(
    N
    - u_0 \pd_\varphi M
    - u_0 \pd_\varphi u_0 \pd_u M
    \nonumber \\&&&
    + \pd_\varphi B
    - 2 B \pd_\varphi \lambda
    - 2 \pd_\varphi^3 u_0
    - 4 \sOfLambda \pd_\varphi u_0
    - 2 u_0 \pd_\varphi \sOfLambda
    \big) Y
  \Big] \dd \varphi
  \,,
\end{alignat}
where the metric coefficients are evaluated at $u = u_0(\varphi)$.
For the flux~\cref{eq:3d-flux} we find
\begin{align}
  F_\xi
  = - \frac{1}{2\kappa} \Big(
    &\pd_u M T
    + \big({
    \pd_u N
    - \pd_\varphi M
    + \pd_u { \pd_\varphi B }
    - 2 \pd_\varphi \lambda \pd_u B
    + 2 \pd_\varphi \sOfLambda
    }\big) Y
  \nonumber \\
  &
    + \big( u \pd_u M + 4 \sOfLambda \big) \pd_\varphi Y
    + 2 \pd_\varphi^3 Y
    \Big) \dd \varphi \dd u
    \,.
\end{align}
For vacuum solutions where $G_{\sia\sib} = 0$ the coefficients in the metric satisfy
\begin{align}
  M
  &= \Theta
  &
  N
  &= \Xi + u \pd_\varphi \Theta
  &
  B
  &= 0
  \label{eq:vacuum}
  \,,
\end{align}
where $\Theta$ and $\Xi$ are functions of $\varphi$ only.
In this case the Hamiltonian functions reduce to
\begin{align}
  H_\xi
  &= \frac{1}{2\kappa} \int_{\bdry\Sigma} \Big[
  \Big( (
    \Theta
    + 2 \sOfLambda
    ) T
    + ( \Xi - 2 u \pd_\varphi \sOfLambda ) Y
    \Big) \dd \varphi
    - \Big(
    2 \pd_\varphi^2 Y
    + 4 \sOfLambda Y
    \Big) \dd u
    \Big]
    \,.
\end{align}
For $\lambda = 0$ and constant $u_0$ this is the same as the expression found by Barnich \& Troessaert~\cite{Barnich:2010eb}.
Even for vacuum solution the Hamiltonian functions corresponding to most superrotations are not conserved.
The flux is given by
\begin{align}
  F_\xi
  &= - \frac{1}{\kappa} \Big(
  Y \pd_\varphi \sOfLambda
  + 2 \pd_\varphi Y \sOfLambda
  + \pd_\varphi^3 Y
  \Big) \dd \varphi \dd u
  \,.
\end{align}

\section{Invariance and Poisson Algebra}
\label{sec:3d-invariance}

By introducing enough background fields, Hamiltonian functions can always be written in a way such that they are invariant under diffeomorphisms when transforming all dynamical and background fields, the slice $\Sigma$ and the vector field $\xi^\sia$.
When transforming only the slice $\Sigma$ we obtain information about the conservation of the Hamiltonian function.
Transforming the dynamical fields leads to the Poisson bracket by~\cref{eq:poisson-bracket-alt}.

In the current case the $\dv$-derivative of the Hamiltonian function $\dv H_\xi$ is by construction independent of any background field, but the Hamiltonian function itself depends on $\conformalFactor$.
Consider diffeomorphisms generated by a BMS transformation $\zeta$ and denote by $\Delta^g_\zeta$, $\Delta^\conformalFactor_\zeta$, $\Delta^\Sigma_\zeta$, $\Delta^\xi_\zeta$ the action of $\zeta$ on the dynamical metric $g_{\sia\sib}$, the background field $\conformalFactor$, the slice $\Sigma$ and the vector field $\xi$, respectively.
Since the Hamiltonian functions are invariant under transformation of all of these we have
\begin{align}
  \Delta^g_\zeta H_\xi
  + \Delta^\conformalFactor_\zeta H_\xi
  + \Delta^\Sigma_\zeta H_\xi
  + \Delta^\xi_\zeta H_\xi
  = 0
  \,.
\end{align}
Consider now a finite change in the conformal factor
\begin{align}
  \otherConformalFactor
  &= e^\gamma \conformalFactor
    \,,
\end{align}
where $\tilde n^\sia \tilde \cd_\sia \gamma \bdryeq 0$ in order to preserve the condition $f \bdryeq 0$ as required by the definition of the Hamiltonian function.
It follows that
\begin{align}
  \bar g_{\sia\sib}
  &= e^{2\gamma} \tilde g_{\sia\sib}
  \\
  \bar n_\sia
  &= e^\gamma \left( \tilde n_\sia + \conformalFactor \tilde \cd_\sia \gamma \right)
    \,.
\end{align}
We can choose $\bar m^\sia$ and $\bar l^\sia$ such that
\begin{align}
  \bar m^\sia
  &\bdryeq e^{-\gamma} \tilde m^\sia
  \\
  \bar l^\sia
  &\bdryeq e^{-\gamma} \tilde l^\sia
    \,.
\end{align}
The decomposition into supertranslations and superrotations is not affected, but since $\xi_T^\sia \bdryeq h \tilde n^\sia \bdryeq \bar h \bar n^\sia $ we have
\begin{align}
  \bar h
  &\bdryeq e^\gamma h
    \,.
\end{align}
To construct the change of the supertranslation Hamiltonian functions we need
\begin{align}
  \bar m^\sia \bar m^\sib \bar S_{\sia\sib}
  &= \bar m^\sia \bar m^\sib \left(
    \tilde S_{\sia\sib}
    - \tilde\cd_\sia { \tilde\cd_\sib \gamma }
    + \tilde \cd_\sia \gamma \tilde \cd_\sib \gamma
    - \frac{1}{2} \tilde g_{\sia\sib} {(\tilde \cd \gamma)}^2
    \right)
  \\
  &\bdryeq e^{-2\gamma} \left(
    \tilde m^\sia \tilde m^\sib \tilde S_{\sia\sib}
    - {\left( \tilde m^\sia \tilde\cd_\sia \right)}^2 \gamma
    + \frac12 {\left( \tilde m^\sia \tilde \cd_\sia \gamma \right)}^2
    \right)
  \,,
\end{align}
where we used~\cref{eq:metric-decomposition} as well as the fact that $\tilde m^\sib \tilde \cd_\sib \tilde m^\sia$ is proportional to $\tilde n^\sia$ on any point of $\scri$.
Since $\tilde m^\sia S_{\sia\sib}$ and $\conformalFactor S_{\sia\sib}$ are both finite at $\scri$ by the boundary conditions it holds that
\begin{align}
  \bar m^\sia \bar m^\sib S_{\sia\sib}
  &\bdryeq e^{-2\gamma} \tilde m^\sia \tilde m^\sib S_{\sia\sib}
  \,.
\end{align}
The change of the Hamiltonian function is now given by
\begin{align}
  \bar H_\xi
  &= H_\xi
    + \frac{1}{\kappa} \int_{\bdry \Sigma} \left(
    h {\left( \tilde m^\sia \tilde\cd_\sia \right)}^2 \gamma
    - \frac{h}{2} {\left( \tilde m^\sia \tilde \cd_\sia \gamma \right)}^2
    + \conformalFactor \tilde l^\sia \tilde l^\sib R_{\sia\sib} \xi^\sic \tilde \cd_\sic \gamma
    \right) \tilde m
    \label{eq:charge-change-omega}
    \,.
\end{align}
The change of $H_\xi$ depends on $\gamma$ at $\scri$, so we see that $H_\xi$ itself depends on the leading order of $\conformalFactor$.
Linearize~\cref{eq:charge-change-omega} and use $\Delta^\conformalFactor_\zeta \conformalFactor = \lied_\zeta \conformalFactor$ to get
\begin{align}
  \Delta^\conformalFactor_\zeta H_\xi
  &= \frac{1}{\kappa} \int_{\bdry \Sigma} \left(
    h {\left( \tilde m^\sia \tilde\cd_\sia \right)}^2 ( \conformalFactor^{-1} \zeta^\sia \tilde n_\sia )
    + \conformalFactor \tilde l^\sia \tilde l^\sib R_{\sia\sib} \xi^\sic \tilde \cd_\sic ( \conformalFactor^{-1} \zeta^\sia \tilde n_\sia )
    \right) \tilde m
    \,.
\end{align}
The action of $\Delta^\Sigma_\zeta$ on the Hamiltonian function is obtained from the flux
\begin{align}
  \Delta^\Sigma_\zeta H_\xi
  &= - \int_{\bdry \Sigma} \zeta \contract F_\xi
  \\
  &= \frac{1}{\kappa} \int_{\bdry \Sigma} \Big({
    \conformalFactor^{-1} \xi^\sia \tilde n^\sib G_{\sia\sib}
    - {\left( \tilde m^\sia \tilde \cd_\sia \right)}^3 \left( \tilde m_\sib \xi^\sib \right)
    }\Big) \zeta^\sic \big( \tilde l_\sic \tilde m - \tilde m_\sic \tilde l \big)
    \,.
\end{align}
The action of $\Delta^\xi_\zeta$ is given by the Lie derivative
\begin{align}
  \Delta^\xi_\zeta H_\xi
  &= H_{\lied_\zeta \xi}
  = H_{[\zeta, \xi]}
  \,.
\end{align}
The action of $\Delta^g_\zeta$ is the same as the action of $X_\zeta$, i.e.
\begin{align}
  \Delta^g_\zeta H_\xi
  &= X_\zeta \contract \dv H_\xi
  \,.
\end{align}
If $H_\xi$ are Hamiltonian functions, as in the case when there are no additional matter fields, we can use \cref{ex:diffeos,eq:poisson-central-extension,eq:poisson-bracket-alt} to rewrite this as
\begin{align}
  \Delta^g_\zeta H_\xi
  &= \pb{H_\zeta}{H_\xi}
  = - H_{[\zeta, \xi]} + C(\zeta, \xi)
  \,,
\end{align}
with some $C$ such that $\dv C(\zeta, \xi) = 0$.
It follows for vacuum solutions ($G_{\sia\sib} = 0$) that
\begin{align}
  C(\zeta, \xi)
  &= - \Delta^\conformalFactor_\zeta H_\xi
  - \Delta^\Sigma_\zeta H_\xi
  \\
  &= \frac{1}{\kappa} \int_{\bdry \Sigma} \bigg(
    {\left( \tilde m^\sia \tilde \cd_\sia \right)}^3 \left( \tilde m_\sib \xi^\sib \right) \zeta^\sic \big( \tilde l_\sic \tilde m - \tilde m_\sic \tilde l \big)
    - h {\left( \tilde m^\sia \tilde\cd_\sia \right)}^2 ( \conformalFactor^{-1} \zeta^\sia \tilde n_\sia ) \tilde m
  \bigg)
  \,.
\end{align}

\section{Additional Background Structure}
\label{sec:3d-addit-backgr-struct}

Until now the Hamiltonian functions required a conformal factor $\conformalFactor$ as background structure to be well-defined.
We found that the superrotation flux is not zero, even for Minkowski space.
One can define Hamiltonian functions such that all fluxes vanish at the cost of introducing additional background structure.
This structure is given by a scalar field $\psi$ satisfying the condition
\begin{align}
  \tilde n^\sia \tilde \cd_\sia \psi
  &\bdryeq 1
  \label{eq:cond-psi}
  \,.
\end{align}
The Hamiltonian~\cref{eq:flat3dhamiltonian} can then be modified by adding a $\dv$-closed term,
\begin{align}
  \hat H_\xi
  &=
    H_\xi - \frac{1}{\kappa} \int_{\bdry\Sigma}
    {\left( \tilde m^\sia \tilde \cd_\sia \right)}^3 \left( \tilde m_\sib \xi^\sib \right) \psi \tilde m
    \,.
\end{align}
The flux of the hatted Hamiltonian functions,
\begin{align}
  \hat H_{\xi,\Sigma_2} - \hat H_{\xi,\Sigma_1}
  &= - \int_{\spatialBdry} \hat F_\xi
  \,,
\end{align}
is given by the simple expression
\begin{align}
  \hat F_\xi
  &= - \frac{1}{\kappa} \conformalFactor^{-1} \xi^\sia \tilde n^\sib G_{\sia\sib} \, \tilde l \wedge \tilde m
    \,,
\end{align}
which vanishes for vacuum solutions to Einstein's equations.
In the coordinate system of~\cref{sec:3d-expr-coord-syst} condition~\cref{eq:cond-psi} is equivalent to the condition that $\psi$ can be written in the form
\begin{align}
  \psi
  &\bdryeq \big( u - \psi_0(\varphi) \big) e^\lambda
  \,.
\end{align}
Using
\begin{align}
  {\left( \tilde m^\sia \tilde \cd_\sia \right)}^3 \left( \tilde m_\sib \xi^\sib \right) \psi \tilde m
  &\bdryeq e^{-\lambda} \pd_\varphi \left( e^{-\lambda} \pd_\varphi \left( e^{-\lambda} \pd_\varphi \left( e^\lambda Y \right) \right) \right) \psi e^\lambda \dd \varphi
  \\
  &\bdryeq \big(
    {\pd_\varphi^3 Y}
    + 2 \pd_\varphi Y \sOfLambda
    + Y \pd_\varphi \sOfLambda
    \big) ( u_0 - \psi_0 ) \dd \varphi
  \,,
\end{align}
we find the Hamiltonian function to be
\begin{alignat}{3}
  \hat H_\xi
  &= \frac{1}{2\kappa} \int_{\bdry\Sigma} &\Big[
    \!&\big( M + 2 \sOfLambda \big) T
    + \big(
    N
    - u_0 \pd_\varphi M
    - u_0 \pd_\varphi u_0 \pd_u M
    \nonumber \\&&&
    + \pd_\varphi B
    - 2 B \pd_\varphi \lambda
    - 2 \pd_\varphi^3 \psi_0
    - 4 \sOfLambda \pd_\varphi \psi_0
    - 2 \psi_0 \pd_\varphi \sOfLambda
    \big) Y
  \Big] \dd \varphi
  \,.
\end{alignat}
The flux can be expressed as
\begin{align}
  \hat F_\xi
  = - \frac{1}{2\kappa} \Big({
    \pd_u M T
    + \big({
    \pd_u N
    - \pd_\varphi M
    + \pd_u { \pd_\varphi B }
    - 2 \pd_\varphi \lambda \pd_u B
    }\big) Y
    + u \pd_u M \pd_\varphi Y
    }\Big) \dd \varphi \dd u
    \,.
\end{align}
For constant $\psi_0$ and constant $\lambda$ the Hamiltonian functions are given by
\begin{alignat}{3}
  \hat H_\xi
  &= \frac{1}{2\kappa} \int_{\bdry\Sigma} &\Big[
    \!&M T
    + \big(
    N
    - u_0 \pd_\varphi M
    - u_0 \pd_\varphi u_0 \pd_u M
    + \pd_\varphi B
    \big) Y
  \Big] \dd \varphi
  \,,
\end{alignat}
which is the same as the Hamiltonian function obtained using the method by Barnich and Brandt~\cite{Barnich:2001jy}.
For constant $u_0$ an equivalent expression was given by Barnich and Troessaert~\cite{Barnich:2010eb}.
For vacuum solutions~\cref{eq:vacuum} the Hamiltonian functions read
\begin{alignat}{3}
  \hat H_\xi
  &= \frac{1}{2\kappa} \int_{\bdry\Sigma} \Big[
    \big( \Theta + 2 \sOfLambda \big) T
    + \big(
    \Xi
    - 2 \pd_\varphi^3 \psi_0
    - 4 \sOfLambda \pd_\varphi \psi_0
    - 2 \psi_0 \pd_\varphi \sOfLambda
    \big) Y
  \Big] \dd \varphi
  \label{eq:3d-vacuum-charge-coord}
  \,.
\end{alignat}

\section{Scalar Field Solutions}
\label{sec:3d-scalar-sol}

We now introduce a scalar field into the theory and show how to solve the equations of motion.
They are given by $\Delta_{\sia\sib} = 0$ and $\Delta = 0$ with
\begin{align}
  \Delta_{\sia\sib}
  &\defn G_{\sia\sib}
  - \kappa T_{\sia\sib}
  \\
  \Delta
  &\defn \pd_\sia \left( \sqrt{-g} \pd^\sia \Phi \right)
  \,,
\end{align}
where
\begin{align}
  T_{\sia\sib}
  &= \pd_\sia \Phi \pd_\sib \Phi
  - \frac{1}{2} (\pd \Phi)^2 g_{\sia\sib}
  \,.
\end{align}
In \cref{sec:3d-expr-coord-syst}, we only used the metric at $\scri$ to construct a coordinate system.
By also using the metric in the bulk of spacetime one can always construct a coordinate system such that the metric has the form
\begin{align}
  g_{\sia\sib} \dd x^\sia \dd x^\sib
  &= y V e^{2\beta} \dd u^2 + 2 y^{-2} e^{2\beta} \dd u \dd y + y^{-2} {( \dd \varphi - U \dd u )}^2
  \,,
  \label{eq:asymptotic-metric}
\end{align}
with $U, V, \beta$ functions of $u, y, \varphi$ satisfying
\begin{align} \label{eq:eom-bcs}
  U &= O(y^2)
  &
  \beta &= O(y)
  &
  V &= O(y^{-1})
  \,.
\end{align}
This is similar to the one given by Bondi~\cite{Bondi:1960jsa} in four dimensions.
In contrast with \cref{sec:3d-expr-coord-syst}, we have fixed additional components of the metric to make the following analysis simpler at the cost that the coordinate system now depends on the metric.

\subsection{Solving the Equations of Motion}

The equations of motion are not all independent.
We will first solve some of them and then use the Bianchi identities to simplify others.
The equations $\Delta_{yy} = \Delta_{y\varphi} = \Delta_{yu} = 0$, with
\begin{align}
  \Delta_{yy}
  &= - 2 y^{-1} \pd_y \beta - \kappa {(\pd_y \Phi)}^2
    \label{eq:eom-main-1}
  \\
  \Delta_{y\varphi}
  &= - \frac{y}{2} \pd_y \left(y^{-1} e^{-2 \beta} \pd_y U \right)
    - y^{-1} \pd_y \left( y \pd_\varphi \beta \right)
    - \kappa \pd_y \Phi \pd_\varphi\Phi
    \label{eq:eom-main-2}
  \\
  \Delta_{yu}
  &= - U \Delta_{y\varphi}
    + \frac{y^3}{2} V \Delta_{yy}
    + \frac{y}{2} \pd_y \left( y V \right)
    + e^{2 \beta} \left(
    \frac{\kappa}{2} {\left(\pd_\varphi\Phi \right)}^2
    + {\left(\pd_\varphi \beta \right)}^2
    + \pd_\varphi^2 \beta
    \right)
  \nonumber \\
  &\quad
    + \frac{1}{4} e^{-2 \beta } {\left(\pd_y U\right)}^2
    + \frac{y^2}{2} \pd_y ( y^{-2} \pd_\varphi U ) \label{eq:eom-main-3}
\end{align}
make it possible to express $\beta$, $U$, and $V$ in terms of $\Phi$ and four functions of $u$ and $\varphi$ serving as integration ``constants''.
From the contracted Bianchi identity ($\cd_\sib G^\sib_\sia = 0$) and because the energy momentum tensor is divergence-less ($\cd_\sia T^\sia_\sib = 0$) it follows that
\begin{align}
  \cd_\sib \Delta^\sib_\sia
  &= \frac{1}{\sqrt{-g}} \pd_\sib (\sqrt{-g} \Delta^\sib_\sia)
  - \Gamma\indices{^\sic_\sib_\sia} \Delta^\sib_\sic
  = 0
  \,.
\end{align}
Using the relation
\begin{align}
  \Gamma\indices{^\sic_\sib_\sia} \Delta^\sib_\sic
  &= - y \Delta_{\varphi \varphi} \dd y_\sia
    \,,
\end{align}
it follows that if $\Delta_{yy} = \Delta_{y\varphi} = \Delta_{yu} = \Delta = 0$, then
\begin{align}
  \Delta_{\varphi\varphi}
  &= 0
    \label{eq:bianchi-consequence-1}
  \\
  \pd_\sia ( \sqrt{-g} \Delta^\sia_u )
  &= 0
    \label{eq:bianchi-consequence-2}
  \\
  \pd_\sia ( \sqrt{-g} \Delta^\sia_\varphi )
  &= 0
    \label{eq:bianchi-consequence-3}
    \,.
\end{align}
From~\cref{eq:bianchi-consequence-3} follows
\begin{align}
  \pd_y ( y^{-1} \Delta_{u\varphi} )
  &= 0
    \label{eq:bianchi-consequence-4}
\end{align}
and if $\Delta_{u\varphi} = 0$,~\cref{eq:bianchi-consequence-2} becomes
\begin{align}
  \pd_y ( y^{-1} \Delta_{uu} )
  &= 0
    \label{eq:bianchi-consequence-5}
    \,.
\end{align}
This means that $y^{-1} \Delta_{u\varphi}$ and $y^{-1} \Delta_{uu}$ vanish if they vanish in the limit $y \to 0$.
We have to solve the four main equations
\begin{align}
  \Delta_{yy} = \Delta_{y\varphi} = \Delta_{yu}
  = \Delta
  &= 0
\end{align}
and the two supplementary conditions
\begin{align}
  \lim_{y \to 0} y^{-1} \Delta_{u\varphi}
  &= 0 \label{eq:supp-1}
  \\
  \lim_{y \to 0} y^{-1} \Delta_{uu}
  &= 0 \label{eq:supp-2}
    \,.
\end{align}
First, we solve~\cref{eq:eom-main-1,eq:eom-main-2,eq:eom-main-3} and obtain
\begin{align}
  \beta
  &= - \frac{\kappa}{2} \int y {(\pd_y \Phi)}^2 \dd y
     + H(u, \varphi)
    \label{eq:sol-beta}
  \\
  U
  &= - \int y e^{2 \beta} \left(
    2 \int \left(
    y^{-2} \pd_y ( y \pd_\varphi \beta )
    + \kappa y^{-1} \pd_y \Phi \pd_\varphi\Phi
    \right) \dd y
    + \bar N(u, \varphi)
    \right) \dd y
  \nonumber \\
  &\quad
    + L(u, \varphi)
    \label{eq:sol-U}
  \\
  V
  &= - y^{-1} \int \Big(
    y^{-1} e^{2 \beta} \left(
    \kappa {\left( \pd_\varphi \Phi \right)}^2
    + 2 {\left(\pd_\varphi \beta \right)}^2
    + 2 \pd_\varphi^2 \beta
    \right)
  \nonumber \\
  &\qquad\qquad\qquad
    + \frac12 y^{-1} e^{-2 \beta } {\left( \pd_y U \right)}^2
    + y \pd_y ( y^{-2} \pd_\varphi U )
    \Big) \dd y
    + y^{-1} \bar M(u, \varphi)
    \label{eq:sol-V}
    \,,
\end{align}
where $H$, $L$, $\bar M$, and $\bar N$ are arbitrary functions of $u$ and $\varphi$.
In accordance with~\cref{eq:scalar-field-falloff} we assume that the scalar field can be expanded as the series
\begin{align}
  \Phi
  &= \sqrt{y} \sum_{k=0}^\infty y^{k} \Phi_k(u, \varphi) \,.
\end{align}
Plugging this into~\cref{eq:sol-U,eq:sol-V,eq:sol-beta} leads to expansions of $\beta$, $U$ and $V$ in powers of $y$ with leading contributions
\begin{align}
\begin{aligned}
  \beta
  &= - \frac{\kappa}{8} y \, \Phi_0^2 + O(y^2)
  \\
  U
  &= - \frac12 y^2 \bar N + O(y^3)
  \\
  V
  &= y^{-1} \bar M + O(1) \,.
\end{aligned} \label{eq:asympt-metric-params}
\end{align}
where we have set $H = L = 0$ to satisfy the falloff conditions~\cref{eq:eom-bcs}.
We find that the expressions for $M$, $N$, and $B$ in~\cref{eq:flat3dMetric} are
\begin{align}
  M
  &= \bar M
  &
  N
  &= \bar N
  &
  B
  &= - \frac{\kappa}{4} \, \Phi_0^2
    \,,
\end{align}
so we drop the bars from $\bar M$ and $\bar N$.
The supplementary conditions are
\begin{align}
  \lim_{y \to 0} y^{-1} \Delta_{u\varphi}
  &= \frac{1}{4} \left( 2\pd_\varphi M - 2 \pd_u N - 3\kappa \pd_u \Phi_0 \pd_\varphi \Phi_0 + \kappa\Phi_0\pd_u { \pd_\varphi \Phi_0 } \right) = 0
  \\
  \lim_{y \to 0} y^{-1} \Delta_{uu}
  &= - \frac{1}{2} \pd_u M - \kappa \, {\left(\pd_u\Phi_0\right)}^2 = 0
  \,,
\end{align}
and are solved by
\begin{align}
  M(u, \varphi)
  &= - 2 \kappa \int {\left(\pd_u\Phi_0\right)}^2 \dd u + \Theta(\varphi)
  \label{eq:sol-M}
  \\
  N(u, \varphi)
  &= \int \left( \pd_\varphi M + \frac{\kappa}{2} \left(- 3 \pd_u \Phi_0 \pd_\varphi\Phi_0 + \Phi_0\pd_u { \pd_\varphi \Phi_0 } \right) \right) \dd u + \Xi(\varphi)
  \label{eq:sol-N} \,,
\end{align}
with two arbitrary functions $\Theta$ and $\Xi$.
With all other equations solved we turn to the equation of motion of the scalar field:
\begin{align}
  \Delta
  &= y^{-1} \pd_u { \pd_y \Phi }
    + \pd_y (y^{-1} \pd_u \Phi + y^{-1} U \pd_\varphi \Phi - y^2 V \pd_y \Phi)
  \\
  &\quad+ y^{-2} \pd_\varphi \left( y e^{2 \beta } \pd_\varphi \Phi + y U \pd_y \Phi \right)
    = 0
    \label{eq:eom-scalar}
    \,.
\end{align}
Inserting the series expansion of $\Phi$ this becomes
\begin{align}
  \Delta = \sqrt{y} \sum_{k=1}^\infty y^k ( 2 k \pd_u \Phi_k + X_k )
  \label{eq:eom-scalar-final}
  \,,
\end{align}
where $X_k$ are expressions containing $\Phi_l$ with $l < k$.
We can solve~\cref{eq:eom-scalar-final} order by order for arbitrary $\Phi_0$.
A function depending on $\varphi$ appears for each order as constant of integration.

\chapter{Linking Past and Future Null Infinity in Three Dimensions}
\label{cha:linking-past-future}

Recently, the rich infrared structure of perturbative quantum gravity in four-dimensional asymptotically flat spacetimes has attracted increased attention.
The asymptotic boundary of these spacetimes contains past and future null infinity denoted by $\scri^-$ and $\scri^+$, respectively.
Both are separately invariant under the infinite-dimensional BMS group.
Surprisingly, this symmetry group is intimately related to both the gravitational memory effect and Weinberg's soft graviton theorem~\cite{Strominger:2013jfa,Strominger:2014pwa,He:2014laa}.
In particular, the latter arises as a Ward identity for BMS invariance of the S-matrix.
To consider the BMS group as a symmetry of the S-matrix one must relate the two --- a priori independent --- symmetry groups at each boundary.

In this chapter a linking between the two asymptotic regions and their symmetries in three-dimensional Einstein gravity~\cite{Prohazka:2017equ} is presented.
In four and higher, even dimensions, the linking was accomplished previously~\cite{Strominger:2013jfa,Kapec:2015vwa} (although for the higher-dimensional case see the objections~\cite{Hollands:2016oma}).
Three-dimensional pure Einstein gravity does not exhibit local degrees of freedom, i.e.\ gravitational waves, but the theory possesses degrees of freedom on the boundary.
Nontrivial scattering in the interior is obtained by coupling the theory to propagating matter.
Due to its technical simplicity, e.g., detailed knowledge of the phase space, the theory then provides a unique testing lab for further studies of the infrared sector of quantum gravity, building upon~\cite{He:2014laa,Strominger:2013jfa}.
We provide a first step toward studying such a setup and its relation to BMS symmetry by breaking the two separate BMS symmetries, ending up with a single global one.

Attempts at a holographic framework of asymptotically flat spacetimes yield another motivation for this work.
Compared to anti-de Sitter (AdS) space, where holography is realized in form of the Anti-de Sitter/conformal field theory (AdS/CFT) correspondence, flat space holography is still poorly understood.
AdS$_3$/CFT$_2$ is one of the prime examples of holography, due to the high level of control over both sides of the correspondence.
Given the conceptual clarity of AdS holography in three dimensions, three-dimensional space suggests itself as a natural testing ground for ideas of flat space holography.

Most of the recent evidence~\cite{Barnich:2012rz,Barnich:2012xq,Barnich:2013yka,Bagchi:2012xr,Bagchi:2012yk,Bagchi:2013lma,Bagchi:2015wna,Campoleoni:2015qrh,Campoleoni:2016vsh,Detournay:2014fva,Garbarz:2015lua,Hartong:2015usd,Bonzom:2015ans,Carlip:2016lnw}
for a field theory dual to Einstein gravity on three-dimensional flat space was focused on one connected component of $\scri$ only.
A holographic framework for flat spacetimes should benefit from considerations involving both null boundary components.
In \cref{sec:flat} we start by providing boundary conditions, asymptotic symmetries and charges for our spacetimes.
Following a discussion of the phase space of vacuum solutions in  \cref{sec:phase-space-and-mapping} we provide a linking of their asymptotic regions in \cref{sec:linking} using symmetry arguments.
In \cref{sec:matter} we argue that the linking can be generalized to hold when matter is present.

\section{Asymptotically Flat Spacetimes}
\label{sec:flat}

We use again the gauge fixed metric~\cref{eq:asymptotic-metric} with $r = y^{-1}$
\begin{align}
\label{eq:1}
  \dd s^2
  &= r^{-1} V^+ e^{2\beta^+} \dd u^2
    - 2e^{2\beta^+}\dd u \dd r
    + r^2(\dd \varphi - U^+\dd u)^2
\end{align}
around $\scri^+$ and similarly around $\scri^-$,
\begin{align}
  \label{eq:2}
  \dd s^2
  &= r^{-1} V^- e^{2\beta^-} \dd v^2
    + 2e^{2\beta^-}\dd v \dd r
    + r^2(\dd \varphi - U^-\dd v)^2
    \, .
\end{align}
We assume the periodicity $\varphi \sim \varphi + 2 \pi$.
Here $u$ and $v$ are retarded and advanced time coordinates.
Diffeomorphisms preserving the form of the metric act as
\begin{align}
\label{eq:bms-diffeos}
\begin{aligned}
  u
  &\to u f'(\varphi) + \alpha(f(\varphi)) + O(r^{-1})
  \\
  r
  &\to r / f'(\varphi) + O(1)
  \\
  \varphi
  &\to f(\varphi) + O(r^{-1})
    \, ,
\end{aligned}
\end{align}
around $\scri^+$ and similarly around $\scri^-$.
The function $f$ is required to be a diffeomorphism on the circle and parametrizes superrotations. Supertranslations are parametrized by $\alpha$.
The diffeomorphisms~\cref{eq:bms-diffeos} can be continued arbitrarily into the bulk.
Moreover, their form around $\scri^+$ is a priori not related to their form around $\scri^-$.
It follows that there is the freedom of choosing the coordinate systems~\cref{eq:1} and~\cref{eq:2} independently.
This freedom is precisely expressed by the BMS group acting on $\scri^+$, which we refer to as $\mathrm{BMS}^+$
and the one acting on $\scri^-$, $\mathrm{BMS}^-$.

Metrics of the form~\cref{eq:1} and~\cref{eq:2}, solving the vacuum Einstein equations, have the form
\begin{align}
  \label{eq:3}
  \dd s^2
  = \Theta^+ \dd u^2
  - 2\dd u \dd r
  + \left(\Xi^+ + u \pd_\varphi \Theta^+ \right) \dd u \dd \varphi
  + r^2 \dd \varphi^2
  \,,
\end{align}
and
\begin{align}
  \label{eq:7}
  \dd s^2
  = \Theta^- \dd v^2
  + 2\dd v \dd r
  + \left(\Xi^- + v \pd_\varphi \Theta^- \right) \dd v \dd \varphi
  + r^2 \dd \varphi^2
  \,,
\end{align}
with arbitrary functions $\Theta^\pm(\varphi)$ and $\Xi^\pm(\varphi)$.
To make contact with previous calculations~\cite{Barnich:2014zoa} we consider the charges~\cref{eq:3d-vacuum-charge-coord} with $s = \psi_0 = 0$, i.e.
\begin{align}
\label{eq:charges}
  H_{T,Y}
  &= \frac{1}{2 \kappa} \int_0^{2\pi} \left(\Theta T + \Xi Y \right) \dd\varphi
    \,.
\end{align}
Here, $T(\varphi)$ and $Y(\varphi)$ parametrize infinitesimal supertranslations and superrotations, respectively.
The energy of a spacetime is given by the charge $H_{1,0}$, its angular momentum by $H_{0,1}$.
Under a finite BMS transformation~\cref{eq:bms-diffeos}, the functions $\Theta$ and $\Xi$ transform as~\cite{Barnich:2012rz}
\begin{align}
\label{eq:finBMS}
\begin{aligned}
  \Theta
  &\to
    (f')^2 \Theta \circ f - 2 S[f]
  \\
  \Xi
  &\to
    (f')^2 \Big[
    \Xi
    + \Theta' \alpha
    + 2 \alpha' \Theta
    - 2 \alpha'''
    \Big] \circ f
   \,,
\end{aligned}
\end{align}
where $S[f]$ denotes the Schwarzian derivative.

In the following sections we derive a mapping between the two asymptotic regions, which then leads to the linking of the symmetry groups $\BMS^+$ and $\BMS^-$.

\section{Phase Space and Validity of the Mapping}
\label{sec:phase-space-and-mapping}

In this section we collect results on the phase space of three-dimensional, asymptotically flat gravity without matter
and clarify under which condition the linking of future and past null infinity presented in the next section is sensible and feasible.

The functions $\Theta$ and $\Xi$ transform, as can be seen from~\cref{eq:finBMS},
in the coadjoint representation of the centrally extended BMS group.
The phase space splits into disjoint orbits of the BMS group.
These orbits were classified in~\cite{Barnich:2015uva}; for a thorough introduction to the topic, consult~\cite{Oblak:2016eij}.
All solutions with different constant $\Theta$ or $\Xi$ belong to separate orbits, which means that these orbits can be uniquely labeled by their constant representative.
Relevant to the discussion are two additional families of orbits that do not admit constant representatives:
First, there is a two-parameter family of orbits with $\Theta = - 1$, but nonconstant $\Xi$.
Second, there are particular orbits without constant $\Theta$ representative, so called ``massless deformation'' orbits~\cite{Barnich:2014zoa}.
All other orbits do not have an energy bounded from below~\cite{Barnich:2014zoa}.
Positivity of the energy is a physically reasonable requirement, so these orbits are not considered in the following.

\begin{figure}
\centering
\includegraphics{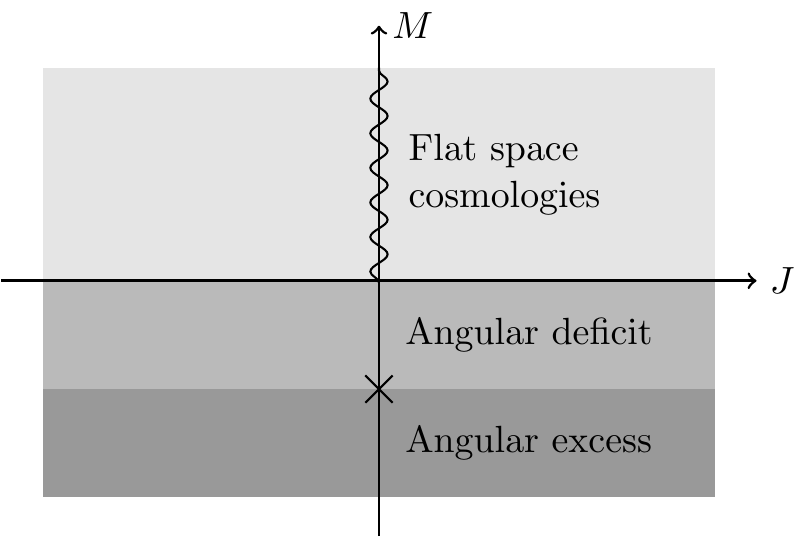}
\caption{The phase space of the spacetimes given in equation~\eqref{eq:constrep}.
The cross at $M=-1, J=0$ is Minkowski space.
The snake line indicates that the linking between past and future null infinity appears nonsensical at $M \ge 0, J = 0$.
The energy of a spacetime with angular excess is not bounded from below when acted upon by BMS transformations.}
\label{fig:MJ}
\end{figure}
\begin{figure}
\centering
\includegraphics{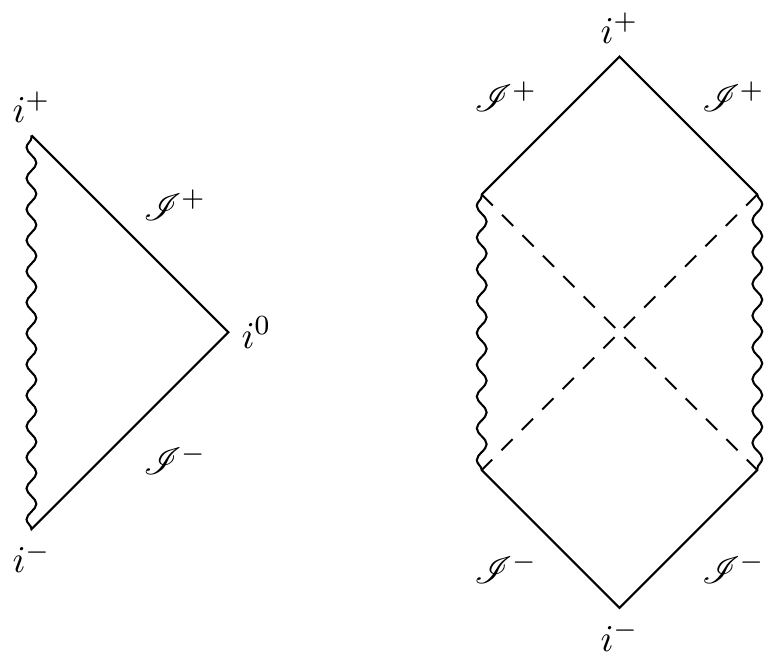}
\caption{Penrose diagrams for spacetimes with $M < 0$ (except $M=-1, J=0$ where there is no singularity) as well as spacetimes with $M = 0, J \neq 0$ (left) and flat space cosmologies (right).}
\label{fig:penrose}
\end{figure}
We take a closer look at orbits with constant representatives $\Theta^+(\varphi) = M$ and $\Xi^+(\varphi) = J$, summarized in \cref{fig:MJ}.
Here, $M$ and $J$ are, up to a factor, mass and angular momentum given by the charges~\cref{eq:charges}.
The factor is introduced to avoid clutter.
To recover true mass and angular momentum, use $M = \frac{\kappa}{\pi} M_{\mathrm{true}}$ and $J = \frac{\kappa}{\pi} J_{\mathrm{true}}$.
Then, at $\scri^+$ the metric is
\begin{align}
\label{eq:constrep}
  \dd s^2
  &= M \dd u^2 - 2 \dd u \dd r + J \dd u \dd \varphi + r^2 \dd \varphi^2
\end{align}
and similarly at $\scri^{-}$.
For strictly positive $M$ and nonvanishing $J$ the metric describes shifted boost orbifolds~\cite{Cornalba:2002fi,Cornalba:2003kd} which are quotients of Minkowski space.
They are also called flat space cosmologies and describe contracting and expanding phases separated by a region behind a cosmological horizon, see \cref{fig:penrose}.
They furthermore arise as a limit~\cite{Cornalba:2002fi} of Bañados-Teitelboim-Zanelli (BTZ) black holes~\cite{Banados:1992wn,Banados:1992gq}.
For vanishing $J$, we arrive at the boost orbifold~\cite{Khoury:2001bz,Seiberg:2002hr} with drastic changes in the geometric structure.
The spacetime where both $M$ and $J$ vanish is called the null-boost orbifold~\cite{Simon:2002ma,Liu:2002kb}.
In the last two cases there is a singularity between future and past infinity (see figures 5 and 9 in~\cite{Cornalba:2003kd}),
so a mapping for $M \ge 0, J = 0$ seems unreasonable.
The ``O-plane''~\cite{Cornalba:2003kd} consists of orbits with $M = 0, J \neq 0$.

For strictly negative mass (left Penrose diagram in \cref{fig:penrose}) we distinguish between angular deficit ($-1 < M <0$) and angular excess ($M < -1$) solutions.
Minkowski space is at $M = -1$, $J = 0$.
While there are no black holes in three-dimensional flat space~\cite{Ida:2000jh},
angular deficit solutions describe point particles (rotating for nonvanishing $J$) and can be seen as the three-dimensional analog to Kerr metrics~\cite{Deser:1983tn} (being axially symmetric vacuum solutions) or cosmic strings~\cite{Gott:1990zr}.

The linking of past and future null infinity presented in this chapter is valid for all spacetimes that admit a constant representative, excluding $M \ge 0, J = 0$ (the snake line in \cref{fig:MJ}).
From the discussion above, we see that this includes nearly all physically relevant spacetimes, with the exception of the two-parameter family of orbits admitting $\Theta = -1$ as well as orbits where $\Theta$ belongs to the massless deformation.

\section{Linking Past and Future Null Infinity}
\label{sec:linking}

We now construct the map between $\scri^+$ and $\scri^-$ for spacetimes discussed in the previous section.
For this purpose we first introduce explicit coordinate systems.
One coordinate system will cover a neighborhood around $\scri^+$, the other one a neighborhood around $\scri^-$.
The map we then construct sends points at $\scri^+$ to points at $\scri^-$.
Since one coordinate system does not cover both of these regions, we describe the position of the point at $\scri^+$ in one coordinate system, and the position of the corresponding point at $\scri^-$ in the other coordinate system.

We consider spacetimes that admit a constant representative at $\scri^+$.
The first coordinate system $(u, r, \varphi)$, that is introduced around $\scri^+$, is required to be such that the metric has the simple form~\eqref{eq:constrep}.
Notice that this coordinate system is defined only up to isometries of the spacetime.
Given this coordinate system we define the second coordinate system $(v, r, \varphi')$ around $\scri^-$ by the following transformations.
\begin{align}
\shortintertext{$M > 0$, $J \neq 0$:}
&\begin{aligned}
  \label{eq:trafoGt0}
  u
  &= \frac{2 r}{M} + v
    - \frac{J}{2 M^{3/2}} \ln \left( 1 + \frac{4 r \sqrt{M}}{J - 2 r \sqrt{M}} \right)
  \\
  \varphi
  &= \varphi'
    + \frac{1}{\sqrt{M}} \ln \left( 1 + \frac{4 r \sqrt{M}}{J - 2 r \sqrt{M}} \right)
\end{aligned} \\
\shortintertext{$M = 0, J \neq 0$:}
&\begin{aligned}
  \label{eq:trafoEq0}
  u
  &= - \frac{8 r^3}{3 J^2} + v
  \qquad
  \varphi
  = \varphi' + \frac{4 r}{J}
\end{aligned}\\
\shortintertext{$M < 0$:}
&\begin{aligned}
  \label{eq:trafoLt0}
  u
  &= \frac{2r}{M} + v - \frac{J}{(-M)^{3/2}} \arctan \left( \frac{J}{2 r \sqrt{-M}} \right)
  \\
  \varphi
  &= \varphi' - \frac{2}{\sqrt{-M}} \arctan \left( \frac{J}{2 r \sqrt{-M}} \right)
\end{aligned}
\end{align}
The coordinate transformations~\cref{eq:trafoGt0,eq:trafoEq0,eq:trafoLt0} are constructed such that the coordinates $(u, r, \varphi)$ cover $\scri^+$, while $(v, r, \varphi')$ cover $\scri^-$.
Apart from that, the form of the coordinate transformations is of no fundamental importance for the argument and they are chosen such that equations later in this section are particularly simple.
The transformations can be easily checked for Minkowski space ($M = -1, J = 0$).
Here, $u = t - r$ and $v = t + r$ are usual retarded and advanced times.
Depending on which one is held fixed, one ends up at either $\scri^+$ or $\scri^-$ as $r$ goes to infinity.
On other spacetimes with $M \neq 0$ this works analogously.
We now discuss the more complicated case of flat space cosmologies ($M > 0$, $J \neq 0$).

\begin{figure}
\centering
\includegraphics{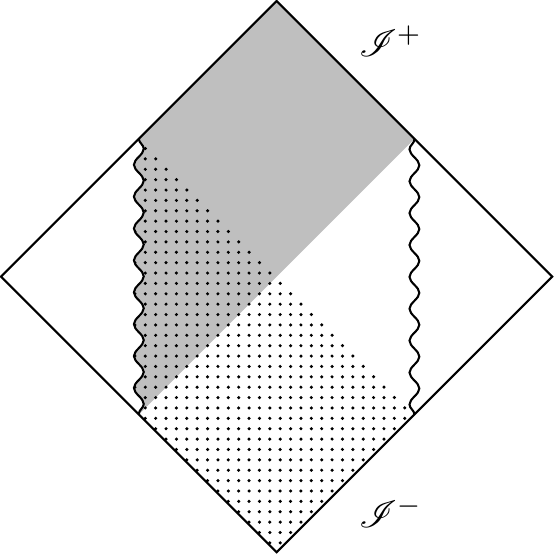}
\caption{Penrose diagram of a constant $Y$ slice of Minkowski space.
  The snake lines indicate where causal singularities develop when taking the quotient to obtain flat space cosmologies.
  The gray and the dotted regions mark different coordinate patches.}
\label{fig:fsc-quotient}
\end{figure}

Flat space cosmologies can be constructed as quotients of Minkowski space.
We use Cartesian coordinates $(T, X, Y)$ and define the coordinates ($u, r, \varphi$) with $r > 0$ by
\begin{align}
  \begin{aligned}
  T
  &= \frac{r}{\sqrt{M}} \cosh \left( \sqrt{M} \varphi \right)
    - \frac{J}{2M} \sinh \left( \sqrt{M} \varphi \right)
  \\
  X
  &= \frac{r}{\sqrt{M}} \sinh \left( \sqrt{M} \varphi \right)
    - \frac{J}{2M} \cosh \left( \sqrt{M} \varphi \right)
  \\
  Y
  &= \frac{1}{\sqrt{M}} \left(
    - r + M u + \frac{J \varphi}{2}
    \right)
    \,.
  \end{aligned}
\end{align}
The coordinates ($u, r, \varphi$) cover the region
\begin{align}
  \begin{aligned}
  - \sqrt{T^2 + \frac{J^2}{4M^2}} < X < T
  &\quad \text{if} \quad J > 0
  \\
  -T < X < \sqrt{T^2 + \frac{J^2}{4M^2}}
  &\quad \text{if} \quad J < 0
    \,,
  \end{aligned}
\end{align}
which, for $J > 0$, corresponds to the gray region in \cref{fig:fsc-quotient}.
The metric in these coordinates is
\begin{align}
  \dd s^2
  &= M \dd u^2 - 2 \dd u \dd r + J \dd u \dd \varphi + r^2 \dd \varphi^2
    \,.
\end{align}
Upon identifying
\begin{align}
  \varphi
  &\sim \varphi + 2 \pi
\end{align}
we end up with flat space cosmologies parametrized by $M$ and $J$.
The identifications are given in Cartesian coordinates as
\begin{align}
  \begin{pmatrix}
    T \\ X \\ Y
  \end{pmatrix}
  &\sim
  \begin{pmatrix}
    T \cosh( 2 \pi \sqrt{M} ) + X \sinh( 2 \pi \sqrt{M} ) \\
    X \cosh( 2 \pi \sqrt{M} ) + T \sinh( 2 \pi \sqrt{M} ) \\
    Y + \frac{\pi J}{\sqrt{M}}
  \end{pmatrix}
  \,,
\end{align}
corresponding to a boost in $X$ direction plus a translation in $Y$ direction.
This is why flat space cosmologies are also referred to as shifted boost orbifolds~\cite{Cornalba:2002fi,Cornalba:2003kd}.
At $r=0$, where $X^2 - T^2 = {\left(\frac{J}{2M}\right)}^2$, null-like separated points become identified, leading to a causal singularity there.

A similar coordinate system ($v, r, \varphi'$) can be defined as
\begin{align}
  \begin{aligned}
  T
  &= - \frac{r}{\sqrt{M}} \cosh \left( \sqrt{M} \varphi' \right)
    - \frac{J}{2M} \sinh \left( \sqrt{M} \varphi' \right)
  \\
  X
  &= - \frac{r}{\sqrt{M}} \sinh \left( \sqrt{M} \varphi' \right)
    - \frac{J}{2M} \cosh \left( \sqrt{M} \varphi' \right)
  \\
  Y
  &= \frac{1}{\sqrt{M}} \left(
    - r - M v - \frac{J \varphi'}{2}
    \right)
    \,,
  \end{aligned}
\end{align}
carefully chosen such that the identifications $\varphi' \sim \varphi' + 2 \pi$ correspond to the ones before.
This coordinate system covers the dotted region in \cref{fig:fsc-quotient}.
The metric becomes
\begin{align}
  \dd s^2
  &= M \dd v^2 + 2 \dd v \dd r + J \dd v \dd \varphi' + r^2 \dd \varphi'^2
    \,.
\end{align}
In the region where the two coordinate systems overlap, we find the coordinate transformation given by~\cref{eq:trafoGt0}.

We have now constructed and related our two coordinate systems.
The first one is defined up to isometries.
The second one is uniquely fixed by~\cref{eq:trafoGt0,eq:trafoLt0,eq:trafoEq0} once the first one is fixed.
We now define how points at $\scri^+$ are sent to points at $\scri^-$.

We send a point $A$ using coordinates $(u, r, \varphi)$ at $\scri^+$ to a point $B$ at $\scri^-$ using coordinates $(v, r, \varphi')$.
Any such map can be written as~\footnote{
This is different to the coordinate transformations~\cref{eq:trafoGt0,eq:trafoEq0,eq:trafoLt0}.
Plugging a point $P$ with the coordinates $(u_{P},r_{P},\varphi_{P})$ into the transformations~\cref{eq:trafoGt0,eq:trafoEq0,eq:trafoLt0} leads to the \emph{same} point just in other coordinates $(v_{P},r_{P},\varphi'_{P})$.}
  \begin{align}
    \label{eq:matching1}
    \begin{aligned}
      v_B &= f_1(u_A, \varphi_A)
      \\
      \varphi'_B &= f_2(u_A, \varphi_A)
      \\
      r_B &= r_A = \infty \,,
    \end{aligned}
  \end{align}
with some functions $f_1$ and $f_2$.
Since the coordinate system $(u, r, \varphi)$ is defined only up to isometries, one has to demand that the outcome of the mapping is independent of any such choice.
All spacetimes under consideration admit at least two isometries: Time translations, and rotations.
Time translations act as $u \to u + a$, and by~\cref{eq:trafoGt0,eq:trafoLt0,eq:trafoEq0}, also as $v \to v + a$.
Similarly, rotations act as $\varphi \to \varphi + b$ and $\varphi' \to \varphi' + b$.
Invariance under these isometries leads to the requirements that
\begin{align}
  \begin{aligned}
    f_1(u, \varphi) + a &= f_1(u + a, \varphi + b)
    \\
    f_2(u, \varphi) + b &= f_2(u + a, \varphi + b) \,,
  \end{aligned}
\end{align}
for all real numbers $a$ and $b$.
This almost fixes $f_1$ and $f_2$ and we find the invertible map
\begin{align}
\label{eq:matching2}
\begin{aligned}
  v_B
  &= u_A + c_1
  \\
  \varphi'_B
  &= \varphi_A + c_2
    \,,
\end{aligned}
\end{align}
with some constants $c_1$ and $c_2$.
The only invariant maps between $\scri^+$ and $\scri^-$ are of this form.

Now we fix the solely remaining freedom in our map, the constants $c_1$ and $c_2$.
To do this we consider Lorentz boosts on Minkowski space.
A Lorentz boost that is generated by a vector field~\footnote{
In Cartesian coordinates, the boost is generated by the vector field $t \pd_x + x \pd_t$, where $t = u + r = v - r$ and $x = r \cos \varphi$.
} $- u \cos \varphi \pd_u - \sin \varphi \pd_\varphi$ at $\scri^+$ is generated by $v \cos \varphi' \pd_v + \sin \varphi' \pd_{\varphi'}$ at $\scri^-$.
The map~\eqref{eq:matching2} is invariant under this boost if and only if $c_1 = 0$ and $c_2 = \pi$.
Considering any other boost leads to the same conclusion.
We find that Minkowski space admits a unique invariant map.
We take $c_1$ and $c_2$ to be independent of $M$ and $J$.
This does not follow from our symmetry considerations and is the only choice in the derivation.
Further investigation is required to determine if this choice is valid.
For now we stick to it due to its simplicity and for a lack of a better alternative.
The mapping prescription for spacetimes admitting constant representatives is then:
\begin{align}
\label{eq:matching3}
\begin{aligned}
  v_B
  &= u_A
  \\
  \varphi'_B
  &= \varphi_A + \pi
    \,.
\end{aligned}
\end{align}
Using symmetry arguments we found an antipodal relation in the angular coordinate as in the four-dimensional case~\cite{Strominger:2013jfa}.
Everything else falls into place.
A finite BMS transformation, parametrized by $\alpha$ and $f$, that acts on $\scri^+$ as
\begin{align}
\label{eq:BMS+m}
  \begin{aligned}
    u &\to u f'(\varphi) + \alpha(f(\varphi))
    \\
    \varphi &\to f(\varphi) \,,
  \end{aligned}
\end{align}
has to act with the same functions $\alpha$ and $f$ on $\scri^-$ as
\begin{align}
\label{eq:BMS-m}
  \begin{aligned}
    v &\to v f'(\varphi' - \pi) + \alpha(f(\varphi' - \pi))
    \\
    \varphi' &\to f(\varphi' - \pi) + \pi \,.
  \end{aligned}
\end{align}
This is the unique map between $\BMS^+$ and $\BMS^-$ that preserves the mapping~\cref{eq:matching3}.

Now we go back to the original goal of mapping asymptotic regions of spacetimes with any metric admitting a constant representative.
We take a metric that is given around $\scri^+$ as~\eqref{eq:3}.
By assumption we can apply a BMS transformation~\cref{eq:finBMS} to bring the metric into constant form~\cref{eq:constrep}.
Then we use the coordinate transformations~\cref{eq:trafoGt0,eq:trafoLt0,eq:trafoEq0} to find the metric around $\scri^-$
\begin{align}
  \dd s^2
  &= M \dd v^2 + 2 \dd v \dd r + J \dd v \dd \varphi' + r^2 \dd \varphi'^2
    \,.
\end{align}
Undoing the BMS transformation using the above relation between~\eqref{eq:BMS+m} and~\eqref{eq:BMS-m}, we finally get
a metric of the form~\eqref{eq:7}
with
\begin{align}
\label{eq:dofmatching}
  \begin{aligned}
    \Theta^+(\varphi) &= \Theta^-(\varphi + \pi)
    \\
    \Xi^+(\varphi) &= \Xi^-(\varphi + \pi) \,.
  \end{aligned}
\end{align}
From the definition of the charges~\cref{eq:charges} we immediately obtain infinitely many conservation laws,
\begin{align}
  H^+_{T,Y}
  &= H^-_{\tilde T, \tilde Y}
    \,,
\end{align}
one for every function $T(\varphi) = \tilde T(\varphi + \pi)$ and $Y(\varphi) = \tilde Y(\varphi + \pi)$.
The mapping is energy preserving: $H^+_{1,0} = H^-_{1,0}$.

\section{Adding Matter}
\label{sec:matter}

Up until now we have restricted ourselves to the vacuum solutions~\eqref{eq:3} and~\eqref{eq:7}.
Here we turn to the classical scattering problem of a massless field coupled to gravity, where initial and final data are prescribed on $\scri^-$ and $\scri^+$.
Both sets of data transform under each BMS group separately.
When considering BMS as a symmetry of the scattering problem, the separate symmetries of $\scri^+$ and $\scri^-$ must be broken to a single one.
Using the results of the vacuum case presented above, a similar mapping of symmetries can be achieved in the presence of matter, as follows.

We require that the solution to the Einstein equations admits some well-defined spacelike infinity $i^0$ and that there is vacuum in a neighborhood of $i^0$.
Thus in this neighborhood around $i^0$, the metric will have the form~\cref{eq:3} and~\cref{eq:7}.
Using the algorithm established above we can find a mapping between $\scri^+$ and $\scri^-$, and consequently a relation between the two respective symmetry groups $\BMS^+$ and $\BMS^-$ according to~\eqref{eq:BMS+m} and~\eqref{eq:BMS-m}.
This mapping is a priori valid only in the neighborhood of $i^0$, in which the coordinate system~\eqref{eq:constrep} is well-defined.
However, a BMS-transformation is determined on the entirety of $\scri^{\pm}$ by prescribing it on one cross section~\cite{Geroch:1977jn}.
The linking of $\BMS^+$ and $\BMS^-$ near $i^0$ is therefore enough to establish a linking on the whole of $\scri$, thus breaking the symmetry $\BMS^+\otimes\BMS^-$ to a single $\BMS$ acting on both $\scri^+$ and $\scri^-$.
In particular, the mapping~\eqref{eq:dofmatching} of the gravitational degrees of freedom near $i^0$ is still valid.
Given the flux of matter through $\scri^{\pm}$, these relations can be used as initial conditions for integrating the constraint equations along $\scri^\pm$, thus providing initial or final data for the scattering problem.

\chapter{Distinct Minkowski Spaces}
\label{cha:dist-mink-spac}

Previous calculations performed in a particular coordinate system suggested that a defect arises when acting with supertranslations on Minkowski space~\cite{Compere:2016jwb}.
In order to preserve the coordinate system the supertranslation generators had to be extended in a particular way into the bulk.
The extension of the generators was not smooth and led to apparent defects in the resulting spacetime.
At least since the coordinate-free formulation of BMS transformations (see \cref{sec:asymptotically-flat}), it is evident (see for example~\cite{Guica:2008mu,Barnich:2010eb}) that if one puts emphasis on the covariant description of the theory and does not fix a particular coordinate system, there is much more freedom in extending the generator.

We now show that defects can be avoided by extending vector fields generating BMS transformations into the interior of Minkowski space such that they are well defined and smooth everywhere~\cite{Scholler:2017uni}.
We first study the case of supertranslations.

\section{Supertranslations}
\label{sec:supertranslations-minkowski}

Consider $n$-dimensional Minkowski space in spherical coordinates with ``retarded time'' coordinate $u = t - r$
and inverse radial distance $y = 1/r$ such that the physical metric reads
\begin{align}
  g_{\sia\sib} \dd x^\sia \dd x^\sib
  &= - \dd u^2 + 2 y^{-2} \dd u \dd y + y^{-2} \dd \nu^2
    \,,
\end{align}
where $\dd \nu^2$ is the metric on the unit $(n-2)$-sphere.
We set the conformal factor $\conformalFactor = y$.
Any supertranslation according to~\cref{eq:bmsDef2} has the form
\begin{align}
  \xi
  &= (T(x^A) + O(y)) \pd_u
    + O(y^2) \pd_y
    + O(y) \pd_A
    \,,
\end{align}
with some smooth function $T$ depending on coordinates $x^A$ of the $(n-2)$-sphere.

\begin{figure}
  \centering
  \includegraphics{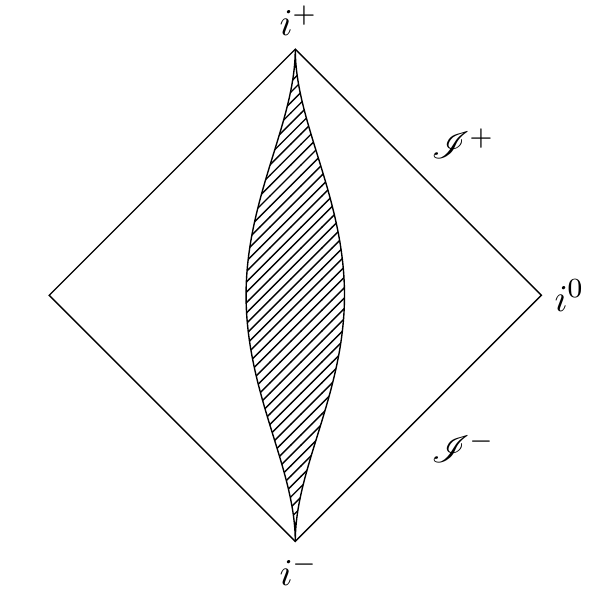}
  \caption{The vector field $\xi^\sia$ is constructed to vanish at the shaded region in Minkowski space.}
  \label{fig:vanishing-xi}
\end{figure}

Now we extend $\xi^\sia$ into the bulk. Since the difference between any two such extensions vanishes at $\scri$, all extensions of the same $\xi^\sia$ are in the same BMS equivalence class. We are free to extend $\xi^\sia$ however we see fit. A convenient choice is
\begin{align}
  \xi
  &= T(x^A) s(r) \pd_t
    \label{eq:xi}
    \,,
\end{align}
given in the coordinate system $(t, r, x^A)$.
Here $s$ is some smooth cutoff function on $\R$, satisfying
\begin{align}
  s(r)
  &=
    \begin{cases}
      0 & r < 1
      \\
      1 & r > 2
      \,,
    \end{cases}
\end{align}
and interpolating in an arbitrary but smooth way between 0 and 1 for $1 \le r \le 2$.
By construction, the vector field $\xi^\sia$ vanishes in some neighborhood around the line $r = 0$ (see \cref{fig:vanishing-xi}).
Therefore, the non-smoothness of $T(x^A)$ at $r = 0$, stemming from the non-smoothness of the coordinates $x^A$ there, is irrelevant.
Since $\pd_t$ is smooth, we conclude that $\xi^\sia$ is smooth everywhere in the bulk.
Since $\xi^\sia$ is time independent, it is simple to integrate, leading to the supertranslation given in coordinates by
\begin{align}
  t'
  &= t + T(x^A) s(r)
    \,.
\end{align}
Supertranslations can therefore be extended to globally well-defined diffeomorphisms.

\section{Lorentz Transformations}

For spacetime dimension bigger than three, the group of BMS transformations is a semidirect product between the Lorentz group and the supertranslations.
If we pick a Lorentz subgroup, we can write any BMS transformation as the product of an element of this Lorentz subgroup and a supertranslation.
On Minkowski spacetime there is a preferred choice to pick a Lorentz subgroup given by the requirement that the transformations are global isometries.
When we pick such a Lorentz subgroup, the combination of a Lorentz transformation and a supertranslation is also globally well-defined.
Since this gives us the whole group of BMS transformations, we conclude that all BMS transformations are globally well-defined diffeomorphisms when acting on Minkowski space with dimension bigger than three.

In three dimension the BMS group is bigger, containing also superrotations.
Generators of the subgroup consisting of supertranslations and the Lorentz transformations of the Minkowski vacuum under consideration are still well-defined in three dimensions.
The construction of globally well-defined superrotations is expected to work similarly to the case of supertranslations and should not cause any additional difficulty.

\chapter{Conclusion}

In \cref{cha:three-dimens-asympt} Hamiltonian functions of BMS symmetries in vacuum of general relativity on three-dimensional asymptotically flat spacetimes were constructed without reference to a coordinate system.
Boundary conditions on the energy momentum tensor were given such that after the introduction of additional matter fields the charges are well defined and obey a simple conservation law.
While it was shown that the $\dv$-derivatives of the charges are conserved for vacuum solutions as well as stationary solutions with a scalar field, the charges themselves are not all conserved.
This is related to the fact that they depend on the conformal factor used in the completion of spacetime.
It was argued that charges cannot be defined without reference to any background structure and that by addition of even more background structure one can make the charges conserved.
This is different to the four-dimensional case, where charges can be defined without introduction of any background structure~\cite{Wald:1999wa}.
It is not clear which of the two choices of background structures considered in this work is preferable.
This warrants further investigation.
While a large part of the derivation was given in a coordinate free manner, some calculations were performed in a coordinate system constructed using the background structure of the theory only.
This means among other things that no metric components in the bulk were fixed.
The fact that the coordinate system is independent of the metric is essential since the $\dv$-derivatives of components in a coordinate system depending on the metric do not reproduce the $\dv$-derivatives of coordinate free expressions.
Using invariance of the Hamiltonian functions under diffeomorphisms, central terms in the Poisson algebra were found for vacuum solutions of the theory.
Finally, it was also shown how the equations of motion for general relativity coupled to a scalar field can be solved order by order at $\scri$.

In \cref{cha:linking-past-future} a linking between future and past null infinity on three-dimensional asymptotically flat spacetimes that admit a constant representative (see \cref{fig:MJ}) was constructed.
The map given by~\cref{eq:matching3} together with~\cref{eq:trafoGt0,eq:trafoLt0,eq:trafoEq0} provides the linking between the two asymptotic regions and their respective symmetry groups.
An immediate consequence of this linking is the existence of an infinite number of conservation laws, expressed in~\eqref{eq:dofmatching}, which corresponds to conservation of energy and angular momentum at every angle.
In the context of flat space holography, the two functions $\Theta$ and $\Xi$ can be seen as components of the stress-tensor of the dual boundary theory~\cite{Barnich:2012aw,Barnich:2013yka,Carlip:2016lnw}.
Due to the matching presented in this work the two boundary theories defined on $\scri^+$ and $\scri^-$ are connected.
It would be of interest to verify the --- somewhat arbitrary --- assumption that $c_1$ and $c_2$ in~\cref{eq:matching2} are independent of $M$ and $J$.
The single $\BMS$ group obtained from the linking can be regarded as a symmetry for the S-matrix of three-dimensional Einstein gravity coupled to matter.
Further study is required to determine to what extent the relations between BMS symmetry, memory effect and soft theorems present in four dimensions~\cite{Strominger:2013jfa,He:2014laa,Strominger:2014pwa} are realized in three dimensions.

In \cref{cha:dist-mink-spac} it was shown that BMS transformations act as smooth diffeomorphisms on Minkowski space.
Each element of the orbit of BMS transformations acting on Minkowski space is therefore isometric to Minkowski space.
The different elements of this orbit can be regarded as different gravitational vacua~\cite{Ashtekar:1981hw,Strominger:2013jfa}.
Since they are all isometric to each other, they are locally indistinguishable from one another.
In three dimensions, however, since the action of BMS transformations on the conformal factor changes it in a way that changes the charges (see \cref{cha:three-dimens-asympt}), they have different superrotation charges.
Similarly in four dimensions~\cite{Barnich:2011mi} their superrotation charges differ.
We are left with multiple different Minkowski spaces, labeled in three and four dimensions by the values of their charges.
It also follows that we cannot expect to find sources for superrotation charges localized anywhere in spacetime.
This is consistent with the fact that, in a covariant formulation, charges can be defined only asymptotically, since there are no nontrivial, conserved $n-2$ forms~\cite{Barnich:1994db}.

\appendix
\addtocontents{toc}{\protect\setcounter{tocdepth}{0}}

\chapter{Differential Forms on Product Manifolds}
\label{cha:diff-forms-prod}

Consider the product $A \times B$ of two manifolds $A$ and $B$.
We define the space of $(r,s)$-forms on $A \times B$ as the tensor product
\begin{align}
  \Omega^{r,s}(A, B)
  &= \Omega^r(A) \otimes_\R \Omega^s(B)
\end{align}
of $r$-forms on $A$ and $s$-forms on $B$.
The exterior derivatives $\dd_A$ and $\dd_B$ of $A$ and $B$, respectively, act in a natural way on $\Omega^{r,s}(A, B)$.
For $\alpha \in \Omega^r(A)$ and $\beta \in \Omega^s(B)$ define
\begin{align}
  \dd_A (\alpha \otimes \beta)
  &=
    \dd_A \alpha \otimes \beta
  \\
  \dd_B (\alpha \otimes \beta)
  &=
    \alpha \otimes \dd_B \beta
    \,,
\end{align}
which by linearity extends to all of $\Omega^{r,s}(A, B)$.
With this convention it follows that the two exterior derivatives commute
\begin{align}
  \dd_A \dd_B = \dd_B \dd_A
  \,.
\end{align}
For $\alpha$ and $\beta$ as above and similarly $\alpha' \in \Omega^{r'}(A)$ and $\beta' \in \Omega^{s'}(B)$, we define
\begin{align}
  (\alpha \otimes \beta) \wedge (\alpha' \otimes \beta')
  &= \alpha \wedge \alpha' \otimes \beta \wedge \beta'
  \,,
\end{align}
so that for $\mu \in \Omega^{r,s}(A, B)$ and $\nu \in \Omega^{r',s'}(A, B)$
\begin{align}
  \dd_A (\mu \wedge \nu)
  &= \dd_A \mu \wedge \nu
    + (-)^r \mu \wedge \dd_A \nu
  \\
  \dd_B (\mu \wedge \nu)
  &= \dd_B \mu \wedge \nu
    + (-)^s \mu \wedge \dd_B \nu
    \,.
\end{align}
We denote by ``$\contract$'' contraction of a vector field with the first appropriate tensor slot.
For $\alpha$ and $\beta$ as above and vector fields $X \in \Gamma(T A)$ and $Y \in \Gamma(T B)$ this amounts to the definition
\begin{align}
  X \contract (\alpha \otimes \beta)
  &= (X \contract \alpha) \otimes \beta
  \\
  Y \contract (\alpha \otimes \beta)
  &= \alpha \otimes (Y \contract \beta)
    \,.
\end{align}
It follows that contraction with a vector field of one of the manifolds $A$ and $B$ commutes with the exterior derivative on the other one:
for $X \in \Gamma(TA)$, $Y \in \Gamma(TB)$, and $\omega \in \Omega^{r,s}(A, B)$
\begin{align}
  X \contract \dd_B \omega
  &= \dd_B ( X \contract \omega)
  \\
  Y \contract \dd_A \omega
  &= \dd_A ( Y \contract \omega)
  \,.
\end{align}
Any other differential operator $D_A \colon \Omega^r(A) \to \Omega^{r'}(A)$, or $D_B \colon \Omega^s(B) \to \Omega^{s'}(B)$ is extended similarly to act on $\Omega^{r,s}(A, B)$.
For example, the Lie derivative along a vector field $X \in \Gamma(TA)$ or $Y \in \Gamma(TA)$ acts as
\begin{align}
  \lied_X (\alpha \otimes \beta)
  &= (\lied_X \alpha) \otimes \beta
  \\
  \lied_Y (\alpha \otimes \beta)
  &= \alpha \otimes (\lied_Y \beta)
    \,,
\end{align}
so that we obtain two versions of Cartan's formula
\begin{align}
  \lied_X \omega
  &= X \contract \dd_A \omega
    + \dd_A (X \contract \omega)
  \\
  \lied_Y \omega
  &= Y \contract \dd_B \omega
    + \dd_B (Y \contract \omega)
    \,.
\end{align}

\chapter{Cohomology of Local Forms}
\label{cha:cohom-local-forms}

\begin{figure}[h]
\centering
\includegraphics{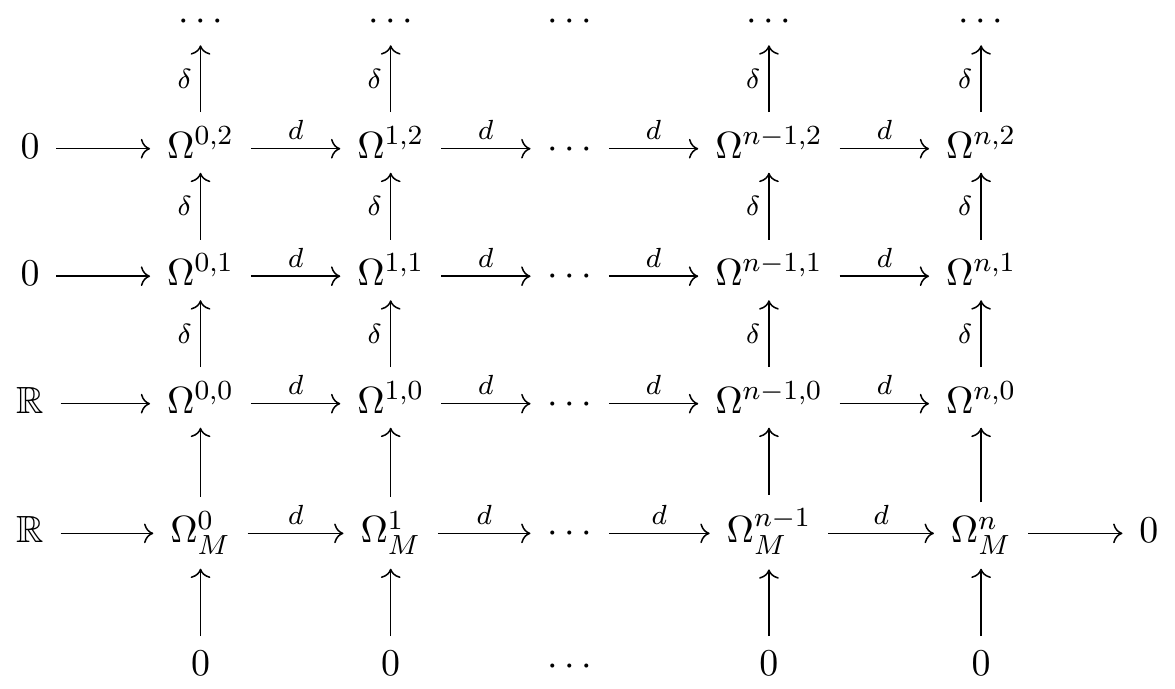}
\caption{The double complex of local forms.}
\label{fig:bicomplex}
\end{figure}

In this chapter, some statements on the cohomology of local forms as introduced in \cref{cha:symmetries} are presented.
For details see~\cite{anderson_bicomplex_1989}.
The two differentials $\dd$ and $\dv$ give rise to the double complex as shown in \cref{fig:bicomplex}.

\begin{theorem}
\label{thm:bicomplex-exact}
All rows and columns of the double complex in \cref{fig:bicomplex} are locally exact.
\end{theorem}
Applied to the lowest row in \cref{fig:bicomplex}, involving $\Omega_M^\ast$, this theorem is the Poincaré lemma.
For rows above, involving $\Omega^{\ast,\ast}$, the theorem is called the algebraic Poincaré lemma.
For rows involving $\Omega^{\ast,s}$ with $s \ge 1$ a stronger, global statement can be made:
\begin{theorem}
\label{thm:horizontal-complex-exact}
For each $s \ge 1$ the horizontal complex
\begin{center}
  \includegraphics{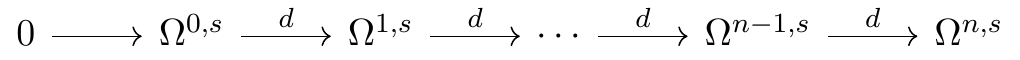}
\end{center}
is globally exact.
\end{theorem}

In other words, \cref{thm:horizontal-complex-exact} says that for any form $\omega \in \Omega^{r,s}$ with $s \ge 1$
\begin{align}
  \dd \omega = 0
  &\Leftrightarrow
  \omega = \dd \kappa
  & &\text{for $0 < r < n$}
  \\
  \dd \omega = 0
  &\Leftrightarrow
  \omega = 0
  & &\text{for $r = 0$}
  \,,
\end{align}
where $\kappa \in \Omega^{r-1,s}$.
The invariant homotopy operator used to prove exactness in \cref{thm:horizontal-complex-exact} when evaluated at a point $\phi \in \confSpace$ depends on $\omega$ at $\phi$ only~\cite{anderson_bicomplex_1989}.
This makes it possible to consider forms only at the space $\solutions$ of solutions to the equations of motion and formulate the following variant of the theorem:
\begin{theorem}
\label{thm:horizontal-complex-exact-onshell}
For any form $\omega \in \Omega^{r,s}$ with $s \ge 1$
\begin{align}
  \text{$\dd \omega = 0$ at $\solutions$}
  &\Leftrightarrow
  \text{$\omega = \dd \kappa$ at $\solutions$}
  & &\text{for $0 < r < n$}
  \\
  \text{$\dd \omega = 0$ at $\solutions$}
  &\Leftrightarrow
  \text{$\omega = 0$ at $\solutions$}
  & &\text{for $r = 0$}
  \,.
\end{align}
\end{theorem}
It is necessary to require that $\dd \omega = 0$ at $\solutions$, which is a stronger condition than $\dd \omega \approx 0$ (see \cref{cha:conventions}).
The weaker condition is not sufficient to conclude that $\omega \approx \dd \kappa$.

\chapter{Conformal Transformations at the Boundary}
\label{cha:conformal-boundary}

We discuss the consequences of Einstein's equation and conformal completion as introduced in \cref{sec:conf-comp}.
Let $\tilde M$ be an $n$-dimensional manifold with boundary and with a metric $\tilde g_{\sia\sib}$.
Denote the interior of $\tilde M$ by $M$ and its boundary by $\scri$.
Pick any smooth function $\conformalFactor$ on $\tilde M$ that vanishes precisely at $\scri$, such that the normal vector
$\tilde n^\sia = \tilde g^{\sia\sib} \cd_\sib \conformalFactor$
is non-vanishing on $\scri$.
On $M$ define the metric $g_{\sia\sib}$ conformally related to $\tilde g_{\sia\sib}$ as
\begin{align}
  \tilde g_{\sia\sib}
  &= \conformalFactor^2 g_{\sia\sib}
    \,.
\end{align}

We denote by ``$\bdryeq$'' equality in the limit as $\scri$ is approached.
Indices of tensors with and without tilde are raised and lowered with $\tilde g_{\sia\sib}$ and $g_{\sia\sib}$, respectively.
From the behavior of the Ricci tensor under conformal transformations
\begin{align}
  \tilde R_{\sia\sib}
  &= R_{\sia\sib}
    - (n - 2) \, \conformalFactor^{-1} \tilde \cd_\sia \tilde n_\sib
    - \conformalFactor^{-1} \tilde g_{\sia\sib} \tilde \cd_\sic \tilde n^\sic
    + (n - 1) \, \conformalFactor^{-2} \tilde g_{\sia\sib} \tilde n_\sic \tilde n^\sic
    \label{eq:ricci-conf} \,,
\end{align}
and its trace
\begin{align}
  \tilde R
  &= \conformalFactor^{-2} R
    - 2 (n - 1) \, \conformalFactor^{-1} \tilde \cd_\sia \tilde n^\sia
    + n (n - 1) \, \conformalFactor^{-2} \tilde n_\sia \tilde n^\sia
    \,,
\end{align}
where $\tilde\cd$ is the Levi-Civita connection with respect to $\tilde g_{\sia\sib}$.
It follows that
\begin{align}
  R
  &\bdryeq
  - n(n-1) \tilde n^\sia \tilde n_\sia
    \label{eq:norm-n-from-R}
  \\
  \conformalFactor^2 R_{\sia\sib}
  &\bdryeq \frac{1}{n} R \, \tilde g_{\sia\sib}
  \\
  \conformalFactor \left( R_{\sia\sib} - \frac{1}{n} R \, g_{\sia\sib} \right)
  &\bdryeq
    (n - 2) \left(
    \tilde\cd_\sia \tilde n_\sib
    - \frac{1}{n} \tilde g_{\sia\sib} \tilde \cd_\sic \tilde n^\sic
    \right)
    \label{eq:bla}
    \,,
\end{align}
so $R$, as well as the traceless part of $\conformalFactor R_{\sia\sib}$ have a smooth limit to $\scri$.
Furthermore:
\begin{align}
  \conformalFactor^{-1} \left(
    \tilde n^\sia \tilde n_\sia
    + \frac{1}{n(n-1)} R
    \right)
  &\bdryeq
    \frac{2}{n} \tilde \cd_\sia \tilde n^\sia
    \label{eq:nnDivN}
\end{align}
We see from~\cref{eq:norm-n-from-R} that $R$ has to be finite at $\scri$ and its value determines if $\scri$ is timelike, spacelike, or null.
Consider spacetimes that obey Einstein's equation with spacetime dimension $n>2$,
\begin{align}
  R_{\sia\sib} - \frac{1}{2} R \, g_{\sia\sib} + \Lambda \, g_{\sia\sib}
  = \kappa T_{\sia\sib}
  \,.
\end{align}
From~\cref{eq:ricci-conf} it follows that the condition that the trace of the energy momentum tensor vanishes at $\scri$, $T \bdryeq 0$, is equivalent to $\conformalFactor^2 T_{\sia\sib} \bdryeq 0$.
If these hold, we find that
\begin{align}
  \tilde n^\sia \tilde n_\sia
  &\bdryeq
    - \frac{2}{(n-2)(n-1)} \Lambda
    \,.
\end{align}
When considering matter for $n=4$ it is normally not unreasonable~\cite{Geroch:1977jn} that $\conformalFactor^{-1} T_{\sia\sib} \bdryeq 0$, although for the scalar field we only have $\conformalFactor T_{\sia\sib} \bdryeq 0$.

If $\conformalFactor T_{\sia\sib} \bdryeq 0$, it follows from~\cref{eq:bla} using the identity
\begin{align}
  R_{\sia\sib} - \frac{1}{n} R g_{\sia\sib}
  = \kappa \left( T_{\sia\sib} - \frac{1}{n} T g_{\sia\sib} \right)
\end{align}
that
\begin{align}
  \tilde\cd_\sia \tilde n_\sib
  \bdryeq \frac{1}{n} \tilde g_{\sia\sib} \tilde \cd_\sic \tilde n^\sic \label{eq:shear-free-follows} \,.
\end{align}

\section{Scalar Field}
\label{sec:scalar-field-conf}

Consider a massless scalar field with action
\begin{align}
  L
  &= - \frac{1}{2} \cd^\sia \Phi \cd_\sia \Phi \, \epsilon
  \,,
\end{align}
and assume $\tilde\Phi$ to be smooth where
\begin{align}
  \tilde\Phi &= \conformalFactor^{1-n/2} \Phi \,.
\end{align}
The energy momentum tensor is given by
\begin{align}
  T_{\sia\sib}
  &=
    \cd_\sia \Phi \cd_\sib \Phi
    - \frac{1}{2} g_{\sia\sib} {(\cd \Phi)}^2
  \\
  &=
    \conformalFactor^{n-4} {\left( \frac{n-2}{2} \right)}^2 \left(
    \tilde n_\sia \tilde n_\sib
    - \frac{1}{2} \tilde g_{\sia\sib} \tilde n^\sic \tilde n_\sic
    \right) \tilde\Phi^2
  \nonumber \\
  &\quad
    + \conformalFactor^{n-3} (n-2) \left(
    \tilde n_{(\sia} \tilde\cd_{\sib)} \tilde\Phi
    - \frac{1}{2} \tilde g_{\sia\sib} \tilde n^\sic \tilde\cd_\sic \tilde\Phi
    \right) \tilde\Phi
  \nonumber \\
  &\quad
    + \conformalFactor^{n-2} \left(
    \tilde\cd_\sia \tilde\Phi \tilde\cd_\sib \tilde\Phi
    - \frac{1}{2} \tilde g_{\sia\sib}  {(\tilde\cd \tilde\Phi)}^2
    \right)
    \,.
\end{align}
For $n=3$ this is not sufficient for~\cref{eq:shear-free-follows} to hold.
The Ricci tensor and scalar curvature obey
\begin{align}
  \frac{1}{\kappa} R_{\sia\sib}
  &\approx
  \cd_\sia \Phi \cd_\sib \Phi
  \\
  &= \conformalFactor^{n-4} {\left( \frac{n-2}{2} \right)}^2 \tilde n_\sia \tilde n_\sib \tilde\Phi^2
  + \conformalFactor^{n-3} (n-2) \tilde n_{(\sia} \tilde\cd_{\sib)} \tilde\Phi \tilde\Phi
  + \conformalFactor^{n-2} \tilde\cd_\sia \tilde\Phi \tilde\cd_\sib \tilde\Phi
  \\
  \frac{1}{\kappa} R
  &\approx \conformalFactor^{n-2} {\left( \frac{n-2}{2} \right)}^2 \tilde n^\sia \tilde n_\sia \tilde\Phi^2
  + \conformalFactor^{n-1} (n-2) \tilde n^\sia \tilde\cd_\sia \tilde\Phi \tilde\Phi
  + \conformalFactor^n \tilde\cd^\sia \tilde\Phi \tilde\cd_\sia \tilde\Phi
  \,.
\end{align}

\chapter{Conventions and Useful Formulae}
\label{cha:conventions}

Greek letters from the middle of the alphabet ($\mu, \nu, \sigma, \dots$) are used as spacetime indices.
Symmetrization and anti-symmetrization of tensors is defined such that these operations do not change already symmetric and antisymmetric tensors, respectively:
\begin{align}
  t^{(\sia\sib)}
  &= \frac{1}{2} ( t^{\sia\sib} + t^{\sib\sia} )
  \\
  t^{[\sia\sib]}
  &= \frac{1}{2} ( t^{\sia\sib} - t^{\sib\sia} )
\end{align}
Contraction of a vector with the leftmost tensor slot of any $p$-form $\omega$ is denoted by a dot:
\begin{align}
  (t \contract \omega)_{\sia_2 \cdots \sia_p}
  &= t^{\sia_1} \omega_{\sia_1 \cdots \sia_p}
\end{align}
The forms $\omega$ and $\sigma$ on the space of field configurations $\confSpace$ are defined to satisfy the relation
\begin{align}
  \omega
  &\approx \sigma
  \,,
\end{align}
if the pullback to the solution subspace $\solutions$ of $\omega$ is equal to the pullback to $\solutions$ of $\sigma$.
We make the important distinction to the condition that $\omega$ equals $\sigma$ at all points of $\solutions$, or
\begin{align}
  \omega
  &= \sigma
  \quad \text{at}
  \quad \solutions
  \,,
\end{align}
which in general is stronger.
For zero-forms, however, the two notions of equality agree.
A similar distinction is made for forms on a conformally completed spacetime.
There the pullback to its boundary $\scri$ is denoted by an underline, i.e.
\begin{align}
  \underline{\omega}
  &= \underline{\sigma}
\end{align}
means that the pullback of $\omega$ to $\scri$ equals the pullback of $\sigma$ to $\scri$.
On the other hand,
\begin{align}
  \omega
  &\bdryeq \sigma
\end{align}
denotes that $\omega$ equals $\sigma$ at points of $\scri$, smoothly extending the domain of $\omega$ and $\sigma$ to $\scri$ if necessary.

\section{General Relativity}

The same sign conventions for the Ricci tensor as in~\cite{wald_general_1984,misner_gravitation_1973,hawking_large_1973} is used throughout the text:
\begin{align}
  R\indices{^\sic_\sid_\sia_\sib}
  &=
    \partial_\sia \Gamma\indices{^\sic_\sid_\sib}
    - \partial_\sib \Gamma\indices{^\sic_\sid_\sia}
    + \Gamma\indices{^\sic_\sie_\sia} \Gamma\indices{^\sie_\sid_\sib}
    - \Gamma\indices{^\sic_\sie_\sib} \Gamma\indices{^\sie_\sid_\sia}
  \\
  R_{\sia\sib}
  &= R\indices{^\sic_\sia_\sic_\sib}
\end{align}

\section{Differential Forms}

\newcommand*{\diffvol}{{\boldsymbol{\epsilon}}}
\newcommand*{\formA}{{\boldsymbol{\alpha}}}
\newcommand*{\formB}{{\boldsymbol{\beta}}}
\newcommand*{\formAdeg}{{a}}
\newcommand*{\formBdeg}{{b}}

In this section we use the following symbols: \\[0.5em]
\begin{tabular}{cl}
  $\formA$ & $\formAdeg$-form \\
  $\formB$ & $\formBdeg$-form \\
  $\diffvol$ & volume form associated with the metric \\
  $\pd_\sia$ & torsion-free covariant derivative \\
  $\cd_\sia$ & torsion-free and metric compatible covariant derivative \\
  $v$ & vector field \\
  $\mathbf{e}^i$ & co-basis (not necessarily orthonormal) \\
  $s$ & number of minuses in the signature of the metric \\
  $n$ & manifold dimension \\
\end{tabular}
\\[1em]
The exterior derivative and the Hodge star operator are defined as follows
\begin{align}
  \left( \formA \wedge \formB \right)_{\sia_1 \cdots\, \sia_\formAdeg \sib_1 \cdots\, \sib_\formBdeg}
  &= \frac{(\formAdeg+\formBdeg)!}{\formAdeg!\formBdeg!} \formA_{[\sia_1 \cdots\, \sia_\formAdeg} \formB_{\sib_1 \cdots\, \sib_\formBdeg]}
  \\
  \left(\dd \formA \right)_{\sia \sib_1 \cdots\, \sib_\formAdeg}
  &= (\formAdeg+1) \pd_{[\sia} \formA_{\sib_1 \cdots\, \sib_\formAdeg]}
  \\
  \left( \hodge \formA \right)_{\sia_1 \cdots\, \sia_{n-\formAdeg}}
  &= \frac{1}{\formAdeg!} \formA^{\sib_1 \cdots\, \sib_\formAdeg} \diffvol_{\sib_1 \cdots\, \sib_\formAdeg \sia_1 \cdots\, \sia_{n-\formAdeg}}
  \,.
\end{align}
The following relations are immediate consequences
\begin{align}
  v \contract ( \formA \wedge \formB )
  &= ( v \contract \formA ) \wedge \formB
    + (-1)^{\formAdeg} \formA \wedge ( v \contract \formB )
  \\
  \left( \hodge^{\scriptscriptstyle-1} \formA \right)_{\sia_1 \cdots\, \sia_{n-\formAdeg}}
  &= (-1)^{s+\formAdeg(n-\formAdeg)} \frac{1}{\formAdeg!} \formA^{\sib_1 \cdots\, \sib_\formAdeg} \diffvol_{\sib_1 \cdots\, \sib_\formAdeg \sia_1 \cdots\, \sia_{n-\formAdeg}}
  \\
  \hodge { \hodge \formA }
  &= (-1)^{s+\formAdeg(n-\formAdeg)} \formA
  \\
  \diffvol^{\sia_1 \cdots\, \sia_n} \diffvol_{\sia_1 \cdots\, \sia_n}
  &= (-1)^s \, n!
  \\
  \diffvol^{\sia_1 \cdots\, \sia_n} \diffvol_{\sib_1 \cdots\, \sib_n}
  &= (-1)^s \, n! \, \delta^{[\sia_1}_{\sib_1} \cdots \delta^{\sia_n]}_{\sib_n}
  \\
  \diffvol^{\sic_1 \cdots\, \sic_r \sia_1 \cdots\, \sia_{n-r}} \diffvol_{\sic_1 \cdots\, \sic_r \sib_1 \cdots\, \sib_{n-r}}
  &= (-1)^s \, (n-r)! \, r! \, \delta^{[\sia_1}_{\sib_1} \cdots \delta^{\sia_{n-r}]}_{\sib_{n-r}}
  \\
  \formA
  &= \frac{1}{\formAdeg!} \formA_{i_1 \cdots\, i_\formAdeg} \mathbf{e}^{i_1} \wedge \dots \wedge \mathbf{e}^{i_\formAdeg}
  \\
  \hodge ( \mathbf{e}^{i_1} \wedge \dots \wedge \mathbf{e}^{i_r} )
  &= \frac{1}{(n-r)!} \diffvol^{i_1 \cdots\, i_r}{}_{j_{r+1} \cdots\, j_n} \mathbf{e}^{j_{r+1}} \wedge \dots \wedge \mathbf{e}^{j_n}
  \\
  \big({ \hodge (\formA \wedge \formB) }\big)_{\sic_1 \cdots\, \sic_{n-\formAdeg-\formBdeg}}
  &= \frac{1}{\formAdeg!\formBdeg!} \formA^{\sia_1 \cdots\, \sia_\formAdeg} \formB^{\sib_1 \cdots\, \sib_\formBdeg} \diffvol_{\sia_1 \cdots\, \sia_\formAdeg \sib_1 \cdots\, \sib_\formBdeg \sic_1 \cdots\, \sic_{n-\formAdeg-\formBdeg}}
  \\
  \big( \hodge^{\scriptscriptstyle-1} ( \formA \wedge \hodge \formB ) \big)_{\sib_1 \cdots\, \sib_{\formBdeg-\formAdeg}}
  &= (-1)^{\formAdeg(\formAdeg+\formBdeg)} \frac{1}{\formAdeg!} \formA^{\sia_1 \cdots\, \sia_\formAdeg} \formB_{\sia_1 \cdots\, \sia_\formAdeg \sib_1 \cdots\, \sib_{\formBdeg-\formAdeg}} \quad (\text{if $\formBdeg \ge \formAdeg$})
  \\
  \big({ \hodge ( \hodge^{\scriptscriptstyle-1} \formA \wedge \formB ) }\big)_{\sib_1 \cdots\, \sib_{\formAdeg-\formBdeg}}
  &= \frac{1}{\formBdeg!} \formA_{\sia_1 \cdots\, \sia_\formBdeg \sib_1 \cdots\, \sib_{\formAdeg-\formBdeg}} \formB^{\sia_1 \cdots\, \sia_\formBdeg} \quad (\text{if $\formAdeg \ge \formBdeg$})
  \\
  \formA \wedge \hodge \formB
  &= \frac{1}{\formAdeg!} \formA^{\sia_1 \cdots\, \sia_\formAdeg} \formB_{\sia_1 \cdots\, \sia_\formAdeg} \diffvol \quad (\text{if $\formBdeg = \formAdeg$})
  \\
  ( \hodge^{\scriptscriptstyle-1} { \dd { \hodge\formA } } )_{\sia_1 \cdots\, \sia_{\formAdeg-1}}
  &= \cd^\sib \formA_{\sia_1 \cdots\, \sia_{\formAdeg-1} \sib}
  \\
  \dd ( v \contract \diffvol )
  &= \pd_\sia ( v^\sia \diffvol )
  = \cd_\sia v^\sia \diffvol
  \,.
\end{align}

\section{Symplectic Geometry}

In order for the map from Hamiltonian functions to vector fields to be a Lie algebra homomorphism instead of an antihomomorphism we define the sign of the Poisson bracket such that
\begin{align}
  \pb{H}{K}
  = X_H(K)
  \,.
\end{align}
The sign of the symplectic structure is fixed by the relation
\begin{align}
  \pb{H}{K}
  &= \symplecticStructure(X_H, X_K)
  \,.
\end{align}
The signs of the Poisson bracket and the symplectic structure are same as used by Woodhouse~\cite{woodhouse_geometric_1997}
but differ from the ones typically used in classical mechanics.
For a point particle with position $q$ and momentum $p$ the Poisson bracket is given by
\begin{align}
  \pb{H}{K}
  &= \frac{\pd H}{\pd p} \frac{\pd K}{\pd q}
    - \frac{\pd H}{\pd q} \frac{\pd K}{\pd p}
  \,.
\end{align}
This definition is also consistent with the sign of the presymplectic structure defined on the covariant phase space in \cref{sec:covar-phase-space} and matches the convention of Lee \& Wald~\cite{Lee:1990nz} as well as Wald \& Zoupas~\cite{Wald:1999wa}.

\bibliographystyle{utphys}
\bibliography{phd-thesis}
\addcontentsline{toc}{chapter}{\bibname}

\end{document}